\documentclass[onecolumn,10pt]{article}
\usepackage{graphicx}
\usepackage{dcolumn}
\usepackage{bm}
\usepackage{amssymb}
\usepackage[colorlinks=true,linktocpage=true]{hyperref}
\usepackage{amsmath,amsfonts,amsthm,mathrsfs}
\usepackage{tikz}
\usepackage{enumerate}
\usetikzlibrary{calc,decorations.markings}

\usepackage{subfig}

\usepackage[titletoc]{appendix}
\usepackage[labelfont=bf]{caption}
\usepackage[affil-it]{authblk}
\usepackage{subfig}
\newcommand{\be}{\begin{equation}}
\newcommand{\ee}{\end{equation}}
\newcommand{\bea}{\begin{eqnarray}}
\newcommand{\eea}{\end{eqnarray}}
\newcommand{\rmd}{\mathrm{d}}
\newcommand{\bes}{\begin{equation*}}
\newcommand{\ees}{\end{equation*}}
\usepackage{color}
\numberwithin{equation}{section}
\def\mathbi#1{\textbf{\em #1}}
\newcommand{\gra}{\alpha}
\newcommand{\grb}{\beta}
\newcommand{\grg}{\gamma}
\newcommand{\grd}{\delta}
\newcommand{\gre}{\epsilon}
\newcommand{\grz}{\zeta}
\newcommand{\grh}{\eta}
\newcommand{\gru}{\theta}

\newcommand{\grk}{\kappa}
\newcommand{\grl}{\lambda}
\newcommand{\grm}{\mu}
\newcommand{\grn}{\nu}
\newcommand{\grj}{\xi}
\newcommand{\grp}{\pi}
\newcommand{\grr}{\rho}
\newcommand{\grs}{\sigma}
\newcommand{\grt}{\tau}
\newcommand{\gry}{\upsilon}
\newcommand{\grf}{\phi}
\newcommand{\grx}{\chi}
\newcommand{\grc}{\psi}
\newcommand{\grw}{\omega}
\newcommand{\grS}{\Sigma}
\newcommand{\grG}{\Gamma}

\newcommand{\nn}{\nonumber}

\title{\vspace{-1cm}\bf{Gravity as a thermodynamic phenomenon}}
\author{M.Sc. Thesis \\\vspace{0.2cm} Dimitris Moustos%
\thanks{Electronic address: \texttt{dmoustos@upatras.gr}}}
\affil{Department of Physics, University of Patras, Greece}
\date{}
\begin{document}

\maketitle
\vspace{-1cm}
\begin{abstract}
The analogy between the laws of black hole mechanics and the laws of thermodynamics led Bekenstein and Hawking to argue that black holes should be considered as real thermodynamic systems that are characterised by entropy and temperature. Black hole thermodynamics indicates a deeper connection between thermodynamics and gravity. We review and examine in detail the arguments that suggest an interpretation of gravity itself as a thermodynamic theory.
\end{abstract}
{\hypersetup{linkcolor=blue}
\tableofcontents}
\begin{section}{Intoduction}
In 1687, Sir Isaac Newton published \textit{Philosophi\ae \ Naturalis Principia Mathematica}. In this three-volume book, Newton introduced the concepts of absolute space and time and formulated the three laws of classical mechanics and the law of universal gravitation. The law of universal gravitation states that every point mass $M$ attracts any other point mass $m$ with a gravitational force $\mathbf{F}$ that is proportional to the product of their masses, and inversely proportional to the square of their distance $\mathbf{r}=r\hat{\mathbf{r}}$:
\be
\mathbf{F}=-G\frac{Mm}{r^{2}}\hat{\mathbf{r}},
\ee
where $G$ is the Newton's gravitational constant. Acting on a particle of mass $m$ the gravitational force accelerates it according to Newton's second law $\mathbf{F}=m\mathbf{a}$. The Newtonian theory of gravity can also be described in a way analogous to electrostatics, if one introduces a gravitational potential $\Phi$. Then, Poisson's equation
\be\label{newtongrav}
\nabla^2\Phi=4\pi G\grr,
\ee
is valid, where $\grr$ is the mass density of an arbitrary continuous distribution of matter that generates the  potential. The acceleration of a body in the potential $\Phi$ is given by the latter's gradient: $\mathbf{a}=-\nabla\Phi$. Newton employed his laws to explain Kepler's laws of planetary motion and formulated the principles of kinematics of terrestrial bodies---with the study of which Galileo had dealt several years earlier.

The formulation by Maxwell in about 1865 of 
the equations of electrodynamics came to challenge Newton's ideas about space and time. Moreover, in 1859 Le Verrier had reported a discrepancy between the observed rate of precession of the perihelion of Mercury's orbit and the theoretically---calculated within the framework of Newtonian theory---expected.
Nowadays, it is known that Newton's laws are inappropriate for use at very small scales and at very high velocities.

The concepts of space and time were unified by Albert Einstein, who introduced {\em spacetime} as a fundamental notion of his theory of Special Relativity. Einstein proposed  Special Relativity in 1905 \cite{SR}. He constructed his new theory  based on two postulates: (i) the speed of light $c$ is the same in any inertial frame and (ii) the laws of physics are invariant in any inertial frame. He also modified properly the Newtonian laws of mechanics so that they would be invariant under  Lorentz transformations--- as was the equations of electrodynamics--- and consistent with the principles of special relativity. 

Inconsistency of Newton's gravitational law with  Special Relativity led Einstein to develop, in 1915 \cite{GR}, the theory of General Relativity. In General Relativity, gravity is no longer regarded as a force, but as a manifestation of the curvature of the spacetime. Spacetime's curvature is generated by the presence of matter. Einstein, in order to formulate his theory, based on two principles: the Equivalence Principle and the Principle of General Covariance. The equivalence principle states that at every spacetime point in an arbitrary gravitational field, a locally inertial coordinate system can be chosen, such that, within a sufficiently small region of this point, all physics laws take the form of those of Special Relativity. The principle of general covariance states that the equations that express the laws of physics should be generally covariant, i.e., they should preserve their form under general coordinate transformations \cite{Carroll,Weinberg}.

The content of General Relativity is summarized as follows \cite{Gravitation, Wald}. Spacetime is a four-dimensional manifold $\mathcal{M}$  endowed with a pseudo-Riemannian metric $g_{\grm\grn}$. The curvature of spacetime is related to the matter distribution existed in it by the Einstein's equation\footnote{We use the metric signature $(-+++)$. Greek indices take the values $\{0,1,2,3\}$, whereas Latin indices denotes spatial coordinates and take the values $\{1,2,3\}$. We use units $G=\hbar=c=k_B=1$ unless otherwise specified.}
\be\label{GR}
G_{\grm\grn}\equiv R_{\grm\grn}-\frac{1}{2}Rg_{\grm\grn}=\frac{8\pi G}{c^4} T_{\grm\grn},
\ee
where $G_{\grm\grn}$ is the Einstein tensor, $R_{\grm\grn}$ is the Ricci tensor, $R$ is the Ricci scalar and $T_{\grm\grn}$ is the stress-energy tensor. In vacuum, the Einstein's equation is reduced to $R_{\grm\grn}=0$. Newtonian gravity's equation (\ref{newtongrav}) is obtained by (\ref{GR}) in the limit of a weak gravitational field and slowly moving matter.

General Relativity gives the correct value of the precession of Mercury's perihelion. This was probably the first success (and confirmation) of the theory. The presence of gravitational fields causes also the bending of light. The correct value---calculated in the framework of Einstein's theory--- of the bending of light was confirmed by Sir Arthur Eddington during the total solar eclipse on May 29, 1919. Other significant results of General Relativity are the prediction of the existence of gravitational waves and the existence of black holes as solutions to Einstein's equation.

A {\em black hole} is a region of spacetime where the gravitational field is so strong that even light cannot escape from its horizon, i.e., the boundary of the black hole. Schwarzschild found, in 1916, the first black hole solution to the Einstein's equation. We note that in the framework of Newtonian gravity, John Michell in 1784 and P. S. Laplace in 1796 had suggested the existence of massive stars whose escape velocity exceeds the speed of light. 

In the 1970s, it was argued that black holes should be considered as real thermodynamic systems. Such systems are described by four laws, in correspondence with the standard laws of thermodynamics. In particular, Bekenstein \cite{Bekenstein73} suggested that black hole's entropy equals to $S=(k_BAc^3)/(4G\hbar)$, where $A$ is its horizon's area. In addition, Hawking \cite{hawkingrad} demonstrated that one can associate a temperature $T=(\hbar\grk)/(2\pi ck_B)$ with a black hole, where $\grk$ is its surface gravity. Black hole thermodynamics suggests a more fundamental connection between thermodynamics and gravity.  This perspective motivated the idea that gravity is a thermodynamic phenomenon. 

In the next sections, we present the laws of black hole mechanics and their correspondence to the standard laws of thermodynamics. We also provide the arguments that led Bekenstein and Hawking to define the entropy and the temperature of a black hole respectively. In addition, we present the Unruh effect: for an observer moving with uniform proper acceleration $a$, the Minkowski vacuum
appears as a heat bath at the Unruh temperature $T=(\hbar a)/(2\pi ck_B)$, where $a$ is the magnitude of the observer's acceleration. Finally, we provide the formulation of the holographic principle, which restricts the number of the degrees of freedom that is required in order for a physical system to be described. The Unruh effect and the holographic principle are basic components of the arguments presented in the next sections.

\end{section}
%================================================================
\begin{subsection}{Black holes}

A black hole is a region of spacetime where the gravitational field is so strong that even light cannot escape. In other words, a black hole is a region of spacetime where escape velocity is greater than the velocity of light. In particular, let $(\mathcal{M}, g_{\grm\grn})$ be an asymptotically flat spacetime. Roughly speaking asymptotically flat is the spacetime that is approximated by the Minkowski spacetime at infinity. A black hole\footnote{For further details see \cite{Wald,townsend,largescale}. } region $\mathcal{B}$ of such a spacetime is defined as
\be
\mathcal{B}\equiv\mathcal{M}-J^-(\mathscr{I}^+),
\ee
where $J^-$ denotes the causal past and $\mathscr{I}^+$ the future null infinity. The boundary of $\mathcal{B}$, $\mathcal{H}\equiv \dot{J}^-(\mathscr{I}^+)$ is called the event horizon of the black hole and is a null hypersurface. Next, we give some definitions concerning black holes.

An asymptotically flat spacetime is said {\em stationary} if there exists a Killing vector field $\grj^{\grm}$ that is timelike near infinity. A spacetime is said {\em static} if it is stationary and invariant under time reversal. A black hole is {\em axisymmetric} if there exists a Killing vector field $\grx^{\grm}$, which is spacelike near infinity, and for which all orbits are closed.

A null hypersurface $\mathcal{N}$ is a {\em Killing horizon} of a Killing vector field $\grj^{\grm}$ if on $\mathcal{N}$, $\grj^{\grm}$ is normal to $\mathcal{N}$. Next, we consider a Killing horizon $\mathcal{N}$. Let $\ell^{\grm}$ be a vector normal to this horizon, so that $\ell^{\grn}\nabla_{\grn}\ell^{\grm}=0$ on $\mathcal{N}$.  On the horizon $\mathcal{N}$, $\grj^{\grm}=f\ell^{\grm}$, for some function $f$. Then, it follows that
\be
\grj^{\grn}\nabla_{\grn}\grj^{\grm}=\grk \grj^{\grm} \big|_{\mathcal{N}},
\ee
where $\grk=\grj^{\grn}\partial_{\grn}\ln |f|$ is called the {\em surface gravity} of the Killing horizon  $\mathcal{N}$. Furthermore, one shows that
\be
\grk=\lim (Va),
\ee
where $a$ is the magnitude of the acceleration of a particle moving on timelike orbits of $\grj^{\grm}$ in the region of the Killing horizon and $V=\sqrt{-\grj^{\grm}\grj_{\grm}}$ is the redshift factor of $\grj^{\grm}$.  The quantity $Va$ corresponds to the force that must be exerted at infinity to hold a unit mass particle at rest near the Killing horizon, i.e., surface gravity is the acceleration that is required in order for a test body to stay at rest closely at horizon---this explains the term surface gravity.
\end{subsection}

%================================================================
\begin{subsection}{The laws of black hole mechanics}\label{bht}

In 1973 Bardeen, Carter and Hawking \cite{fourlaws} formulated the four laws of black hole mechanics. The mathematical resemblance between the laws of black hole mechanics and the laws of thermodynamics was obvious. However, the authors considered this resemblance as a mere analogy.

The {\em zeroth law} of black hole mechanics  \cite{Waldthermo} states that if Einstein's equation holds and the matter stress-energy tensor obeys the dominant energy condition\footnote{The stress-energy tensor $T_{\grm\grn}$ satisfies the dominant energy condition if for all future-directed timelike vector fields $\gry$, the vector field 
\be
j(\gry)\equiv -\gry^{\grm}T_{\grm}^{\grn}\partial_{\grn}
\ee
is future-directed timelike, null or zero. Physically this implies that the speed of energy flow cannot be observed to be greater than the speed of light.}, then the surface gravity $\grk$ is constant on the future event horizon of a stationary (i.e., in equilibrium) black hole.

According to the {\em first law} of black hole mechanics, a stationary black hole of mass $M$, charge $Q$ and angular momentum $J$, with a future event horizon of surface gravity $\grk$, electric surface potential $\Phi_H$ and angular velocity $\Omega_H$ is related to a nearby black hole (perturbed) solution with mass $M+\grd M$, charge $Q+\grd Q$ and angular momentum $J+\grd J$ by the relation
\be\label{firstlaw}
\grd M=\frac{\grk}{8\pi}\grd A+\Omega_H\grd J+\Phi_H\grd Q.
\ee

The {\em second law} of black hole mechanics is identified with the Hawking's area theorem \cite{areatheorem}. According to this theorem, if the Einstein's equation holds (i) with the matter satisfying the null energy energy condition (i.e., $T_{\grm\grn}k^{\grm}k^{\grn}\geq 0$ for all null $k^{\grm}$) and (ii) the spacetime is strongly asymptotically predictable (i.e., if there is a globally hyperbolic region containing $ J^-(\mathscr{I}^+)\cup\mathcal{H}^+$), then the area of the future event horizon is a non-decreasing function of time. In addition, according to the area theorem, if two black holes coalesce, the area of the formed black hole will exceed the total area of the original black holes. 

According to the Planck-Nerst formulation of the third law of thermodynamics, the entropy $S$ tends to zero as the temperature $T$ approaches absolute zero. However, this formulation of the third law does not hold for black holes. There exist the so-called {\em extremal black holes} that have a zero temperature but a nonzero entropy. Nevertheless, the third law of thermodynamics is considered a consequence of statistical physics. It is not required for the complete thermodynamic description of a system \cite{Callen}. It is often not considered to be a fundamental law of thermodynamics.

Nevertheless, Bardeen, Carter and Hawking formulated a third law of black hole mechanics. The {\em third law} states that it is impossible for the surface gravity of the horizon $\grk$ to be reduced to zero by a finite sequence of processes. The third law of black hole mechanics corresponds to a weaker version of the Planck-Nerst formulation of the third law. According to this weaker version, it is impossible for one to reach the absolute zero temperature by any finite number of processes (unattainability of absolute zero) \cite{Kubo,Reichl}.

The above laws of black holes mechanics bear a close resemblance to the laws of thermodynamics (see table \ref{bhthermo}). The role of the internal energy $E$ is played by the mass $M$ of a black hole, the role of the temperature $T$ by the surface gravity $\grk$ of its horizon and the role of the entropy $S$ by its horizon's area $A$. Bardeen, Carter and Hawking considered this resemblance to be a mere mathematical analogy. The claim that a black hole has entropy, and therefore temperature, is inconsistent with the inherent definition of a black hole. If a black hole has temperature, it should radiate as a black body, while nothing can escape from it. 

At about the same time that the four laws of black hole mechanics were proposed, Bekenstein  observed that the area theorem--- formulated by Hawking in 1971--- resembles the second law of thermodynamics. Unlike Bardeen, Carter and Hawking, Bekenstein argued that this was not just a mere analogy. He proposed that a black hole should have an entropy that is proportional to its area. In addition, Bekenstein proposed a generalized second law.

\begin{table}[t!]
\begin{center}
\begin{tabular}{|c||p{5cm}|p{5cm}|}
\hline
 &  {\bf Black hole mechanics}& {\bf Thermodynamics} \\
\hline \hline
{\bf Zeroth law}   & Surface gravity $\grk$ constant on the horizon of a stationary black hole & Temperature $T$ constant throughout a body at thermal equilibrium \\
\hline
{\bf First law}  & $\grd M=\frac{\grk}{8\pi}\grd A+\Omega_H\grd J+\Phi_H\grd Q$ & $dE=TdS $ + work terms \\
\hline
{\bf Second law }&  $\grd A\geq 0$ in any process & $\grd S \geq 0$ in any process\\
\hline
{\bf Third law}& $\grk=0$ cannot be reached by any process & $T=0$ cannot be reached in any process \\
\hline
\end{tabular}
\end{center}
\caption{The correspondence between black hole mechanics and thermodynamics. For the validity of the third law, see the related discussion in the text.}
\label{bhthermo}
\end{table}
\end{subsection}

%================================================================

\subsection{Black hole entropy and the generalized second law}

At the beginning of the 1970's decade, Pensrose and Floyd \cite{PenroseFloyd} noted that in a Penrose process \cite{Penroseprocess}---a process employed for the extraction of rotational energy from a Kerr black hole and its conversion to mechanical energy of particles--- the horizon area of a black hole never decreases. Christodoulou \cite{Christodoulou,ChristodoulouRuffini}, examining the efficiency of a Penrose process, demonstrated  that the {\em irreducible mass}
\be
M_{\text{ir}}^2=\frac{A}{16\pi}
\ee 
of a black hole (where $A$ is the area of the horizon) cannot be reduced by any black hole transformation produced by the capture of small particles by the black hole (i.e., $\grd M_{\text{ir}}\geq 0$). The most efficient processes to extract energy are those associated with reversible transformations of the black hole that holds the irreducible mass constant. The less efficient processes are associated with irreversible transformations that increases the irreducible mass. This closely resembles thermodynamics, where the reversible processes are the most efficient ones.
Eventually, Hawking \cite{areatheorem} gave the general mathematical proof that the area of a black hole horizon never decreases with time (i.e., $dA\geq 0$). This is known as the Hawking's area theorem and constitutes (as we saw in \ref{bht}) the second law of black hole mechanics.

During that time (early 70's), the so-called {\em uniqueness theorems} \cite{Israel68,Carter70,Hawking72} were also formulated. The uniqueness theorems state that the most general stationary black hole spacetime belongs to the three-parameter Kerr-Newman family and is determined uniquely by the black hole's mass $M$, charge $Q$ and  angular momentum $L$. The black holes have no hair (no other independent characteristics), as Wheeler paraphrased \cite{Gravitation}. 

The fact that black holes have no hair creates a paradox \cite{Bekenstein80}. Consider the thought experiment where a a cup of hot coffee characterised by some entropy is dropped into a stationary black hole. When the cup passes through the event horizon, any information concerning it is lost for an exterior to the black hole observer. Since the exterior observer cannot know the amount of entropy inside the black hole, she can never be sure that the total entropy in the universe has not decreased. The second law of thermodynamics is violated for her.

Motivated by the works of Christodoulou and Hawking, Bekenstein \cite{Bekenstein73,Bekenstein72} suggested that a black hole should have an entropy that is proportional to its horizon area:
\be
S_{BH}=\grh\frac{A k_B }{\ell^2_p},
\ee
where $\grh$ is a dimensionless constant, $A$ the horizon area, $k_B$ the Boltzmann's constant and $\ell_p=\sqrt{\hbar G/c^3}$ the Planck length (introduced for dimensional reasons). Furthermore, to resolve the paradox concerning the decrease of entropy, Bekenstein proposed the {\em generalised second law} of thermodynamics. According to the generalised second law, the sum of the black hole entropy $S_{BH}$, and the ordinary entropy of the matter $S_{\text{matter}}$ in the exterior to the black hole region  never decreases with time, i.e.,
\be
\grd\left(S_{BH}+S_{\text{matter}}\right)\geq 0.
\ee
Thus, a decrease in the entropy in the region exterior to the black hole is at least compensated by an increase in the entropy of the black hole.

The decisive step towards the establishment of the black hole thermodynamics was Hawking's  publication in 1974 \cite{hawkingrad}, where he showed that the temperature of a black hole is  non-zero. In particular, Hawking demonstrated that taking into account the quantum phenomena near the horizon, a black hole radiates to infinity all kind of particles, with a black body spectrum at temperature
\be
T_{BH}=\frac{\hbar\grk}{2\pi c k_B},
\ee
where $\grk$ is the black hole's surface gravity. Then, from the first law of black hole thermodynamics (\ref{firstlaw}) for a Schwarzschild black hole (where $\grk=(4M)^{-1}$), the proportionality constant in Bekenstein's entropy is found to be $1/4$. Thus, the black hole entropy is
\be\label{BHentropy}
S_{BH}=\frac{A k_B c^3}{4G\hbar}.
\ee
Note that Hawking radiation decreases the black hole's area.  As a consequence, the classical area law is violated. Nevertheless, the generalized second law still holds. 

In statistical mechanics, the entropy of a system is equal to the logarithm of the number of the microstates available to the system at given values of energy. Then, one can naturally ask what are the quantum degrees of freedom responsible for the value of the black hole entropy (\ref{BHentropy}) and where they reside. Of course, a complete microscopic description of the black hole entropy requires a fully formulated quantum theory of gravity. The origin of the black hole entropy is one of the important open issues in the research concerning the black hole thermodynamics. There are many theories that aim to answer the above questions. Some calculations of the black hole entropy have been performed in the framework of string theory (e.g. \cite{stringentropy}), loop quantum gravity (e.g. \cite{LQGRovelli,LQGAshtekar}) and induced gravity \cite{BHinducedgravity}. According to another view, the black hole entropy is related to the entanglement entropy (see \cite{Solodukhin} for a review), resulting from correlations between quantum field degrees of freedom residing on the different sides of the horizon.

%================================================================

\subsection{Hawking radiation}
In 1974, Hawking  \cite{hawkingrad} demonstrated that if the quantum phenomena in the vicinity of its horizon are taken into account, a black hole radiates to infinity all kind of particles, with a black body spectrum at temperature
\be\label{introhawk}
T_{BH}=\frac{\hbar\grk}{2\pi c k_B},
\ee
where $\grk$ is the surface gravity of the black hole. Thus, the temperature of a black hole is non-zero. In this section, we introduce Bogolyubov transformations and provide the derivation of the Hawking radiation. We follow \cite{townsend} in part. For further details, we also refer the reader to \cite{birrelldavies,WaldQFTcurved}.

\medskip

{\em Bogoliubov transformations}. We consider a scalar field $\grf (x)$ in a globally hyperbolic spacetime $\mathcal{M}$. The field satisfies the Klein-Gordon equation
\be\label{KleinGordon}
(\square-m^2)\grf (x)=0.
\ee
For a spacelike hypersurface $\Sigma$, the inner product between the solutions of equation (\ref{KleinGordon}) is 
\be\label{innerproduct}
(\grf_1,\grf_2)=-i\int_{\Sigma} (\grf_1\nabla_{\grm}\grf^*_2-\grf_2\nabla_{\grm}\grf^*_1)d \Sigma^{\grm}.
\ee
One takes the hypersurface $\Sigma$ to be a Cauchy surface. The value of this inner product does not depend on the choice of $\Sigma$.

Next, one introduces a basis $\{\grc_i\}$ of the solutions of the Klein-Gordon equation that are orthonormal in the inner product  (\ref{innerproduct}), i.e., 
\begin{align}\label{orthonormal}
(\grc_i,\grc_j)&=\grd_{ij},\nonumber \\ (\grc_i,\grc^*_j)&=(\grc^*_i,\grc_j)=0,\\(\grc^*_i,\grc^*_j)&=-\grd_{ij}.\nonumber
\end{align}
Then, we expand the field as
\be
\grf (x)=\sum_i\left(a_i\grc_i+a_i^{\dagger}\grc^*_i\right),
\ee
where $\{a_i\}$ are operators in a Hilbert space $\mathcal{H}$. Their Hermitian conjugates are $a_i^{\dagger}$ and satisfy the commutation relations
\begin{align}
[a_i,a_j]&=0, \nonumber \\
[a_i^{\dagger},a_j^{\dagger}]&=0, \\
[a_i,a_j^{\dagger}]&=\grd_{ij}. \nonumber
\end{align}
One chooses the Hilbert space to be the Fock space built from a vacuum state $|0\rangle$ that satisfies
\be
a_i|0\rangle=0, \qquad \forall i.
\ee
Thus, the basis of the Hilbert space $\mathcal{H}$ is the $\{|0\rangle,a_i^{\dagger}|0\rangle,a_i^{\dagger}a_j^{\dagger}|0\rangle,\dots\}$.

In a general spacetime, the choice of an orthonormal basis of solutions of (\ref{KleinGordon}) is not unique. A different choice of an orthonormal basis implies a different notion of the vacuum, and hence a different notion of particles. In general, there are not any timelike Killing vectors and thus the classification of modes in positive or negative frequency is not possible. However, in a stationary spacetime, the existence of a timelike Killing vector $k^{\grm}\partial_{\grm}$ allows the choice of an orthonormal basis $\{u_i,u_i^*\}$, with the positive frequency modes satisfying the equation
\be
k^{\grm}\partial_{\grm}u_i=-i\grw_iu_i, \qquad \grw_i>0,
\ee
and the negative frequency modes the equation
\be
k^{\grm}\partial_{\grm}u^*_i=i\grw_iu^*_i, \qquad \grw_i>0.
\ee
Then, the vacuum state is the state of the lowest energy. A state $(a_i^{\dagger})^n|0\rangle$ is a state of $n$ particles. One defines the particle number operator as
\be
N=\sum_ia_i^{\dagger}a_i.
\ee

We consider now a new basis $\{u_i\}$ that obeys the corresponding relations (\ref{orthonormal}). The scalar field solution of Klein-Gordon equation is expanded in terms of the new basis as
\be
\grf(x)=\sum_i\left(a_i'u_i+{a_i'}^{\dagger}u_i^*\right).
\ee
 In this case, there is a vacuum state $|0'\rangle$ that satisfies
\be
a'_i|0'\rangle=0, \qquad \forall i.
\ee
This vacuum state builds a new Fock space.
One writes each mode in terms of the other as
\be
\begin{aligned}\label{modesrelation}
u_i&=\sum_j\left(A_{ij}\grc_j+B_{ij}\grc_j^*\right),\\
\grc_i&=\sum_j\left(A^*_{ji}u_j-B_{ji}u_j^*\right).
\end{aligned}
\ee
The corresponding relation between the operators is
\be
\begin{aligned}
a_i&=\sum_j\left(A_{ji}a_j'+B_{ji}^*{a_j'}^{\dagger}\right), \\
a_i'&=\sum_j\left(A^*_{ij}a_j-B^*_{ij}a_j^{\dagger}\right).
\end{aligned}
\ee
The above transformations are known as Bogoliubov transformations. The matrices $A_{ij}$ and $B_{ij}$ are called Bogoliubov coefficients. The Bogoliubov coefficients satisfy the conditions
\begin{align}\label{bogolyubovrelations}
\sum_k\left(A_{ik}A^*_{jk}-B_{ik}B_{jk}^*\right)=\grd_{ij}\nonumber &\Leftrightarrow AA^{\dagger}-BB^{\dagger}=\mathbb{I},\\
\sum_k\left(A_{ik}B_{jk}-B_{ik}A_{jk}\right)=0&\Leftrightarrow AB^{\intercal}-BA^{\intercal}=0.
\end{align}

Next, we consider a stationary submanifold $\mathcal{S}$ of the spacetime $\mathcal{M}$. In this part of spacetime, one expands the solution of the Klein-Gordon equation in terms of an orthonormal basis $\{\grc_i\}$. We consider also a different submanifold $\mathcal{S}'$. In $\mathcal{S}'$, one expands the field solution in terms of an orthonormal basis $\{u_i\}$. The particle number operator for the $i$\textsuperscript{th} mode in the submanifolds $\mathcal{S}$  and $\mathcal{S}'$ is respectively
\be
N_i={a_i}^{\dagger}a_i, \qquad N'_i={a'_i}^{\dagger}a'_i.
\ee
The vacuum state $|0\rangle$ in $\mathcal{S}$ is the state with no particles. The expectation value of the number operator in the vacuum state $|0'\rangle$  in $\mathcal{S}'$ is
\be\label{particlenumber}
\langle N_i'\rangle= \langle 0'|{a_i'}^{\dagger}a'_i|0'\rangle 
=\sum_j B_{ji}B_{ij}^{\dagger}=\left(B^{\dagger}B\right)_{ii}.
\ee
This is generally a non-zero value. The two vacuum states coincide only if the matrix $B$ vanishes.

%================================================================
\medskip

{\em Hawking radiation}. We consider, for simplicity, a massless scalar field $\Phi$ in a Schwarzchild black hole spacetime. Since the spacetime is spherically symmetric, one expands the solutions of the Klein-Gordon equation in spherical harmonics $Y_{\ell m}(\gru ,\grf)$, i.e., 
\be
\Phi_{\grw\ell m}=\frac{R_{\grw\ell}(r)}{r}Y_{\ell m}(\gru ,\grf)e^{-i\grw t}.
\ee
The function $R_{\grw\ell}$ satisfies the radial equation
\be
\left[\frac{\rmd^2}{\rmd {r^*}^2}+\left[\grw^2-\left(1-\frac{2M}{r}\right)\left(\frac{\ell (\ell +1)}{r^2}+\frac{2M}{r^3}\right)\right]\right]R_{\grw\ell}(r)=0,
\ee
where $r^*=r+2M\ln |r/2M-1|$ is the Regge-Whealer radial coordinate and $M$ is the mass of the black hole. In the region near the horizon $(r\!\rightarrow\! 2M)$, the Klein-Gordon equation possesses positive frequency outgoing asymptotic solutions $e^{-i\grw u}$ and ingoing solutions $e^{-i\grw \gry}$ respectively, where $u=t-r^*$ are the outgoing and $\gry=t+r^*$ are the ingoing null coordinates.

\begin{figure}[t!]
\centering
\includegraphics[scale=0.35]{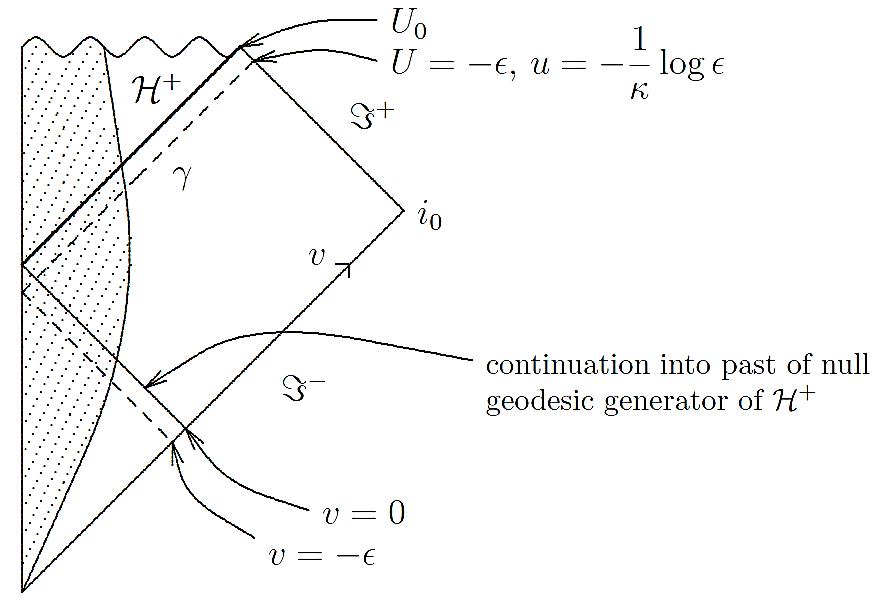}
\caption{A Penrose diagram showing the formation of a black hole by the gravitational collapse of a star (shaded region). The future event horizon $(u=\infty)$ is denoted by $\mathcal{H}^+$. The light ray $\grg$ (dashed line) corresponds to a particle's worldline in the geometric optics approximation. This ray is traced back from future null infinity $\mathscr{I}^+$ to past null infinity $\mathscr{I}^-$. A constant affine parameter $u$ corresponds to null rays escaping to $\mathscr{I}^+$. Null rays with constant $\gry$ pass through the horizon. The figure is taken from \cite{townsend}.
} 
\label{hawk}
\end{figure}
Next, we consider a positive frequency outgoing mode. The mode has the form
\be
\Phi_{\grw} \sim e^{-i\grw u}
\ee
near $\mathscr{I}^+$. Then, one employs a geometric optics approximation, where the wordline of a particle is a null ray $\grg$ of constant phase $u$. The ray $\grg$ is traced backwards in time from $\mathscr{I}^+$. The later it reaches $\mathscr{I}^+$, the closer it approaches the future event horizon $\mathcal{H}^+$.

The ray $\grg$ is one of a family of rays whose limit, as $t\rightarrow \infty$, is a null geodesic generator $\grg_H$ of $\mathcal{H}^+$. It is specified by its affine distance from $\grg_H$ along an ingoing null geodesic through $\mathcal{H}^+$ (see Fig. \ref{hawk}). Let this distance be a small and positive constant $\epsilon$. The affine parameter on an ingoing null geodesic at $\mathscr{I}^+$ is the Kruskal coordinate $U$ and is related to the affine parameter $u$ by the relation $U=-\exp (-\grk u)$. Hence, $U=-\epsilon$ and 
\be
u=-\frac{1}{\grk}\ln \epsilon
\ee
on $\grg$ near $\mathcal{H}^+$; $\grk$ is the surface gravity. Then,
\be
\Phi_{\grw}\sim \exp\left(\frac{i\grw}{\grk}\ln\epsilon\right) 
\ee
near $\mathcal{H}^+$.
This oscillates rapidly at later times near $\mathcal{H}^+$ (where $\epsilon \rightarrow 0$). The geometric optics approximation is justified.

Now, one wishes to relate the $\Phi_{\grw}$ to a solution near past null infinity $\mathscr{I}^-$. Let $x$ be an arbitrary point on the event horizon. Let also $l^{\grm}$ be a null vector tangent to the horizon. In addition, let $n^{\grm}$ be a future directed and normal to the horizon null vector at $x$ . Then, one parallely transports the vectors $l^{\grm}$ and $n^{\grm}$  along the continuation of $\grg_H$ back to $\mathscr{I}^-$. Let this continuation of $\grg_H$ intersect $\mathscr{I}^-$ at $\gry=0$. The continuation of the ray $\grg$ meet $\mathscr{I}^-$ at an affine distance $\epsilon$ along an outgoing null geodesic on $\mathscr{I}^-$ (see figure \ref{hawk}). The affine parameter on outgoing null geodesics in $\mathscr{I}^-$ is $\gry$ (since $ds^2=dud\gry+r^2d\Omega^2$ on $\mathscr{I}^-$). Thus, $\gry=-\epsilon$ on the ray $\grg$ and
\be\label{solatpast}
\Phi_{\grw}(\gry)=
\begin{cases}
\qquad 0, & \text{for}\ \ \gry >0,\\
 \exp\left[\frac{i\grw}{\grk}\ln(-\gry)\right], & \text{for}\ \  \gry <0.
\end{cases}
\ee
Ingoing null rays with affine parameter $\gry>0$ from $\mathscr{I}^-$ pass through $\mathcal{H}^+$. As a consequence, they do not reach $\mathscr{I}^+$.

The Fourier transform of the field solution (\ref{solatpast}) is
\be
\tilde{\Phi}_{\grw} = \int^{\infty}_{-\infty}\! e^{i\grw ' \gry}\Phi_{\grw}(\gry)d\gry =\int^{0}_{-\infty}\! \exp\left[i\grw ' \gry +\frac{i\grw}{\grk}\ln (-\gry)\right]d\gry.
\ee
It can be shown \cite{townsend} that 
\be
\tilde{\Phi}_{\grw}(-\grw')=-\exp\left(-\frac{\pi\grw}{\grk}\right)\tilde{\Phi}_{\grw}(\grw')
\ee
for $\grw'>0$.
Hence, a mode of positive frequency $\grw$ on $\mathscr{I}^+$ at late times matches onto mixed positive and negative modes on $\mathscr{I}^-$. One identifies
\be
A_{\grw\grw'} = \tilde{\Phi}_{\grw}(\grw'), \quad
B_{\grw\grw'} =\tilde{\Phi}_{\grw}(-\grw') 
\ee
as the Bogoliubov coefficients. They are related by 
\be
B_{ij}=-e^{-\frac{\pi\grw_i}{\grk}}A_{ij}.
\ee
Furthermore, the matrices $A$ and $B$ satisfy the Bogoliubov relation (\ref{bogolyubovrelations}), i.e.,
\bea
\grd_{ij}&=& \left(AA^{\dagger}-BB^{\dagger}\right)_{ij}\nonumber\\
&=& \sum_{k}\left(A_{ik}A_{jk}^*-B_{ik}B_{jk}^*\right)\nonumber\\
&=& \left[\exp\left(\frac{\pi(\grw_i+\grw_j)}{\grk}\right)-1\right]\sum_{k}B_{ik}B_{jk}^*.
\eea
Taking $i=j$ one has
\be
\left(BB^{\dagger}\right)_{ii}=\frac{1}{\exp\left(\frac{2\pi \grw_i}{\grk}\right)-1}.
\ee

Next, one takes the inverse Bogoliubov coefficients corresponding to a positive frequency mode on $\mathscr{I}^-$ that matches onto mixed positive and negative frequency modes on $\mathscr{I}^+$. According to  equation (\ref{modesrelation}), the inverse coefficient is 
\be
B'=-B^{\intercal}.
\ee
Hence, the late time particle flux through $\mathscr{I}^+$, given a vacuum on $\mathscr{I}^-$, is
\be
\langle N_i\rangle_{\mathscr{I}^+}=\Big((B')^{\dagger}B'\Big)_{ii}=\Big(B^*B^{\intercal}\Big)_{ii}= \Big(BB^{\intercal}\Big)_{ii}^*.
\ee
However, $\left(BB^{\intercal}\right)_{ii}$ is real, so
\be
\langle N_i\rangle=\frac{1}{\exp\left(\frac{2\pi\grw_i}{\grk}\right)-1},
\ee
that corresponds to a Planck spectrum for a black body radiation at the Hawking temperature
\be
T_H=\frac{\grk}{2\pi}.
\ee

%================================================================
\subsection{Unruh effect}\label{Santa}
\begin{figure}[t!]
\centering
\includegraphics[scale=0.4]{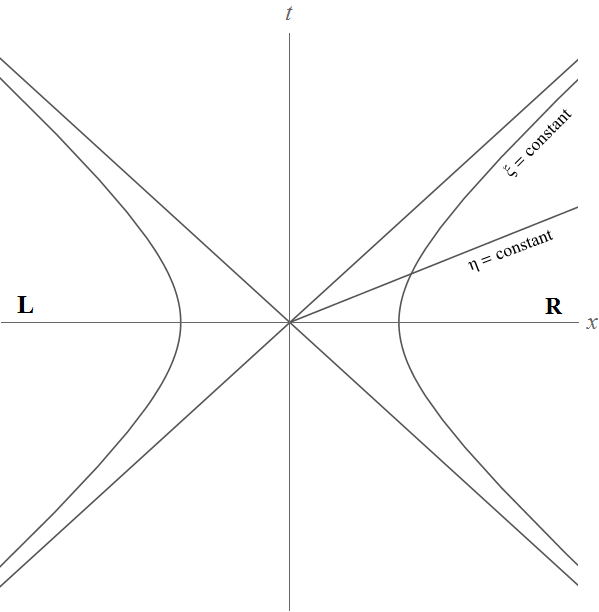}
\caption{Minkowski spacetime in Rindler coordinates in $t-x$ plane. A hyperbola of constant $\grj$ corresponds to a uniformly accelerated worldline with proper acceleration $a=\grj^{-1}$. The bifurcate Killing horizon acts as an event horizon for a Rindler observer in the right Rindler wedge $R$. 
} 
\label{rindlerfig}
\end{figure}
One of the most celebrated results of Quantum Field Theory in curved spacetime is the {\em Unruh effect} \cite{birrelldavies,Unruh}: for an observer moving with uniform proper acceleration $a$, the  Minkowski vacuum appears as a heat bath at the Unruh temperature
\be\label{TU}
T_U=\frac{\hbar a}{2\pi ck_B}.
\ee
Unruh effect is a mathematically similar phenomenon to Hawking radiation. In this section, we introduce the notion of Rindler horizon and use the concept of particle detectors to derive acceleration temperature.
\medskip

{\em Rindler horizon}. We consider the four-dimensional Minkowski spacetime line element written in Cartesian coordinates in the $t-x$ plane
\be\label{Minkelement}
ds^2=-dt^2+dx^2+dx_{\perp}^2,
\ee
where $dx_{\perp}^2=dy^2+dz^2$ is the line element in the transverse space. An observer moving along the $x$-direction with a uniform acceleration of magnitude $a$ follows the hyperbolic trajectory
\be\label{hyptraj}
 t(\grt)=a^{-1}\sinh(a\grt) ,\quad x(\grt)=a^{-1}\cosh(a\grt), \quad y(\grt)=z(\grt)=0
\ee 
where $\grt$ is its proper time. In place of $(t,x,y,z)$ coordinates, one can introduce the so-called  Rindler coordinates $(\grh,\grj,y,z)$  through the coordinate transformation
\be\label{Rindmetric}
t=\grj\sinh(\grk\grh), \qquad x=\grj\cosh(\grk\grh).
\ee
Then, the line element (\ref{Minkelement}) takes the form
\be
ds^2=-\grk^2\grj^2d\grh^2+d\grj^2+dx_{\perp}^2,
\ee
which is known as the Rindler metric. The Rindler coordinates cover only the subregion of the Minkowski space where $x>|t|$. This region is called the (right) Rindler wedge (see figure \ref{rindlerfig}).

The vector $\partial_{\grh}$ is a timelike Killing vector that generates the boost Lorentz symmetry. Its orbits with constant $\grj$ are hyperbolae that represent worldlines of uniformly accelerated observers with proper acceleration $a=\grj^{-1}$. The acceleration increases as $\grj\to 0$. The boost Killing vector field generates a bifurcate Killing horizon\footnote{A {\em bifurcate Killing horizon} is a pair of Killing horizons that intersect in a $(n-2)$ dimensional spacelike surface on which the Killing field vanishes.} at $\grj=0$. This horizon is often called Rindler horizon. The right Rindler wedge is bounded by the Rindler horizon. An observer at this wedge is causally separated from observers at the other wedges. The Killing horizon is an event horizon for him. Note that the Killing vector field in the right Rindler wedge is timelike and, thus, allows the construction of a quantum field theory for this wedge, viewed as spacetime in its own.

\medskip

{\em Particle detectors}. The Unruh effect is usually demonstrated by means of the {\em Unruh-DeWitt detector} \cite{birrelldavies,Dewitt}. An Unruh-DeWitt detector is an ideal particle detector coupled to a quantum field with a monopole interaction and moving along a trajectory $x^{\mu}(\tau)$ in Minkowski spacetime, where $\tau $ is the proper time of the detector.

We consider for simplicity a pointlike particle with  two energy eigenstates $H_0|E_0\rangle=E_0|E_0\rangle$ and $H_0|E_1\rangle=E_1|E_1\rangle$ and respective energies $E_0<E_1$; $H_0$ is the free-particle Hamiltonian. The detector interacts with a massless scalar  field $\hat{\phi}$ through the interaction Hamiltonian (we work in the interaction picture)
\be
\hat{H}_{I,\text{int}}(\grt)=g\hat{m}(\grt)\otimes\hat{\grf}\left[x^{\grm}(\grt)\right],
\ee
where $g$ is a coupling constant and $\hat{m}(\grt)=e^{i\hat{H}_0\grt}\hat{m}(0)e^{-i\hat{H}_0\grt}$ is the detector's monopole moment operator.

We assume that both the particle and the field are initially in their ground states $|E_0\rangle$ and $|0\rangle$ respectively. Our aim is to calculate the probability that the detector undergo, for a general trajectory, a transition from its ground state to the excited state. The probability amplitude for the transition is
\bea
\mathcal{A}_n&=&\langle E_1|\otimes \langle n| \hat{U}_I(\grt,\grt_0) |E_0\rangle\otimes|0\rangle,
\eea
where $|n\rangle$ is an excited state of the field and
\be
\hat{U}_I(\grt,\grt_0)=\mathcal{T}e^{-i\int_{\grt_0}^{\grt}d\grt \hat{H}_I(\grt)},
\ee
is the evolution operator; $\mathcal{T}$ is the time-ordering operator. Assuming that the coupling is small ($g\to0$), first order-perturbation theory implies
\be
\hat{U}_I(\grt,\grt_0)=\hat{1}-i\int_{\grt_0}^{\grt}d\grt \hat{H}_I(\grt)+O(g^2).
\ee
The probability amplitude is
\bea
\mathcal{A}_n&=&-ig\int_{-\infty}^{+\infty} d\grt \langle E_1|\hat{m}(\grt)|E_0\rangle \langle n|\hat{\phi}[x^{\mu}(\grt)]|0\rangle+ O(g^2)\nn\\
&=&-ig\langle E_1|\hat{m}(0)|E_0\rangle\int_{-\infty}^{+\infty}d\tau e^{i\Delta E\tau}\langle n|\hat{\phi}[x^{\mu}(\grt)]|0\rangle +O(g^2),
\eea
 where $\Delta E=E_1-E_0$.
 
The transition probability is obtained by taking the square modulus of the probability amplitude and summing over the complete set of the field states, i.e.,
\bea
\mathcal{P}&=&\sum_n|\mathcal{A}_n|^2\nn\\ &=& g^2|\langle E_1|\hat{m}(0)|E_0\rangle|^2\int_{-\infty}^{+\infty}d\tau
\int_{-\infty}^{+\infty}d\tau' e^{-i\Delta E\Delta\tau}\Delta^+(\tau,\tau'),
\eea
where $\Delta\tau=\tau-\tau'$ and $\Delta^+(\tau,\tau')=\langle 0|\hat{\phi}(\grt)\hat{\phi}(\grt')|0\rangle$ is the positive frequency Wightman function. For a massless scalar field  the Wightman function is
\be\label{Wightmanscalar}
\Delta^+(x,x')=-\lim_{\epsilon \rightarrow 0^+}\frac{1}{4\pi^2}\frac{1}{(\grt-\grt'-i\epsilon)^2-|\mathbf{x}-\mathbf{x}'|^2}.
\ee
It is generally  useful to consider the  transition rate $w$ defined as the transition probability per unit of proper time, i.e.,
\be\label{probw}
w=g^2|\langle E_1|\hat{m}(0)|E_0\rangle|^2\int_{-\infty}^{+\infty}d(\Delta\tau) e^{-i\Delta E\Delta\tau}\Delta^+(\tau,\tau')
\ee

In the case of an inertial detector following the trajectory $x^{\mu}(\tau)=(\grt,0,0,0)$, the Wightman function (\ref{Wightmanscalar}) becomes
\be
\Delta^+(\grt,\grt')=-\frac{1}{4\pi^2(\Delta \grt -i\epsilon)^2}.
\ee
Then, the transition rate is
\be\label{fnl}
w=-\frac{g^2|\langle E_1|\hat{m}(0)|E_0\rangle|^2}{4\pi^2}\int_{-\infty}^{+\infty}d(\Delta \tau) \frac{e^{-i\Delta E\Delta\tau}}{(\Delta\tau-i\gre)^2}.
\ee
Since $\Delta E>0$, the contour integral (\ref{fnl}) is calculated by closing the contour in an infinite semicircle in the lower-half $\Delta\tau$ plane. The integrand has only a pole of  second order at $\Delta\tau=i\gre$ in the upper-half $\Delta\tau$ plane (Fig. \ref{aaaa}) and thus the integral is zero.  As expected, an inertial detector does not detect any particles in the Minkowski vacuum.

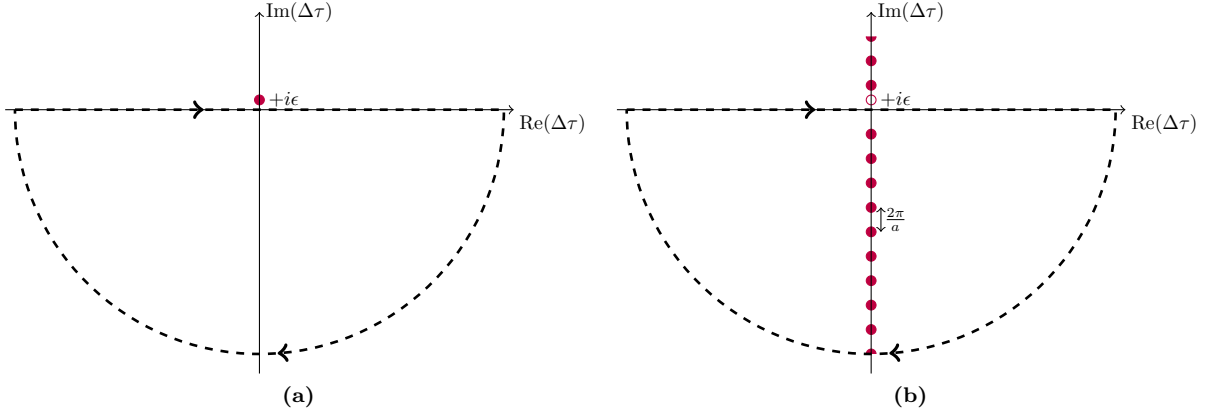
\begin{figure}[t!]
\subfloat[]{\begin{tikzpicture}[scale=0.65, every node/.style={scale=0.75}]
%\draw [help lines] (-8,-8) grid (8,8);
\node at (6,-0.3){$\text{Re}(\Delta\tau)$};
\node at (0.82,2) {$\text{Im}(\Delta\tau)$};
\node at (0.5,0.22){$+i\epsilon$};
\draw [fill=purple,purple] (0,0.2) circle [radius=0.1];
\draw [thin,->] (-5.2, 0) -- (5.2,0);
\draw [thin,->] (0, -5.4) -- (0, 2);

 \draw[dashed,line width=1pt,decoration={ markings, mark=at position 0.15 with {\arrow[line width=1.5pt]{>}},mark=at position 0.68 with {\arrow[line width=1.5pt]{>}}},postaction={decorate}]%xshift=-30pt
  (-5,0)--(5,0) arc (0:-180:5) (-5,0);
\end{tikzpicture}}\hspace{0.2cm}
\subfloat[]{\begin{tikzpicture}[scale=0.65, every node/.style={scale=0.75}]
%\draw [help lines] (-8,-8) grid (8,8);
\node at (6,-0.3){$\text{Re}(\Delta\tau)$};
\node at (0.82,2) {$\text{Im}(\Delta\tau)$};
\node at (0.5,0.22){$+i\epsilon$};
\node at (0.5,-2.25){$\frac{2\pi}{a}$};
\draw [purple] (0,0.2) circle [radius=0.1];
\draw [fill=purple,purple] (0,-0.5) circle [radius=0.1];
\draw [fill=purple,purple] (0,-1) circle [radius=0.1];
\draw [fill=purple,purple] (0,-1.5) circle [radius=0.1];
\draw [fill=purple,purple] (0,-2) circle [radius=0.1];
\draw [fill=purple,purple] (0,-2.5) circle [radius=0.1];
\draw [fill=purple,purple] (0,-3) circle [radius=0.1];
\draw [fill=purple,purple] (0,-3.5) circle [radius=0.1];
\draw [fill=purple,purple] (0,-4) circle [radius=0.1];
\draw [fill=purple,purple] (0,-4.5) circle [radius=0.1];
\draw [fill=purple,purple] (0.1,-5) arc (0:180:0.1);
\draw [fill=purple,purple] (0,0.5) circle [radius=0.1];
\draw [fill=purple,purple] (0,1) circle [radius=0.1];
\draw [fill=purple,purple] (0.1,1.5) arc (0:-180:0.1);
\draw [thin,<->] (0.2, -2) -- (0.2,-2.5);
\draw [thin,->] (-5.2, 0) -- (5.2,0);
\draw [thin,->] (0, -5.4) -- (0, 2);
 \draw[dashed,line width=1pt,decoration={ markings, mark=at position 0.15 with {\arrow[line width=1.5pt]{>}},mark=at position 0.68 with {\arrow[line width=1.5pt]{>}}},postaction={decorate}]%xshift=-30pt
  (-5,0)--(5,0) arc (0:-180:5) (-5,0);
\end{tikzpicture}}
\caption{Singularities and integration contours of transtion rate's contour integrals in the case of (a) an  inertial and (b) a uniformly accelerated particle detector.
} 
\label{unruh_poles}
\end{figure}

We next consider  a uniformly accelerated detector that follows the hyperbolic trajectory
(\ref{hyptraj}).
The  corresponding correlation function is
\be\label{Wacc}
\Delta^{+}(\grt,\grt')= -\lim_{\epsilon \rightarrow 0^+} \frac{a^2}{16\pi^2\sinh^2[a(\Delta \grt - i\gre)/2]}.
\ee
We use the formula \cite{Ryzhik}
\be
\frac{1}{\sinh^2(z)}=\sum_{\grh=-\infty}^{+\infty}
\frac{1}{(z+i\pi\grh)^2}
\ee
 to write the transition rate as
\be
w=-\frac{g^2|\langle E_1|\hat{m}(0)|E_0\rangle|^2}{4\pi^2}\sum_{\grh=-\infty}^{+\infty}\int_{-\infty}^{+\infty}d(\Delta \tau) \frac{e^{-i\Delta E\Delta\tau}}{(\Delta\tau+i\frac{2\pi\grh}{a}-i\gre)^2}.
\ee
The integral is again evaluated by closing the contour
in the lower half-plane. The only contribution to the integral is from the second order infinite series of poles $\Delta\tau=-i\frac{2\pi\grh}{a}+i\gre$ for $n>0$  in the lower half plane, as shown Fig. \ref{bbbb}. Employing Cauchy's residue theorem we find a transition rate
\bea
w&=&\frac{g^2|\langle E_1|\hat{m}(0)|E_0\rangle|^2\Delta E}{2\pi}\sum_{\grh=1}^{\infty}e^{-\frac{2\pi\grh}{a}\Delta E }\nn\\
&=&\frac{g^2|\langle E_1|\hat{m}(0)|E_0\rangle|^2\Delta E}{2\pi}\frac{1}{e^{\frac{2\pi}{a}\Delta E}-1}
\eea
From the Planck factor we conclude that the detector is in a thermal equilibrium at the Unruh temperature $T_U=\frac{a}{2\pi}$. Thus, a uniformly accelerating detector perceives the
Minkowski vacuum as a thermal bath at the Unruh temperature.

%================================================================
\subsection{Entropy bounds}

The validity of the generalised second law imposes certain bounds in the entropy content of ordinary matter. In the section, we present the Bekenstein's entropy bound \cite{Bekenstein81}, the spherical entropy bound \cite{Susskind95} and the covariant entropy bound \cite{covariantbound}. Further details are found in \cite{Bousso}, which we follow in part, and in \cite{Waldthermo,Bekenstreview}.

We consider an arbitrary weakly gravitating matter system of total energy $E$. Let $R$ be the radius of the smallest sphere that circumscribes the system. Next, one employs a Geroch process, i.e., the system is dropped into a black hole from the vicinity of the horizon. Let the black hole be described by the Schwarzschild metric. When the system is absorbed by the black hole, one founds that the surface area of the black hole increases at least by $8\pi ER$. The generalised second law, then, implies that this increase of the surface area at least compensate the entropy of the system that is lost for an exterior to the black hole observer, i.e., $\grd S_{\text{BH}}-S_{\text{matter}}\geq 0$. Consequently, 
\be
S_{\text{matter}}\leq 2\pi E R.
\ee
This entropy bound, called the Bekenstein or the {\em universal upper bound} \cite{Bekenstein81}, is valid for any weakly gravitating matter system in asymptotically flat spacetime. In conventional units, the entropy bound is $S_{\text{matter}}\leq 2\pi k_B E R/(\hbar c)$.

Susskind \cite{Susskind95} proposed the {\em spherical entropy bound}. It is often referred as the {\em holographic bound}, since it is closely related to the formulation of the holographic principle. In order to derive the spherical entropy bound one employs a Susskind process. In this process, a system is evolved to a black hole. 
In particular, we consider a spherically symmetric and weakly gravitating system in a spacetime $\mathcal{M}$, where the formation of black holes is permitted. Let $E$ be the energy of the system, $R$ its radius and $A$ the corresponding area of the sphere. The mass of the system is less than the mass $M$ of a black hole of the same surface area, so that the system is gravitationally stable. Now, by collapsing a shell of mass $M-E$ onto the system, it is evolved to a black hole of area $A$. The total initial entropy is $S_{\text{initial}}=S_{\text{matter}}+S_{\text{shell}}$. The entropy of the final state is that of a black hole, i.e., $S_{\text{final}}=A/4$. The generalised second law holds only if
\be\label{sphericalbound}
S_{\text{matter}}\leq \frac{A}{4},
\ee
which is the spherical entropy bound. The Susskind's bound is weaker than the Bekenstein bound, in situations where both can be applied. We note that a black hole saturates the bound. Thus, a black hole is the most entropic object that can be put inside a given spherical surface.

The spherical entropy bound is not applied in cases where the system lacks spherical symmetry, or the enclosed system is not gravitationally stable. In addition, it is not applied in cosmology. Bousso introduced the so-called {\em covariant entropy bound} \cite{covariantbound} in order to generalize the spherical entropy bound and make it broader valid. The covariant entropy bound is formulated as follows. Let $A(B)$ be the area of an arbitrary $D-2$ dimensional spatial surface $B$. A $D-1$ dimensional hyperfurface $L$ is called a  light-sheet of $B$ and is generated by orthogonal to the surface null geodesics with non-positive expansion $\gru$. The hypersurface L is not allowed to contain caustics, where $\gru$ changes sign from $-\infty$ to $+\infty$. Let, also, $S$ be the entropy on any light-sheet of $B$. The covariant entropy bound states that the entropy $S[L(B)]$ on any light-sheet of $B$ does not exceed a quarter of the area of $B$, that is,
\be
S[L(B)]\leq \frac{A(B)}{4}.
\ee

%==============================================================
\subsection{Holographic principle}

The holographic principle formulated by 't Hooft \cite{thooft} and Susskind \cite{Susskind95} as a direct consequence of the spherical entropy bound. In this section, we derive the holographic principle, following the arguments presented in  \cite{Bousso}.

We consider a finite region of space of volume $V$, bounded by a surface $\partial V$ of area $A$. We assume, for the time being, that gravity is weak. Hence, the above quantities are well defined. We also assume that the spacetime is asymptotically flat. In order to apply the spherical  entropy bound, we assume that the defined region is spherical and that its metric is not strongly time dependent. No restrictions on the enclosed matter content are imposed.

We regard the defined region as a quantum mechanical system. We suppose that the number $N$ of the degrees of freedom of a quantum mechanical system is defined as the logarithm of the dimension $\mathcal{N}$ (which describes all the possible states that the system can be in) of the system's Hilbert space $\mathcal{H}$, that is,
\be
N=\ln \mathcal{N} =\ln \text{dim} (\mathcal{H}).
\ee
The number of the degrees of freedom is equal--- up to a factor of $\ln 2$--- to the number of the bits of information that are needed to characterise a state. For instance, a system of 100 spins has $\mathcal{N}=2^{100}$ states, $N=100\ln 2$ degrees of freedom and can store 100 bits of information. Then, one can ask what is the number of the degrees  of freedom or  the amount of information that describes all possible physics confined to the previously specified region at the most fundamental level. We call the studied system, which is composed of the constituents of a fundamental theory,  the fundamental system.

We suppose that the fundamental system is a local quantum field theory on curved spacetime. This is the usual approximate framework used to combine gravity with quantum fields. This theory naturally demands a UV cut-off that is identified with the Planck length $\ell_p$. The theory also demands a IR cut-off, which is the Planck mass $m_p$. The Planck mass corresponds to the largest amount of energy that can be localized to a Planck volume without producing a black hole. Then, one discretises the space into a Planck grid and assumes that each Planck volume has one harmonic oscillator. Each harmonic oscillator has a finite number of states $n$. Consequently, the total number of oscillators is $V$ (in Planck units). Thus, the total number of independent quantum states in the specified region is
\be
\mathcal{N}\sim n^V.
\ee
The number of degrees of freedom is
\be
N\sim V\ln n\gtrsim V,
\ee
i.e., it grows with the volume.

The statistical interpretation of entropy implies that the number of all possible quantum states of a system is $e^S$. Employing the spherical entropy bound (\ref{sphericalbound}) for the specified fundamental system, one concludes that the number of states obey the relation
\be
\mathcal{N}\leq e^{A/4}.
\ee 
The equality, of course,  applies in the case of a black hole that fits the specified region. Then, for the number of the degrees of freedom
\be
N\leq \frac{A}{4}.
\ee
Thus, the number of the degrees of freedom depends on the area of the boundary rather than the volume of the specified region. We note a  contradiction between the result obtained applying the spherical entropy bound and that obtained from local quantum field theory. In the latter case, the number of the degrees of freedom is much larger.

In the naive field theory estimate, the IR cut-off implies that most of the entropy comes from modes of very high energy. We assumed that each Planck volume contains at most one Planck mass. The mass $M$ contained in a spherical region of radius $R$ obeys the relation $M\lesssim R$, since it cannot contain more mass than a black hole  (the mass of a black hole is given by its radius) of the same area. The imposed UV cut-off obeys this relation at the smallest scale. However, at larger scales $M\sim R^3$ and the formation of a black hole is not prevented. Thus, most of the states included by the field theory estimate are too massive to be gravitationally stable. Long before the quantum fields are excited to such a level, a black hole would form. If this black hole is still to be contained within a specified sphere of area A, its entropy saturates but not exceeds the spherical entropy bound. Consequently, the naive field estimate fails when gravity is included. If one takes gravity into consideration, less number of degrees of freedom is used to generate entropy.  

The fact that the degrees of freedom scale with area is also realised from the view of unitarity, i.e., the fact that the quantum mechanical evolution preserves information. In particular, we suppose that a region is described by a Hilbert space of dimension $e^V$. We let this region to evolve into a black hole. When a black hole is formed, the region is described by a Hilbert space of dimension $e^A$. The number of states has been decreased. Unitarity is violated. Thus, one should start with a Hilbert space of dimension $e^A$.

The arguments presented in this section led 't Hooft and Susskind to formulate the {\em holographic principle}\footnote{The term ``holographic'' is originated from the fact that the principle is a reminiscent of holography, the optical technique by which a three-dimensional image is stored on a two-dimensional surface as a interference pattern. }. There are various formulation of this principle.  According to \cite{Bousso} the holographic principle is formulated as follows.
\medskip

{\bf Holographic principle.} {\em A region bounded by a surface of area $A$ is fully described by no more than $A/4$ degrees of freedom. A fundamental theory, unlike local field theory, should incorporate this counterintuitive result.}
\medskip

Of course, one can also formulate the holographic principle employing the covariant entropy bound. The holographic principle suggests that a given volume of space is fully described by the degrees of freedom associated with its boundary.

The most explicit manifestation of holography is probably the AdS/CFT correspondence. As AdS/CFT  correspondence is denoted the equivalence between a string theory in a $(d+1)$-dimensional Anti-de Sitter (AdS) spacetime and a conformal field theory (CFT) in $d$-dimensions. The CFT is formulated on the boundary of the AdS spacetime. Malcadena \cite{Malcadena} gave the first example of such a correspondence. In particular, Malcadena demonstrated that  a type IIB string theory in $\text{AdS}_5\times \text{{\bf S}}^5$ spacetime is equivalent to a CFT without gravity, the $\mathcal{N}=4$ supersymmetric Yang-Mills theory.

%================================================================
\subsection{Gravity as a thermodynamic phenomenon}

In the previous sections, we saw how the analogy between the laws of black hole mechanics and the laws of thermodynamics led Bekenstein and Hawking to argue that the black holes should be considered as real thermodynamic systems that are characterised by entropy and temperature. In particular, Bekenstein argued that the entropy of a black hole equals to $S=(k_BAc^3)/(4G\hbar)$, where $A$ is the area of its horizon. Furthermore, Hawking showed that the temperature of a black hole is $T=(\hbar\grk)/(2\pi ck_B)$, where $\grk$ is its surface gravity.

The correspondence between the laws of black hole mechanics and that of thermodynamics suggests a deeper connection between thermodynamics and gravity. This perspective motivated several ideas that suggest the interpretation of gravity as a thermodynamic phenomenon. The original idea is due to Jacobson \cite{Jacob95}, who demonstrated that the Einstein's equation can be viewed as an equation of state. Padmanabhan \cite{PadInsights} also showed that in several cases the gravitational equations are interpreted in terms of thermodynamics. More recently, Verlinde \cite{Verlinde} argued that gravity is an entropic force. The above arguments are components of the broader view--- first formulated by Sakharov \cite{Sakharov}--- that gravity is not a fundamental force, but an emergent one.  It arises as the limit of some underlying--- yet unknown--- microscopic theory, in the same sense that hydrodynamics or elasticity emerge from molecular physics.

In the present thesis, we examine the arguments of Jacobson, Padmanabhan and Verlinde that suggest the interpretation of gravity as a thermodynamic theory. Our primary aim is to provide a deeper understanding of the arguments, the methods and the notions used by the authors. The study of the several thermodynamic aspects of gravity brings out various intrinsic features of gravity. Such features had not been pinpointed until nowadays, while their interpretation is not possible in the standard approaches of gravity. The conclusions that one draws from the thermodynamic interpretation of gravity may offer a new window in the understanding of the nature of a possible quantum theory of gravity. 

In chapter \ref{chapter2}, we present the interpretation of the Einstein's equation as an equation of state, proposed by Jacobson. In chapter \ref{chapter3}, we present the programme of Padmanabhan in the interpretation of gravity as a thermodynamic and  an emergent theory. In chapter \ref{chapter4}, we examine Velrinde's arguments suggesting the interpretation of gravity as an entropic force. Finally, in chapter \ref{chapter5}, we summarise  our conclusions.

%===================CHAPTER 2=========================
\section{Thermodynamics of spacetime}\label{chapter2}

The original idea of the interpretation of gravity as a thermodynamic theory is due to Jacobson \cite{Jacob95}. Jacobson demonstrated  that the Einstein's equation can be viewed as an equation of state. The idea is summarised as follows.

In any point of spacetime, one introduces local Rindler horizons, as they are perceived by uniformly accelerated observers. A thermodynamic system is defined as the degrees of freedom residing in the region of the spacetime just beyond one of these horizons. The Einstein's equation is obtained from the demand the Clausius relation $\grd Q=T\rmd S$ to hold for all local Rindler horizons, and the conservation of energy. One takes the entropy $S$ to be proportional to the horizon's area. Furthermore, the heat $\grd Q$ and the temperature $T$ are the energy flux and the Unruh temperature respectively, as these are perceived by an accelerated observer just beyond the horizon. In this way, the Einstein's equation can be viewed as an equation of state. 

If one assumes that the entropy is also proportional to a function of the Ricci scalar, the approach of the non equilibrium thermodynamics is required \cite{Eling06, Chirco09}. One obtains the field equation of $f(R)$ gravity from the entropy balance condition $\rmd S=\grd Q/Τ+\rmd^i S$, where $\rmd^i S$ is the entropy produced inside the system. Such an entropy production term is allowed in the case of Einstein's gravity as well.

In this chapter we examine the above arguments.
%====================================================
\subsection{The Einstein Equation of state}

We consider an arbitrary point $p$ in a generic spacetime $(\mathcal{M},g_{\mu\nu})$. Invoking the equivalence principle, one defines a local inertial frame within an infinitesimal neighborhood around $p$---  the Riemann normal coordinates are $\{x^\gra\}$, such that $p$ stays at the origin $x^\gra=0$--- i.e.,
\be
g_{\grm\grn}(x^\gra)=\grh_{\grm\grn}(x^\gra)+\mathcal{O}[(x^\gra)]^2,
\ee
where $\grh_{\grm\grn}$ is the metric of the Minkowski spacetime. The first partial derivatives of $g_{\grm\grn}$ at $p$ vanish. Then, one introduces a local Rindler frame by employing the standard coordinate transformations (\ref{Rindmetric}). Let $\grj^{\grm}$ be the boost Killing vector field that generates the Rindler horizon.

One wishes to have a consistent thermodynamic description of spacetime. To this end, an appropriate thermodynamic system is necessary to be defined at first. A local causal horizon at a spacetime point $p$ is defined as the one side of the boundary of the past of a spacetime 2-surface patch $\mathcal{P}$ including $p$. Near $p$, this boundary is a congruence of null geodesics orthogonal to $\mathcal{P}$. Then, one defines the thermodynamic system as the degrees of freedom beyond the Rindler horizon of $\mathcal{P}$.

Next, one specifies the thermodynamic macroscopic variables that characterise the defined thermodynamic system. The temperature of the system is taken to be the Unruh temperature 
\be\label{sptemp}
T=\frac{\hbar \grk}{2\grp},
\ee
where $\grk$ is the acceleration of the orbit of the Killing vector $\grj^\grm$. Motivated by the Bekenstein-Hawking definition for the black hole entropy, one assumes that the horizon entropy\footnote{It is commonly believed (e.g. \cite{Parentani, PadInsights}) that not only black hole horizons, but all horizons have temperature and entropy.} is proportional to the area of the horizon, i.e.,

\be
S=\gra A,
\ee
where $\gra$ is a dimensional constant and $A$ the area of the horizon. Furthermore, heat is defined as the energy that flows across a horizon. In this point, a key assumption is employed. One assumes that all energy flow across the horizon is heat. The conserved boost energy current of matter is $J_{\grn}=T_{\grm\grn}\grj^{\grm}$, where $T_{\grm\grn}$ is the energy-momentum tensor. The heat flux to the past of $\mathcal{P}$ is
\be
\grd Q=\int_{\mathcal{H}} \! T_{\grm\grn}\grj ^{\grm}\mathrm{d}\Sigma ^{\grn},
\ee
where the integration is over the horizon. One chooses the direction of $\grj^{\grm}$ to be future pointing to the past of $\mathcal{P}$. 

In addition, one assumes that the Clausius relation 
\be\label{clausius}
\grd Q=T\rmd S
\ee
is valid for all local causal horizons. In this way, the energy flux across the Rindler horizon, the entropy and the temperature are related. A vector $k^{\grm}$ tangent to the horizon is defined such that $\grj ^{\grm}=-\grk \grl k^{\grm}$ for an affine parameter $\grl$. Then, the volume element is $\mathrm{d}\Sigma^{\grm}=k^{\grm}\rmd A\rmd \grl$, where $\rmd A$ is a cross sectional area element of the horizon.  The heat flux is written as
\be\label{heatflux}
\grd Q=-\grk \int_{\mathcal{H}} \! \grl T_{\grm\grn}k^{\grm}k^{\grn}\rmd \grl \rmd A.
\ee

The entropy variation associated with a piece of the horizon is $\rmd S=\gra \grd A$, where $\grd A$ is the variation of the area. The expansion $\gru$ of the horizon's generators is defined as
\be\label{expansion}
\gru=\frac{1}{\grd A }\frac{\rmd (\grd A)}{\rmd  \grl}.
\ee
Thus, the area variation is written as
\be\label{areavariation}
\grd A=\int_{\mathcal{H}} \! \gru\rmd \grl \rmd A.
\ee
The evolution of the null geodesic congruence that generates the horizon is given by the Raychaudhuri equation
\be\label{Raychaudhuri}
\frac{\mathrm{d}\theta}{\mathrm{d}\lambda}=-\frac{1}{2}\theta^2-\sigma^2-R_{\mu\nu}k^{\mu}k^{\nu},
\ee
where $\gru ^2$ is the square of the expansion, $\sigma^2=\grs_{\grm\grn}\grs^{\grm\grn}$ is the square of the shear, and $R_{\grm\grn}$ is the Ricci tensor. Since the null geodesic congruence is hypersurface orthogonal, the rotation term $\omega _{\mu\nu}$ in (\ref{Raychaudhuri}) vanishes according to Frobenius' theorem.

One wishes to employ equilibrium thermodynamics. Hence, the application of local equilibrium conditions is required. To this end, one assumes that the expansion and shear vanish in a first order neighborhood of the point $p$. Then, the thermodynamic system is further specified as the degrees of freedom beyond the local Rindler horizon of $\mathcal{P}$. It is, thus, in local equilibrium at $p$. In this sense, equilibrium refers to the notion of local. Non-equilibrium thermodynamics refers to the notion of non-local and more precisely to the case that horizons have shear. 

The higher order $\gru ^2$ and $\grs^2$ terms are  neglected. The integration of equation (\ref{Raychaudhuri}) yields $\gru=-\grl R_{\grm\grn}k^{\grm}k^{\grn}$. Substituting this in equation (\ref{areavariation}) for the area variation one finds that
\be
\grd A=-\int_{\mathcal{H}} \! \grl R_{\grm\grn}k^{\grm}k^{\grn} \rmd \grl \rmd A.
\ee
Therefore,
\be\label{entropyvar}
\rmd S=-\gra \int_{\mathcal{H}} \! \grl R_{\grm\grn}k^{\grm}k^{\grn} \rmd \grl \rmd A.
\ee

Finally, the substitution of the Unruh temperature (\ref{sptemp}), the equation (\ref{heatflux}) for the heat flux and the equation (\ref{entropyvar}) for the entropy variation into the Clausius relation (\ref{clausius}) yields
\be
\int_{\mathcal{H}} \! T_{\grm\grn}k^{\grm}k^{\grn}\rmd \grl \rmd A=\frac{\hbar \gra}{2\grp}\int_{\mathcal{H}} \! R_{\grm\grn}k^{\grm}k^{\grn} \rmd \grl \rmd A.
\ee
This equation is valid only if
\be
T_{\grm\grn}k^{\grm}k^{\grn}=\frac{\hbar \gra}{2\grp}R_{\grm\grn}k^{\grm}k^{\grn}
\ee
for all null vectors $k ^{\grm}$. Since for any null vector $g_{\grm\grn}k^{\grm}k^{\grn}=0$, the above equation implies that
\be\label{htrht}
\frac{2\grp}{\hbar \gra}T_{\grm\grn}=R_{\grm\grn}+fg_{\grm\grn}
\ee
for some arbitrary function $f$. Next, one assumes the local conservation of energy and momentum, i.e., $\nabla ^{\grm}T_{\grm\grn}=0$. Then, one takes the divergence of (\ref{htrht}) and uses the contracted Bianchi identity $\nabla ^{\grm}R_{\grm\grn}=\frac{1}{2}\nabla _{\grn}R$ to find that $f=-\frac{1}{2}R+\Lambda$, where $\Lambda$ is some arbitrary constant. Eventually, one ends up with the Einstein's equation
\be\label{EEeqn}
R_{\grm\grn}-\frac{1}{2}Rg_{\grm\grn}+\Lambda g_{\grm\grn}=\frac{2\grp}{\hbar\gra}T_{\grm\grn},
\ee
with some undetermined cosmological constant $\Lambda$. The Newton's constant is determined as $G=(4\hbar\gra)^{-1}$ (recall equation (\ref{GR})). In this way, Einstein's equation can be thought as an equation of state. We note that since one can construct local Rindler horizons in all null directions in any spacetime point, Einstein's equation holds everywhere in spacetime.

%===========================================================
\subsection{Non-equilibrium thermodynamics of spacetime}

In the previous section, we presented the derivation of Einstein's equation from the equilibrium thermodynamics of spacetime. In this derivation, the entropy functional was taken to be by some constant proportional to the horizon's area. A question that arises, then, is whether possible curvature corrections to the entropy lead to a corresponding field equation with higher curvature terms. Indeed, in \cite{Eling06, Chirco09} it  is demonstrated that in the case that the entropy is proportional to a function of the Ricci scalar, one obtains the field equation of $f(R)$ gravity. However, in this case, the approaches of non-equilibrium thermodynamics of spacetime are required. In fact, it is possible for one to employ the non-equilibrium approach not only to the case of $f(R)$ gravity, but also to that of Einstein's gravity. In this section we present the above arguments.

To begin with, we consider that the entropy is proportional to the horizon area not only by some constant $\gra$, but also by a function $f(R)=1+\mathcal{O}(R)$ of the Ricci scalar. Then, the entropy variation is
\be\label{netentropyvar}
\grd S=\gra\int \! (\gru f+\dot{f})\rmd \grl \rmd A,
\ee
where $\dot{f}=\rmd f/\rmd \grl$ and the definition (\ref{expansion}) of expansion $\gru$ is employed. One notes that if the expansion $\gru$ vanishes at $p$, the above integral is non-zero, since $\dot{f}=f'(R)k^{\grm}\partial _{\grm}R$. The prime denotes differentiation with respect to the Ricci scalar.
As a consequence, the equilibrium  is not reached. In this case, one does not achieve the matching of the integral of (\ref{netentropyvar}) with that of $\grd Q/T$, since the latter is of order $\grl$ (see equation (\ref{entropyvar})). Thus, one redefines the equilibrium condition as
\be\label{equilibcond}
(\gru f +\dot{f})_p=0.
\ee
The 2-surface patch at $p$ must satisfy it. From the above equation, one concludes that if $f$ is not a function of the Ricci scalar,  but some constant, the entire approach is the same as that described in the previous section.

One finds the $\mathcal{O}(\grl)$ term in the equation (\ref{netentropyvar}) from the Taylor expansion of $(\gru f+\dot{f})$ around the point $p$
\be
(\gru f+\dot{f})= (\gru f+\dot{f})\big|_p+\grl (\dot{\gru}f-f^{-1}\dot{f}^2+\ddot{f})\big|_p + \mathcal{O}(\grl ^2) ,
\ee
where  equation (\ref{equilibcond}) was used to write $\gru=-f^{-1}\dot{f}$. Next, one employs the Raychaudhuri equation (\ref{Raychaudhuri}) and the geodesic equation $k^{\grm}\nabla _{\grm}k^{\grn}=0$ to write the $\mathcal{O}(\grl)$ term  as
\be
-k^{\grm}k^{\grn}(fR_{\grm\grn}-\nabla _{\grm}\nabla _{\grn}f+f^{-1}\partial _{\grm}f\partial _{\grn}f)-\frac{1}{2}f\gru^2-f\grs^2,
\ee
since $\rmd/ \rmd \grl=k^{\grm}\nabla_{\grm}$.
Then, employing one more time the equation (\ref{equilibcond}) one expresses the term $\gru ^2$ as $\gru^2=f^{-2}\dot{f}^2$. The $\mathcal{O}(\grl)$ is rewritten as
\be
-k^{\grm}k^{\grn}(fR_{\grm\grn}-\nabla _{\grm}\nabla _{\grn}f+\frac{3}{2}f^{-1}\partial _{\grm}f\partial _{\grn}f)-f\grs^2.
\ee
Finally, the $\mathcal{O}(\grl)$ term of the entropy variation (\ref{netentropyvar}) is written as
\be\label{netvariation}
\grd S=-\gra \grl \int \! k^{\grm}k^{\grn}(fR_{\grm\grn}-\nabla _{\grm}\nabla _{\grn}f+\frac{3}{2}f^{-1}\partial _{\grm}f\partial _{\grn}f)\rmd \grl \rmd A -\gra \grl \int \! f \grs^2 \rmd\grl \rmd A.
\ee

Let for the moment the shear of the horizon vanishes.  Then, the substitution of the equation (\ref{netvariation}) for the entropy variation,  the equation (\ref{heatflux}) for the heat flux and the equation (\ref{sptemp}) for the Unruh temperature into Clausius relation (\ref{clausius}) yields
\be\label{ofjhjg}
fR_{\grm\grn}-\nabla _{\grm}\nabla _{\grn}f+\frac{3}{2}f^{-1}\partial _{\grm}f\partial _{\grn}f+\Psi g_{\grm\grn}=\frac{2\grp}{\hbar \gra}T_{\grm\grn}.
\ee
One requires the conservation of energy and momentum to be valid. Hence, the divergence of the left hand side of equation (\ref{ofjhjg}) must vanish. It is
\be
\nabla ^{\grm}(fR_{\grm\grn}-\nabla _{\grm}\nabla _{\grn}f)= \partial _{\grn}(\frac{1}{2} \mathcal{L}-\square f),
\ee
where the relation for the commutator of covariant derivatives acting on a one-form $\grw_\grm$, i.e., $[\nabla^{\grm},\nabla_{\grn}]\grw_{\grm}=R_{\grm\grn}\grw^{\grm}$, and the contracted Bianchi identity $\nabla ^{\grm}R_{\grm\grn}=\frac{1}{2}\nabla _{\grn}R$ are employed. The Lagrangian $\mathcal{L}$ of the theory  is defined by $f=\rmd \mathcal{L} /\rmd R$. Now, in order for the left hand side of (\ref{ofjhjg}) to vanish it must 
\be
\Psi=\square f-\frac{1}{2}\mathcal{L}-\Theta,
\ee
where the gradient of $\Theta$ matches with 
\be
\partial _{\grn}\Theta=\nabla ^{\grm}\left(\frac{3}{2}f^{-1}\partial _{\grm}f\partial _{\grn}f\right).
\ee
However, the right hand side of the equation above is not the gradient of a scalar quantity. This contradiction implies that all the above arguments are inconsistent with energy conservation. It seems, then, that one cannot apply the thermodynamic argument of the previous section in the case considered here.

We note that the relation  between the affine time parameter $\grl $ and the Killing time parameter $\gry $ on a bifurcate Killing horizon is 
\be
\grl=-e^{-\grk \gry}.
\ee
Hence, the point $p$ corresponds to infinite Killing time parameter. Since $\rmd \grl / \rmd \gry=-\grk \grl $, for the killing vector $\grj ^{\grm}=-\grk \grl k^{\grm}$ it is
\be
\grj ^{\grm}=\left(\frac{\rmd \grl}{\rmd \gry}\right) k^{\grm}.
\ee
Then, the expansion in terms of the Killing parameter $\gry$ is 
\be
\tilde{\gru}=\left(\frac{\rmd \grl}{\rmd \gry}\right)\gru=\grk e^{-\grk \gry}\gru=-\grk \grl \gru.
\ee
The Taylor expansion of the Killing expansion $\tilde{\gru}$ around the point $p$ that stays in equilibrium is
\be
\tilde{\gru}=\tilde{\gru}_p+\grl\frac{\rmd \tilde{\gru}}{\rmd \grl}\Big|_p+\mathcal{O}(\grl ^2).
\ee
This shows that the Killing expansion vanishes as $\sim e^{-\grk\gry}$ if $\gru _p\neq 0$, while it vanishes as $\sim e^{-2\grk\gry}$ if  $\gru _p=0$. Since in the general case of a non vanishing expansion the approach to the equilibrium occurs at a slower rate, equilibrium thermodynamics and  Clausius relation may not hold. Non-equilibrium thermodynamics and the corresponding entropy balance equation
\be\label{entrbaleq}
\rmd S=\rmd ^eS+\rmd^i S
\ee
may be more appropriate. In this, $\rmd ^eS=\grd Q/T$ is the term concerning the exchange of entropy with the surroundings of a system, and $\rmd^i S$ is the entropy produced within it (see Appendix \ref{appendix a}). The latter term is of order $\grl$, since it vanishes at the equilibrium point $p$.

Indeed, the aforementioned problem is resolved if the entropy balance equation (\ref{entrbaleq}) holds, and there is an entropy production term
\be
\rmd^iS=-\frac{3}{2}\int \! \gra \grl f^{-1}\dot{f}^2 \rmd \grl \rmd A=-\frac{3}{2}\int \! \gra \grl f \gru ^2 \rmd \grl \rmd A
\ee
in it that cancels the problematic term in equation (\ref{ofjhjg}). If one substitutes the expansion $\gru$ by the Killing expansion $\tilde{\gru}$, the above entropy production term is written as
\be\label{cgghd}
\rmd^iS=\frac{3\gra}{2\grk}\int \! f \tilde{\gru}^2 \rmd \gry \rmd A.
\ee
This term closely resembles the entropy production term for a fluid system at temperature $T$ with coefficient
\be
\frac{\grz}{T}=\frac{3}{2\grk}\gra f,
\ee
where $\grz$ is the bulk viscosity (see equation (\ref{entropyproductiocterm})). This suggests that $\grz=3\hbar \gra f/4\grp$. 

Finally, with equation (\ref{cgghd}) being the entropy production term, the entropy balance equation yields
\be
fR_{\grm\grn}-\nabla _{\grm}\nabla _{\grn}f+\left(\square f-\frac{1}{2}\mathcal{L}\right)g_{\grm\grn}=\frac{2\grp}{\hbar\gra}T_{\grm\grn}.
\ee
The Newton's constant is again determined by $G=(4\hbar\gra)^{-1}$. This is the field equation of $f(R)$ gravity.

In the above derivation of the field equation, one assumed that the horizon's shear $\grs$ vanishes. However, if the shear is a non-zero quantity, there should be an additional entropy production term 
\be
\rmd^iS=-\int \! \gra \grl f \grs ^2 \rmd \grl \rmd A
\ee
in the entropy balance equation that cancels the corresponding term. In terms of the Killing shear $\tilde{\grs}$, that is,
\be
\tilde{\grs}=\left(\frac{\rmd \grl}{\rmd \gry}\right)\grs=\grk e^{-\grk \gry}\grs=-\grk \grl \grs
\ee
this entropy production term is written as
\be
\rmd^iS=\frac{\gra}{\grk}\int \! f \tilde{\grs}^2 \rmd \gry \rmd A.
\ee
This term, now, closely resembles the entropy production term for the fluid system with coefficient
\be
\frac{2\grh}{T}=\frac{\gra f}{\grk},
\ee
where $\grh=\gra f\hbar/4\grp$ is the shear viscosity. 

We note that a non-vanishing shear is also allowed in the case of general relativity presented in the previous section. Then, in order for one to derive the Einstein's equation, the non-equilibrium thermodynamics approach is required. In the case of general relativity, the internal entropy production term in the entropy balance equation that cancels the corresponding shear term is
\be
\rmd^iS=-\gra\grl\int \! \grs ^2\rmd \grl \rmd A.
\ee
In terms of the Killing shear the above term is written as
\be\label{tidalheat}
\rmd^iS=\frac{\gra}{\grk}\int \! \tilde{\grs}^2 \rmd \gry \rmd A.
\ee
As before, this corresponds to the entropy production term of equation (\ref{entropyproductiocterm}), with a shear viscosity
\be
\frac{2\grh}{T}=\frac{\gra}{\grk}.
\ee
This yields $\grh=a\hbar/4\grp=1/16\grp G$, which is identical to the value of shear viscosity found in the case of black hole horizons \cite{Price}. It also coincides with the universal relation for the viscosity to entropy density ratio found in the AdS/CFT context \cite{SonStarinets}. We note that in the case of general relativity a bulk viscosity does not appear (since $\gru_p=0$), in contrast to the case of black hole horizons, where one finds a negative bulk viscosity. 

Finally, we note that the equation (\ref{tidalheat}) can be written in the form
\be
T\rmd^iS=\frac{1}{8\grp G}\int \! \tilde{\grs}^2 \rmd \gry \rmd A.
\ee
This expression coincides with the Hartle-Hawking formula for the tidal heating of a black hole. The shear is related with the distortion of the horizon generators due to applied tidal fields. Thus, one can argue that the entropy production term, in this case, is directly associated with the work done on the horizon by a perturbative tidal field \cite{Chirco09}.

%=====================================================
\subsection{Summary and remarks}\label{section 2.3}

In this section, we presented the derivation of the gravitational field equations from the thermodynamics of spacetime in the cases of Einstein and $f(R)$ gravity. At first, we demonstrated the derivation of Einstein's equation from local equilibrium thermodynamics of spacetime. The key assumptions employed in the derivation are summarized as follows:
\begin{itemize}
\item One takes the temperature of the thermodynamic system--- which is defined as the degrees of freedom of a spacetime region beyond a local Rindler horizon--- to be the Unruh temperature.
\item As in the case of black hole entropy, one takes the entropy of the horizon to be proportional to its area.
\item All boost energy flow across the local Rindler horizon is  considered as heat flux.
\item Clausius relation $\grd Q=T\rmd S$ holds for all local causal horizons.
\end{itemize}
Then, one derives Einstein's equation from the Clausius relation along with the conservation of energy. In this way, Einstein's equation is viewed as an equation of state. Next, we showed that an allowed curvature correction---corresponding to the Ricci scalar--- to the entropy function leads to the field equations of $f(R)$ gravity. However, in the latter case, the non-equilibrium thermodynamics approach is required. Furthermore, we showed that the non-equilibrium approach is applied in the case of Einstein's gravity as well. 

Finally, we conclude with some remarks.

\begin{enumerate}

\item
A basic assumption in the derivation of the Einstein's equation (\ref{EEeqn}) in the equilibrium thermodynamics context is that the horizon shear vanishes. This argument seems to be restrictive, in order for one to consider the result general. Hence, the non-equilibrium approach, referring to the case of a local horizon with shear, should be thought as a more general case. However, this notion of non-equilibrium seems to be different to the corresponding notion of standard thermodynamics. The interpretation of Einstein's equation as an equation of state may differ from the familiar notion of equation of state in standard thermodynamics. In any case, the analogy between the Einstein's equation and an equation of state is required just by the Clausius relation along with the conservation of energy. It is not apparent that this analogy can be pushed further.

\item We saw that if one takes the entropy to be a function of the Ricci scalar, the field equations of $f(R)$ gravity are derived from local spacetime thermodynamics. Generally, Einstein's equation is thought to be the lowest order approximation of a field equation having higher curvature terms. One may expect, therefore, that such corrections to the field equations could be obtained by beginning with corrections to the horizon entropy, constructed from local curvature tensor and horizon geometry. Despite many attempts (see \cite{GuedJacob} and references therein), one finds the above argument to be correct only in the special case that entropy is a function of the Ricci scalar. However, there are physical reasons,
concerning ambiguities in the definition of approximate local boost Killing vectors, that suggest this failure of the thermodynamic argument. It seems, then, that one should not expect to understand corrections to Einstein gravity in this way \cite{Jacob12}.

\item The thermodynamic interpretation of Einstein's equation certainly suggests that gravity should not be considered as a fundamental theory, but rather as an emergent one. Then, the quantisation of gravity, in the sense of the canonically quantisation of the Einstein's equation, does not seem appropriate \cite{Jacob95,Hu,Pad27}.  In a similar sense, the quantisation, for example, of the macroscopic collective variables in the Navier-Stokes equation of hydrodynamics does not lead to a quantum theory of matter, despite the fact that the microscopic degrees of freedom (molecules) are described quantum mechanically. The thermodynamic argument suggests that the notion of quantum gravity should refer to a theory of the microscopic degrees of freedom of the spacetime, and not to the quantisation of a classical field.

\item In the interpretation of Einstein's equation as an equation of state, one assumes that the entropy is proportional to the horizon area. Next, we suppose that the entropy is defined as the entanglement entropy \cite{Solodukhin, BKLS, Srednicki}. Entanglement entropy results from the short distance correlations between the field degrees of freedom residing on the different sides of the horizon's boundary surface. This entropy is an infinite quantity in quantum field theory. However, if one introduces a UV cut-off, this quantity becomes finite. Then, the entanglement entropy is proportional to the area of the boundary surface (area law). It also depends on the number and the nature of the various quantum fields.

Bearing the features of entanglement entropy in mind, one makes an intriguing observation. If we assume  that the entanglement entropy is somehow rendered finite by some UV physics, the thermodynamic argument implies that the Einstein's equations can be regarded as an equation of state. One, also, founds that the Newton's constant is
\begin{equation}\label{newconst}
G=\frac{1}{4\hbar \gra}.
\end{equation}
Hence, the entropy density is $\gra=1/4G\hbar=1/4\ell_p^2$, in Planck length, and coincides with the Bekenstein-Hawking entropy $S_{\text{BH}}=A/4\ell_p^2$. Then, a theory without gravity certainly suggests an infinite entropy, whereas a finite entropy is consistent with the existence of gravity. Therefore, it seems that if the thermodynamic argument is valid, gravity somehow renders the entropy finite \cite{Jacob12}. 

The similarity between the entanglement entropy and the Bekenstein-Hawking entropy arises the question if the former is the microscopic origin of the latter. Nevertheless, two main discrepancies between these two definitions for the entropy are immediately apparent. First, entanglement entropy is a UV divergent quantity. Bekenstein-Hawking entropy, by contrast, is finite. Second, entanglement entropy, unlike BH entropy, is proportional to the number of field species existing in nature. This is often referred as the species problem. 

We note that the entanglement entropy may be an important quantity if gravity is an emergent theory. The reason is that one can determine entanglement entropy from non gravitational degrees of freedom. Even the entanglement entropy of a quantum field in flat spacetime obey the area law. It is also a purely geometrical quantity determined by the geometry of the boundary surface.

In \cite{Uglum}, it is argued that the entanglement entropy is a quantum correction to the Bekenstein-Hawking entropy. Then, there is a correspondence between the UV divergences in the entanglement entropy and the UV divergent terms in the effective gravitational action. The UV divergences in the entanglement entropy are absorbed in the renormalization of the couplings. Consequently, the Newton's constant appearing in the Bekenstein-Hawking formula is the renormalized one. 

The suggestion of \cite{Uglum} motivated a considerable amount of work\footnote{See for example \cite{Solodukhin, Solod94, FursaevSolod, Kabat, Solod95, Wilczek, Jacob13} and the references therein.}.
In \cite{Cooperman}, for example, it is shown the interpretation of the entanglement entropy as the Bekenstein-Hawking entropy in the special class of spacetimes possessing a bifurcate Killing horizon. However, the various regularization methods used in the literature do not always imply the desirable result. There are cases (for example, in the presence of gauge fields) that there is a mismatch between the renormalization of Newton's constant and that of entanglement entropy. The interpretation, also, of a bare gravitational constant is unclarified and seems not to have a microscopical origin. In any case, the entanglement entropy
\be
S\sim \frac{NA}{\ell _c(G_0,N)^2}
\ee
scales as the Bekenstein-Hawking entropy, provided that the renormalization works out such
\be
\frac{N}{[\ell _c(G_0,N)]^2}\sim\frac{1}{G_0\hbar },
\ee
where $\ell _c$ is the cut off scale, $N$ the number of field species and $G_0$ the low energy Newton constant  \cite{Jacob12}. 

In induced gravity \cite{Sakharov,Visser}, the Newton's constant is zero at tree level and the Einstein's gravity emerges from quantum field theory on a pseudo-Riemannian manifold at one loop order. In the case of induced gravity, all Bekenstein-Hawking entropy is interpreted as entanglement entropy \cite{JacobInd}. Hence, equation (\ref{newconst}) makes sense. However, there are induced gravity models that the Bekenstein-Hawking entropy is not equal to the entanglement entropy (see, for example, \cite{Frolov97, Frolov98}).  

The thermodynamic argument, by contrast, implies that if the entanglement entropy is finite, it is always equal to the Bekenstein-Hawking entropy. The gravitational coupling in the Bekenstein-Hawking formula is, then, the low energy Newton constant appearing in Einstein's equations. As a result, the thermodynamic argument seems  to reinforce the interpretation of Bekenstein-Hawking entropy as the entanglement entropy. In any case, the dependence of the entanglement entropy on the UV cut off does not allow its complete interpretation without the knowledge of the UV theory of gravity.
\end{enumerate}
%===================CHAPTER 3====================================
%=============================================================
\section{Thermodynamic aspects of gravity}\label{chapter3}

A few years after Jacobson's idea, Padmanabhan develops --- in a series of publications--- a programme for the interpretation of gravity as a thermodynamic and an emergent theory. Through his programme, Padmanabhan pinpoints several thermodynamic aspects of gravity. These aspects bring out several intrinsic features of the nature of gravity.

At first, Padmanabhan shows that in static spherically symmetric spacetimes, the Einstein's equation evaluated on the horizon, is viewed as the thermodynamic identity $T\rmd S=\rmd E+P\rmd V$ \cite{Pad02a, Pad06}. Then, he notices that the Einstein-Hilbert Lagrangian for gravity is decomposed into a surface and a bulk term that are holographically related. This means that there is a way for one to obtain the full Lagrangian of the bulk, only from the knowledge of the boundary term \cite{Pad3,Pad13,Pad14,Pad25}. Padmanabhan demonstrates that the full Einstein-Hilbert action represents the free energy of the spacetime \cite{Pad4}, while the surface term of the action, when evaluated on a horizon, represents its entropy \cite{Pad9}. Furthermore, he shows that (i) if one introduces Rindler horizons everywhere in spacetime and (ii) demands the entropy to be proportional to their horizons' area, then the gravitational action is determined in a uniquely way \cite{Pad4,Pad1}. It is also shown that the microscopic degrees of freedom residing on an horizon obey the equipartition law of energy \cite{Pad6,Pad24,Padequipartion}, and that the field equations of gravity can be viewed as an entropy balance condition \cite{PhysicalInterpretation}.  Finally, assuming that the spacetime is compared to a solid, Padmanabhan demonstrates that one obtains the Einstein's equation from the extremisation of  entropy function of the spacetime. The definition of this entropy function is motivated by the standard elasticity theory of solids \cite{Pad7,Pad16,Pad19}. In this section, we present the programme of Padmanabhan on the interpretation of gravity as a thermodynamic theory. For some reviews on the results of this programme, we also refer the reader to \cite{PadInsights,Pad21,Pad23,Pad27,Pad32}.

%==============================================================
\subsection{Einstein equation as a thermodynamic identity on the horizon}

We consider an arbitrary static and spherically symmetric spacetime with the line element 
\be\label{sphericalmetric}
\rmd s^2=-f(r)c^2 \rmd t^2+\frac{1}{g(r)}\rmd r^2+r^2(\rmd \gru ^2+\sin^2 \gru \rmd\grf^2).
\ee
The spacetime possesses an horizon at some location $r=a$, which is determined by the vanishing of the function $f(a)$. One associates a temperature $T$
\be
k_BT=\frac{\hbar c\sqrt{f'(a)g'(a)}}{4\pi}
\ee
with this horizon (see appendix \ref{imaginarytime}).

The conditions $g(a)=0$ and $f'(a)=g'(a)$ ensure that the singular behavior of the components of the metric at $r=a$ is not due to a true singularity of the spacetime geometry but due to a coordinate singularity.
As a result of these conditions, the expression for the temperature of the horizon takes the form $k_BT=\hbar cg'(a)/4\pi$. Furthermore, the components of the stress-energy tensor on the horizon satisfy the relations
\be
T_t^t|_{r=a}=T_r^r|_{r=a}, \qquad T_{\gru} ^{\gru}|_{r=a}=T_{\grf}^ {\grf}|_{r=a}.
\ee
Taking into account the above considerations, the Einstein's equation for the metric (\ref{sphericalmetric}) is given by $(1-g)-rg'(r)=-(8\pi G/c^4)Pr^2$ (see appendix \ref{appendix b}). Evaluated on the horizon $r=a$,
this equation is written as
\be\label{EEonhoriz}
\frac{c^4}{G}\left[\frac{1}{2}g'(a)a-\frac{1}{2}\right]=4\pi Pa^2,
\ee
where $T_r^r=P$ is the radial pressure of the source.

Next, we consider two such solutions at two different radii $a$ and $a+da$ for the horizon. Then, the multiplication of (\ref{EEonhoriz}) by $da$ and the introduction of an $\hbar$ factor in its first term allow one to write it in the form
\be\label{horizonequation}
\underbrace{\frac{\hbar cg'(a)}{4\pi}}_{k_B T}\underbrace{\frac{c^3}{\hbar G}d \left(\frac{1}{4}4\pi a^2\right)}_{k_B ^{-1}d S}\underbrace{-\frac{1}{2}\frac{c^4d a}{G}}_{-d E}=\underbrace{Pd\left(\frac{4\pi}{3}a^3\right)}_{Pd V}.
\ee
The braces underneath the above equation indicate the interpretation of each term. Thus, the Einstein's equation evaluated on the horizon (locally) is viewed as the thermodynamic identity  $TdS=dE+PdV$. Furthermore, from equation (\ref{horizonequation}), one reads off the expressions for the entropy $S$ and the energy $E$ of the horizon. These expressions are respectively
\be
S=\frac{1}{4\ell _p^2}(4\pi a^2)=\frac{1}{4}\frac{A}{\ell _p^2};\quad  E=\frac{c^4}{2G}a=\frac{c^4}{G}\left(\frac{A}{16\pi}\right)^{1/2},
\ee
where $A$ is the horizon area and $\ell _p=\sqrt{\hbar G/c^3}$ is the Planck length. The entropy associated with the horizon is one quarter of its area. The expression for the energy $E$ is identified with that of the  irreducible mass of a black hole. One can consider that the equation (\ref{horizonequation}) describes the connection between two quasi-static equilibrium states that both are spherically symmetric solutions of Einstein's equation. Both solutions have the same source $T_{\grm\grn}$ and temperature $T$. The radius of their horizon differs by $da$.

We note that while the temperature scales as $\hbar$ and the entropy as $1/\hbar$, the combination $TdS$ is independent of $\hbar$. This fact closely resembles standard thermodynamics where  the temperature scales as $k_B$ and the entropy as $1/k_B$, making $TdS$ independent of $k_B$. The effects emerging from the microstructure are suggested by the $k_B$ factor in the case of statistical mechanics and by the $\hbar$ factor in the case of (quantum) gravity respectively. However, this dependence disappears in the continuum limit of the thermodynamic description of the corresponding microstructure. The above conclusion seems to reinforce the emergent perspective of gravity. The usefulness of the thermodynamic description of spacetime resides in the fact that any possible quantum gravity model must be consistent with it, in an appropriate continuum limit.

The result presented in this section is not restricted only to the case of spherically symmetric spacetime in Einstein's gravity. In fact, the gravitational field equations, when evaluated on the horizon, reduce to a thermodynamic identity  in a wide class of models. Some examples of such models are the stationary axisymmetric horizons and the evolving spherically symmetric horizons in Einstein's  gravity \cite{Pad06}, the static spherically symmetric horizons in Lanczos-Lovelock gravity \cite{Pad06b}, and the  Friedmann-Robertson-Walker cosmological models in various gravity theories (e.g. \cite{CaiKim,Akbar,CaiCao,AkbarCai,Gong}).

%==============================================================
\subsection{Holographic structure of gravitational action}\label{holographicaction}

We consider the action functional $I$ given by the equation
\be\label{graction}
I= \frac{1}{16\pi}\int \! \rmd^4x \sqrt{-g} R +\int \! \rmd^4x \sqrt{-g} \mathcal{L}_{\text{matter}}(\grf^{\grm},\nabla_{\grl}\grf^{\grm}; g_{\grm\grn}),
\ee
where $R$ is the Ricci scalar and $\grf^{\grm}=\grf^{\grm}(x)$ a set of matter fields. The variation of the above action with respect to the metric yields \cite{Dirac}
\be
\grd I=\frac{1}{16\pi}\int \! \rmd^4x \sqrt{-g}\left[R_{\grm\grn}-\frac{1}{2}Rg_{\grm\grn}-8\pi T_{\grm\grn}\right]\grd g^{\grm\grn}.
\ee 
Then, employing the principle of least action $\grd I/\grd g^{\grm\grn}=0$, one obtains the Einstein's equation
\be
R_{\grm\grn}-\frac{1}{2}Rg_{\grm\grn}=8\pi T_{\grm\grn}.
\ee
The stress-energy tensor is defined as
\be
T_{\grm\grn}=-\frac{2}{\sqrt{-g}}\frac{\grd (\sqrt{-g}\mathcal{L}_m)}{\grd g^{\grm\grn}}.
\ee
Thus, one obtains the Einstein's gravitational equations from a variational principle, with an action given by (\ref{graction}). The term
\be\label{Einst-Hilb}
I_{\text{EH}}=\frac{1}{16\pi}\int \! \rmd^4x \sqrt{-g} R 
\ee
is known as the Einstein-Hilbert action. The variation of the Einstein-Hilbert action gives the Einstein's equations in vacuum.
%================================================================
%================================================================

We consider the Einstein-Hilbert Lagrangian of equation (\ref{Einst-Hilb}). If one replaces the general expression for the Ricci scalar $R=g^{\grm\grn}R_{\grm\grn}$, the Lagrangian is written in the form
\bea\label{decompose}
\sqrt{-g}R&=&\sqrt{-g}g^{\gra\grb}\left(\Gamma_{\grr\gra}^{\grm}\Gamma_{\grm\grb}^{\grr}
-\Gamma_{\gra\grb}^{\grm}\Gamma_{\grm\grr}^{\grr}\right)
+\partial_{\grs}\left[\sqrt{-g}\left(g^{\grm\grn}\Gamma_{\grm\grn}^{\grs}-g^{\grm\grs}\Gamma_{\grm\grl}^{\grl}\right)\right] \nonumber \\
&\equiv&\sqrt{-g}L_{\text{bulk}}+L_{\text{sur}},
\eea
where $L_{\text{bulk}}$ is a bulk term, which is quadratic in the Christoffel symbols $\grG^{\grm}_{\grn\grr}$ (or equivalently quadratic in the first derivatives of the metric $g^{\grm\grn}$), and $L_{\text{sur}}$ is a divergence term that can lead to a surface term by integration. The bulk and the surface term are related by the holographic relation
\be\label{holographicrelation}
\sqrt{-g}L_{\text{sur}}=-\partial_{\grs}\left[g_{\grm\grn} \frac{\partial \sqrt{-g}L_{\text{bulk}}}{\partial(\partial_{\grs}g_{\grm\grn})}\right].
\ee
The term ``holographic'' is used to denote that given the form of the action on a surface $\partial\mathcal{V}$, there is a way of obtaining the full action of the bulk $\mathcal{V}$. Indeed, one obtains the Einstein's equations from an action principle that uses only the surface term \cite{Pad9}.

In general, any scalar gravitational Lagrangian  of the form
\be\label{generalgrlagrangian}
\sqrt{-g}L=\sqrt{-g}Q_{\grm}^{\ \, \grn\grr\grs}R^{\grm}_{\ \, \grn\grr\grs}
\ee
is decomposed into a surface and  a bulk term
\be
\sqrt{-g}L=2\sqrt{-g}Q_{\grm}^{\ \, \grn\grr\grs}\grG^{\grm}_{\grs\grl}\grG_{\grn\grr}^{\grl}+ 2\partial_{\grr}\left[\sqrt{-g}Q_{\grm}^{\ \, \grn\grr\grs}\grG_{\grn\grs}^{\grm}\right]\equiv L_{\text{bulk}}+L_{\text{sur}}
\ee
that are related holographically (in D-dimensions)
\be
L_{\text{sur}}=-\frac{1}{[(D/2)-1]}\partial_{\grs}\left[g_{\grm\grn} \frac{\partial L_{\text{bulk}}}{\partial(\partial_{\grs}g_{\grm\grn})}\right].
\ee
 
The tensor $Q_{\grm}^{\ \, \grn\grr\grs}$ is made from the metric and the curvature tensor. It also has all the symmetries of the curvature tensor and zero divergence on all indices $\nabla_{\grm}Q^{\grm\grn\grr\grs}=0$.
The Lagrangian (\ref{generalgrlagrangian}) is the most general Lagrangian for gravity in D-dimensions that is consistent with the principle of equivalence (which allows gravity to be described by a metric), general covariance (which fixes the generic form of the action) and the requirement a well-defined variational principle to exist (which requires $\nabla_{\grm}Q^{\grm\grn\grr\grs}=0$)\cite{Pad13}.
%===============================================================
%===============================================================
\subsection{Surface term and horizon entropy}

If a solution of the Einstein's equation possesses a bifurcate horizon, the surface term in the gravitational action is related to the entropy of the horizon. One shows this relation by evaluating $I_{\text{sur}}$ on a surface infinitesimally away from the origin and taking the appropriate limit. 

Near the horizon, the spacetime is described by the Euclidean Rindler metric 
\be
ds^2=\grk^2\grj^2d\grt^2+d\grj^2+dx_{\perp}^2,
\ee
where one maps the horizon to the origin. Let also a surface $\grj=\epsilon$ be infinitesimally away from the origin in the $\grj -\grt$ plane. One obtains the contribution from the surface term 
\be
L_{\text{sur}}=\partial_{\grs}\left[\sqrt{-g}V^{\grs}\right]
\ee
of equation (\ref{decompose}), with
\be\label{vc}
V^{\grs}=\left(g^{\grm\grn}\Gamma_{\grm\grn}^{\grs}-g^{\grm\grs}\Gamma_{\grm\grl}^{\grl}\right)=-\frac{1}{g}\partial_{\grr}(gg^{\grr\grs}),
\ee
if integrates the term $\sqrt{h}n_\grs V^\grs$, where $\sqrt{h}=\grk\epsilon\sqrt{\grs}$ ($\grs$ is the determinant of the metric in the transverse coordinates), and $n_\grs$ the normal vector to the boundary. Then, one takes the limit $\epsilon \rightarrow 0$. We note that one obtains the second equality of (\ref{vc}) using the identities (\ref{Christoffel symbols}) concerning the Christoffel symbols. 

One finds that
\be
V^{\grj}\overset{\grj\to \epsilon}{=}-\frac{2}{\epsilon}.
\ee
Then, the integral of the surface term is
\bea
16\pi I_{\text{sur}}&=&\int_{\grj=\epsilon} \! \rmd^3x\sqrt{h}n_cV^c=\int^{2\pi/\grk}_{0}\! d\grt\int \! \rmd^2x (\grk\epsilon\sqrt{\grs})\left(-\frac{2}{\epsilon}\right) \nonumber \\
&=& -4\pi A,
\eea
where one takes the range of $\grt$ to be $(0,2\pi /\grk)$. Equivalently, 
\be
-I_\text{{sur}}=\frac{A}{4}.
\ee
Thus, the surface term in the gravitational action represents the entropy of the horizon. There is no ambiguity with the minus sign in the above equation. This sign arises from Euclidean continuation. 

%===============================================================
\subsection{Free energy of spacetime}

We consider a globally hyperbolic spacetime $(\mathcal{M},g)$ foliated by a family of spacelike hypersufaces $\Sigma_t$ \cite{Gourgoulhon}
\be
\mathcal{M}=\bigcup_{t\in \mathbb{R}}\Sigma_t.
\ee
Let $\grg$ and $\mathbf{K}$ be the induced metric and the extrinsic curvature respectively of the hypersurface $\Sigma_t$. Let also $n$ be the timelike unit vector normal to the leaf $\Sigma_t$. The 3+1 decomposition of the Ricci scalar $^4R$\footnote{The prefix 4 denotes a four-dimensional quantity.} is \cite{Gourgoulhon}
\be
^4R=R+K_{ij}K^{ij}-K^2-2\nabla_{\grm}(Kn^\grm)-\frac{2}{N}D_iD^iN,
\ee
where $K$ is the trace of the extrinsic curvature and $a^{\grm}=(0,a^i)$, with $a^i=D^iN/N$, is the acceleration of $\mathbf{x}=\text{constant}$ wordlines. Then, the Einstein-Hilbert action (\ref{Einst-Hilb}) is written as
\bea\label{action3+1}
I_{EH}= \frac{1}{16\pi}\int_{\mathcal{V}}\! \rmd^4x\sqrt{-g}\, ^4R&=&\frac{1}{16\pi}\int_{\mathcal{V}}\! \mathcal{L}\rmd^4x\sqrt{-g}
-\frac{1}{8\pi}\int_{\mathcal{V}}\! \nabla_\grm(Kn^\grm)\rmd^4x\sqrt{-g} \nonumber \\
&&-\frac{1}{8\pi}\int_{\mathcal{V}}\! \frac{1}{N}D_iD^iN\rmd^4x\sqrt{-g},
\eea
where one identifies $\mathcal{L}= N\sqrt{\grg}[R+K_{ij}K^{ij}-K^2]$ as the gravitational Lagrangian density of the ADM (Arnowitt-Deser-Misner) Hamiltonian formulation, with $N$ being the lapse function. The four volume $\mathcal{V}$ is defined as the part of $\mathcal{M}$ bounded by two hypesurfaces $\Sigma _{t_1}$ and $\Sigma _{t_2}$, i.e.,
\be
\mathcal{V}=\bigcup_{t=t_1}^{t_2}\Sigma _t.
\ee

We restrict ourselves to a static spacetime with an horizon. In the case of a static spacetime, the extrinsic curvature vanishes and the integration over the time coordinate $t$ becomes multiplication by $\grb=(2\pi)/\grk$. The divergent term takes the form
\be\label{spacetimeentropy}
-\frac{1}{8\pi}\int_{\partial \mathcal{V}}N(a^iu_i)\sqrt{\grs}\rmd^2x\rmd t=\frac{\grk}{8\pi}\int_0^{\grb}\! \rmd t
 \int \! \sqrt{\grs}\rmd^2x=\frac{1}{4}A,
\ee
where $\grs$ is the metric of the two-dimensional surface $\mathcal{S}_t=\partial \mathcal{V}\, \cap\, \Sigma_t$, i.e., the intersection of the hypersurface $\Sigma_t$ and the timelike boundary hypersurface $\partial \mathcal{V}$. The unit normal to this boundary is $u^i$. As the boundary approaches the horizon, the quantity $N(a^iu_i)$ tends to $-\grk$, with $\grk$ being the surface gravity of the horizon.  The horizon area is $A$.

In the Euclidean sector, the first term of equation (\ref{action3+1}) gives $\grb E$, where $E$ is the energy, in the sense of the integral of the ADM Hamiltonian over the spatial volume. The surface term gives the entropy of the horizon. Finally, equation (\ref{action3+1}) takes the form
\be
I_{\text{EH}}^{\text{Eucildean}}=\frac{1}{4}A-\grb E=(S-\grb E)=-\grb F,
\ee
where $F$ is the free energy (Helmholtz potential).  The minus sign arises from the Euclidean continuation of the action. Thus, we conclude that for any static spacetime with periodicity in Euclidean time, the gravitational action represents the free energy of the spacetime. Minimizing the Einstein-Hilbert action, one obtains the Einstein's equation. Equivalently, in the thermodynamic description, the Einstein's equation is obtained by the minimization of the free energy.
%=================================================================
\subsection{Gravity from spacetime thermodynamics}

In non-relativistic quantum mechanics, the  amplitude for a particle to travel from a point $(q_1,t_1)$ to another $(q_2,t_2)$ is given by
\be
\grc(q_2,t_2)=\int \! \rmd q_1K(q_2,t_2;q_1,t_1)\grc (q_1,t_1),
\ee
where the kernel (or propagator) is
\be
K(q_2,t_2;q_1,t_1)=\sum_{\text{paths}}\text{exp}\left[\frac{i}{\hbar}\int \! \rmd tL_q(q,\dot{q})\right].
\ee
The sum is over all possible paths connecting the two points. In the momentum space, the amplitude for a particle to go from  a point $(p_1,t_1)$ to another $(p_2,t_2)$ is determined by the Fourier transform
\be
G(p_2,t_2;p_1,t_1)=\int \!\rmd q_1 \rmd q_2K(q_2,t_2;q_1,t_1)\text{exp}\left[-\frac{i}{\hbar}(p_2q_2-p_1q_1)\right].
\ee
Hence,
\bea
G(p_2,t_2;p_1,t_1)&=&\sum_{\text{paths}}\int \!\rmd q_1 \rmd q_2 \, \text{exp}\left[\frac{i}{\hbar}\left\{\int \! \rmd t L_q-(p_2q_2-p_1q_1)\right\}\right]\nonumber \\
&=& \sum_{\text{paths}}\int \!\rmd q_1 \rmd q_2 \, \text{exp}\left[\frac{i}{\hbar}\int \! \rmd t \left\{L_q-\frac{\rmd}{\rmd t}(pq)\right\}\right]\nonumber \\
&=& \sum_{\text{paths}}\int \!\rmd q_1 \rmd q_2 \, \text{exp}\left[\frac{i}{\hbar}\int \! \rmd t L_p(q,\dot{q},\ddot{q})\right],
\eea
where
\be\label{momentlang}
L_p\equiv L_q-\frac{\rmd}{\rmd t}\left(q\frac{\partial L_q}{\partial \dot{q}}\right).
\ee
Thus, given any Lagrangian $L_q(q,\dot{q})$ involving  only up to first derivatives of the dynamical variables, it is possible the construction of another Lagrangian $L_q(q,\dot{q},\ddot{q})$ involving up to second derivatives, such that the latter describes the same dynamics. The only difference is the boundary conditions. In the latter case, one keeps the momenta fixed at the endpoints and not the coordinates.

Next, we consider that the dynamics of gravity are described by some--- unknown--- generally covariant action functional
\be
I=\int \! \rmd ^4x\sqrt{-g}L(g,\partial g)\equiv \int \! \rmd ^4x\sqrt{-g}L(g,\partial \Gamma).
\ee
The Lagrangian is a function of the metric $g_{\grm \grn}$ and its first derivatives $\partial _{\grs}g_{\grm \grn}$ (or equivalently of the Christoffel symbols $\Gamma ^{\grr}_{\grk \grl}$). According to (\ref{momentlang}), the equations of motions are also obtained from the action functional 
\bea\label{I'action}
I' &=& \int \! \rmd ^4x\sqrt{-g}L_\text{{bulk}}-\int \! \rmd ^4x\partial _{\grs}\left[g_{\grm \grn}\frac{\partial \sqrt{-g}L_\text{{bulk}}}{\partial (\partial _{\grs}g_{\grm\grn})}\right] \nonumber \\
&\equiv& I_\text{{bulk}} - \int \! \rmd ^4x\partial _{\grs}(\sqrt{-g}V^{\grs}) \equiv I_\text{{bulk}} - \int \! \rmd ^4x\partial _{\grs}P^{\grs},
\eea
where $V^{\grs}$ is constructed of $g_{\grm \grn}$ and $\Gamma^{\grr}_{\grk \grl}$. Furthermore, since the original Lagrangian is quadratic in the first derivatives of the metric, $V^{\grs}$ must be linear in the Christoffel symbols. We note that since the  Christoffel symbols vanish in a local inertial frame and the metric reduces to its Lorentzian form, the action $I_\text{{bulk}}$ cannot be generally covariant. However, $I'$ involves second derivatives of the metric and turns out to be generally covariant.

To proceed further, one needs to determine the quantity $V^{\grs}$. This quantity is linear in the Christoffel symbols and has a single index $\grs$. Hence, $V^{\grs}$ is obtained by the contraction of two of the indices on $\Gamma^{\grr}_{\grk \grl}$. Consequently, the most general choice for $V^{\grs}$ is the linear combination
\be
V^{\grs}=a_1g^{\grs\grl}\Gamma ^{\grz}_{\grl\grz} + a_2g^{\grr\grl}\Gamma ^{\grs}_{\grr\grl},
\ee
where $a_1$ and $a_2$ are some numerical constants. Then, using  the identities 
\be\label{Christoffel symbols}
\Gamma ^{\grz}_{\grl\grz}=\partial _{\grl}(\ln\sqrt{-g}), \quad g^{\grr\grl}\Gamma ^{\grs}_{\grr\grl}=-\frac{1}{\sqrt{-g}}\partial_{\grn}(\sqrt{-g}g^{\grn\grs}),
\ee 
the $P^{\grs}=\sqrt{-g}V^{\grs}$ is written in the form
\be
P^{\grs}=c_1g^{\grs\grn}\partial_{\grn}\sqrt{-g}+c_2\sqrt{-g} \partial_{\grn}g^{\grn\grs},
\ee
where one sets $c_1=a_1-a_2$ and $c_2=-a_2$ as numerical constants. 

Next, we consider a static spacetime, where $g_{\grm\grn}(t,\mathbf{x})=g_{\grm\grn}(\mathbf{x})$ and $g_{0\grn}=0$. Equivalence principle allows the construction of a local Rindler frame around any event $\mathcal{P}$ in spacetime. The acceleration of the observers is $a^i=(0,\mathbf{a})$. The most general static Rindler metric (the acceleration is chosen to be along $x-$axis) is written in the form
\bea\label{generalrindler}
\rmd s^2&=&-2a l\rmd t^2+\frac{\rmd l^2}{2a l}+(\rmd y^2+\rmd z^2)\nonumber \\
&=&-2a l(x)\rmd t^2+\frac{l'^2}{2a l(x)}\rmd x^2+(\rmd y^2+\rmd z^2),
\eea
where $l(x)$ is an arbitrary function and $l'\equiv (\rmd l/ \rmd x)$. In the second line, one makes a coordinate transformation from $l$ to some other variable $x$. The Rindler frame has an horizon at $l(x)=0$. This horizon is endowed with a temperature $T=a/2\pi$, where $a$ is the magnitude of the acceleration. Finally, one postulates that the horizon in the local Rindler frame has also an entropy that is proportional to its area, i.e.,
\be\label{entropypostulate}
\frac{dS}{dA_{\bot}}=\frac{1}{4\mathcal{A}_P},
\ee
where $\mathcal{A}_P$ is a fundamental constant with the dimensions of area. This finite constant represents the minimum area required to hold unit amount of information.

The surface term of the gravitational action (\ref{I'action}) in the static Rindler frame is
\be
I_{\text{sur}}=\int \! \rmd^4x\partial_{\grs}P^{\grs}=\int_{0}^{\grb} \! \rmd \grt \int_{\mathcal{V}} \rmd^3 x \nabla \cdot \mathbf{P}=\grb \int_{\partial \mathcal{V}} \! \rmd^2 x_{\bot}\hat{\mathbf{n}}\cdot \mathbf{P},
\ee
where--- dictated by the periodicity of the imaginary time--- the time integration is restricted to the interval $(0,\grb)$ with $\grb=2\pi/ a$. For the metric (\ref{generalrindler}), the only nonzero component of the surface term $P^{\grs}$ is
\be
P^{x}=2a \left[c_2+\frac{ll''}{l'^2}(c_1-2c_2)\right].
\ee
Hence, the surface term takes the form
\bea
I_{\text{sur}}&=&\grb P^x \int_{\partial \mathcal{V}} \! \rmd^2 x_{\bot}=\grb P^x A_{\bot}\nonumber \\
&=&4\pi A_{\bot}\left[c_2+\frac{ll''}{l'^2}(c_1-2c_2)\right]\nonumber\\ &\equiv& -S.
\eea
As aforementioned, this term is related to the entropy (see also \cite{Pad02a}). The minus sign arises from the Euclidean continuation. Then, using the postulate (\ref{entropypostulate}) for the entropy, one gets the condition
\be
\left[c_2+\frac{ll''}{l'^2}(c_1-2c_2)\right]=-\frac{1}{16\pi \mathcal{A}_P}.
\ee
The right hand side of the above equation is finite if $c_1=2c_2$. Thus, $c_2=-(16\pi \mathcal{A}_P)^{-1}$. 

Now, one substitute the two estimated constants $c_1$ and $c_2$ in the $P^{\grs}$ term. Then, going back to the Lorentzian sector, one gets
\bea
P^{\grs}&=&\frac{1}{16\pi \mathcal{A}_P}(2g^{\grs \grn}\partial_{\grn}\sqrt{-g}+\sqrt{-g}\partial_{\grn}g^{\grs\grn})=\frac{\sqrt{-g}}{16\pi \mathcal{A}_P}(g^{\grs \grl}\Gamma^{\grz}_{\grl\grz}-g^{\grr\grl}\Gamma^{\grs}_{\grr\grl})\nonumber\\
&=& -\frac{1}{16\pi \mathcal{A}_P}\frac{1}{\sqrt{-g}}\partial_{\grn}(gg^{\grn\grs}).
\eea
The second equality is obtained with the use of the identities (\ref{Christoffel symbols}).

The first order Lagrangian density is obtained by the solution of the equation
\be
\left(\frac{\partial \sqrt{-g}L}{\partial_{\grs}g_{\grm\grn}}g_{\grm\grn}\right)=P^{\grs}=\frac{1}{16\pi \mathcal{A}_P}(2g^{\grs \grn}\partial_{\grn}\sqrt{-g}+\sqrt{-g}\partial_{\grn}g^{\grs\grn}).
\ee
This equation is satisfied by the Lagrangian \cite{Padbook}
\be
\sqrt{-g}L_{\text{bulk}}=\frac{1}{16\pi \mathcal{A}_P}\left[\sqrt{-g}g^{\grr\grl}(\Gamma^{\grz}_{\grr\grh}\Gamma^{\grh}_{\grl\grz}-\Gamma^{\grh}_{\grr\grl} \Gamma^{\grz}_{\grh\grz})\right],
\ee
which is sometimes called the $\Gamma\Gamma$ Lagrangian for gravity. According to (\ref{I'action}), the final form of the Lagrangian is
\be
\sqrt{-g}L_{\text{grav}}=\sqrt{-g}L_{\text{bulk}}-\frac{\partial P^{\grs}}{\partial x^{\grs}}=\left(\frac{1}{16\pi \mathcal{A}_P}\right)\sqrt{-g}R.
\ee
This is the standard Einstein-Hilbert Lagrangian for gravity.  It is obtained by the postulate of entropy being proportional to the area of the horizon, the general covariance, the principle of equivalence and the quantum theory in the Rindler frame. We conclude that the surface term dictates the form of the Einstein-Hilbert Lagrangian in the bulk. This is an interesting realization of the holographic principle \cite{Pad4}.

Finally, we note that the differential geometric identity
\be\label{diffequat}
L_{\text{grav}}=L_{\text{bulk}}-\nabla_{\grs}\left[g_{\grm\grn} \frac{\partial \sqrt{-g}L_{\text{bulk}}}{\partial(\partial_{\grs}g_{\grm\grn})}\right]
\ee
implies that the important degrees of freedom in gravity are indeed the surface degrees of freedom. 
At any given event of spacetime, equivalence principle allows one to choose a local inertial frame.  In this inertial frame $L_{\text{bulk}}\sim \grG^2$ vanishes. However, the surface term in the right hand side of (\ref{diffequat}) does not vanish. This term depends on the second derivatives of the metric. Consequently, the left hand side of (\ref{diffequat}) does not vanish as well. In the local inertial frame, all the geometrical information is preserved by the surface term. The relevant to gravity degrees of freedom for a volume $\mathcal{V}$ reside in its boundary. Gravity is intrinsically holographic.

%===============================================================
\subsection{Equipartition of energy in the horizon degrees of freedom}

In thermodynamics, the equipartition theorem connects the number of the microscopic degrees of freedom with the macroscopic thermodynamic variables. In the simplest context of a gas, the equipartition theorem is written as
\be\label{standardequipart}
E=\frac{1}{2}k_B\int dn T,
\ee
where $dn$ is the number of the microscopic degrees of freedom in a certain amount of the gas at temperature $T$. The integral allows the microscopic degrees of freedom at different parts of gas to have different temperature. If the thermodynamic paradigm of gravity is correct, one may expect that there is a corresponding relation that connects the energy, the temperature, and the number of the degrees of freedom in some region of the spacetime, when some condition similar to thermodynamic equilibrium is satisfied. Next, we show that indeed one can find such a relation.

Foliation of spacetime implies the differential geometry relation \cite{horizons}
\be
Ru^\grm u^\grn=\nabla_\grr (Ku^\grr+a^\grr)-K_{\grm\grn}K^{\grm\grn}+K^2,
\ee
where $K_{\grm\grn}$ is the extrinsic curvature and $K$ its trace. In static spacetimes, this equation reduces to the
\be\label{gmst}
Ru^\grm u^\grn=\nabla_\grr a^\grr.
\ee
Employing the Einstein's equation, one writes  equation (\ref{gmst}) as
\be
\nabla_\grr a^\grr=8\pi G\left(T_{\grm\grn}-\frac{1}{2}Tg_{\grm\grn}u^{\grm}u^{\grn}\right).
\ee
Next, one integrates this equation over a four-dimensional region of spacetime. The three-dimensional spatial region is taken to be some compact volume $\mathcal{V}$, with boundary $\partial \mathcal{V}$. The time integration is restricted to the range $[0,\grb]$. One gets 
\be\label{equipartition}
S=\frac{1}{2}\grb E,
\ee\label{equipart}
where $E$ is the Tolman-Komar mass-energy defined as
\be
E=2\int_{\mathcal{V}}\! \rmd^3x\sqrt{\grg}N(T_{\grm\grn}-\frac{1}{2}Tg_{\grm\grn})u^{\grm}u^{\grn},
\ee
and $S$ is the gravitational entropy (\ref{spacetimeentropy}) for any static spacetime with a horizon, i.e.,
\be
S=\frac{\grb}{8\pi G}\int_{\partial \mathcal{V}}N(a^iu_i)\sqrt{\grs}\rmd^2x.
\ee
Equation (\ref{equipartition}) is written as
\be\label{gmst2}
E=\frac{1}{2}k_B\int_{\partial \mathcal{V}}\! \frac{\sqrt{\grs}\rmd^2x}{\ell ^2_p}\left\{\frac{Na^in_i}{2\pi}\right\}.
\ee
The equation (\ref{gmst2}) has the form of the equipartition law of energy (\ref{standardequipart}), i.e.,
\be
E=\frac{1}{2}k_B\int_{\partial \mathcal{V}}\! \rmd nT_{\text{loc}} \, ; \quad d n\equiv\frac{\sqrt{\grs}\rmd^2x}{\ell ^2_p},
\ee
where $T=NT_{\text{loc}}=(Na^in_i/2\pi)$ is the local Tolman temperature, and $d n$ the number density of microscopic degrees of freedom. We see--- once again--- that gravity is holographic, in the sense that the microscopic degrees of freedom scale as the proper area $\sqrt{\grs}d^2x$ of the boundary of the region and not as the volume. Such an equipartition law arises in any diffeomorphism invariant theory of gravity whenever the field equations hold \cite{Padequipartion}. 

We consider a quantum theory of gravity with a minimum quantum of length or area of the order of the Planck length $\ell^2_p\equiv G\hbar/c^3$. Then, a patch of a horizon with area $A$ is divided in 
\be
n=\frac{A}{c_1\ell^2_p}
\ee
microscopic cells, where $c_1$ is some numerical factor. If one supposes that each cell has $c_2$ degrees of freedom, the total number of states is $c_2^n$. The entropy of the patch of the horizon is
\be
S=n\ln c_2=4\frac{\ln c_2}{c_1}\frac{A}{4\ell^2_p}.
\ee
We note that if one sets $4\ln c_2/c_1=1$, the standard result $S=A/(4\ell^2_p)$ is recovered. In addition, according to the equipartition law of thermodynamics, the total energy of the cells is
\be
\mathcal{E}\equiv \frac{1}{2}nT=\frac{1}{2}\frac{ST}{\ln c_2}=\frac{2ST}{c_1}=\frac{E}{c_1},
\ee
where the condition $c_1=4\ln c_2$ and equation (\ref{equipart}) are used. Then, one concludes that if the patches are of the size $\ell^2_p$ (i.e., if one sets $c_1=1$), the equipartition energy of the horizon matches with the gravitational mass that produces the horizon. If one attributes an energy $(1/2)T$ to each patch of area $\ell^2_p$, the match between the equipartition energy and the gravitational mass is obtained even for other choices of $c_1$.

The use of the equipartition law in the non relativistic limit provides an interesting thermodynamic interpretation of gravity. We consider a spherical surface of area $A$ around a massive spherical body of mass $M$. One associates an entropy $S=A/(4\ell^2_p)$ with the degrees of freedom residing on the surface. The equipartition of the energy for the degrees of freedom is
\be
\mathcal{E}=\frac{1}{2}\frac{A}{\ell^2_p}k_BT=\frac{1}{2}\frac{A}{\ell^2_p}\frac{\hbar \grk}{2\pi c}=\frac{A}{4\pi}\frac{c^2\grk}{G},
\ee
where the expression of the temperature $k_BT=(\hbar \grk)/(2\pi c)$ is used. The energy of the massive body is $Mc^2$. Thus, the acceleration induced on a test body at rest on a surface of area $A$ is 
\be
\grk=GM\left(\frac{4\pi}{A}\right)=\frac{GM}{r^2}=\left(\frac{\mathcal{A}_pc^3}{\hbar}\right)\frac{M}{r^2}.
\ee
This is the Newton's law of gravity. The gravitational force is determined by the Planck area $\mathcal{A}_p=\ell^2_p$. If one keeps $\mathcal{A}_p$ constant and takes the limit $\hbar \to 0$, the coupling constant diverges. This divergence implies that gravity is intrinsically a quantum phenomenon. In a next section, we will see that Verlinde employs the equipartition law along with the holographic principle--- in a similar approach--- to show that gravity is an entropic force.

%==============================================================
\subsection{Gravitational equations as entropy balance condition}

We consider a generally covariant theory of gravity in $D$-dimensions described by the action
\be
I=\int \! \rmd^Dx\sqrt{-g}\left[L(g_{\grm\grn}, R_{\grm\grn\grr\grs})+L_{\text{matter}}(g_{\grm\grn},q_A)\right],
\ee
where the Lagrangian $L$ is some scalar built from the metric and the curvature tensor, and $L_{\text{matter}}$ is the Lagrangian of the matter. The matter Lagrangian depends on the metric and some matter variables denoted by $q_A$. For convenience, one assumes that $L$ does not involve derivatives of the curvature tensor. The equations of motion obtained by the variation of the above action with respect to the metric tensor are \cite{Padgravitation}
\be
 2E_{\grm\grn}=T_{\grm\grn},
\ee
 where
\be\label{emn}
E_{\grm\grn}=P_{\grm}^{\ \grr\grs\grl}R_{\grn\grr\grs\grl}-\frac{1}{2}Lg_{\grm\grn}-2\nabla^{\grr}\nabla^{\grs}P_{\grm\grr\grs\grn},
\ee
with
\be
P^{\grm\grn\grr\grs}\equiv \frac{\partial L}{\partial R_{\grm\grn\grr\grs} }.
\ee
The tensor $P^{\grm\grn\grr\grs}$ has the algebraic symmetries of curvature tensor. When  $P^{\grm\grn\grr\grs}$ obeys the additional condition $\nabla_{\grm}P^{\grm\grn\grr\grs}=0$, the field equations reduces to that of the Lanczos-Lovelock theory of gravity. The Lanczos-Lovelock theory of gravity is the most general extension of Einstein's gravity for which the equations of motion do not contain derivatives of the metric higher than second order.

For the considered generally covariant theories of gravity, the infinitesimal coordinate transformations $x^{\grm}\rightarrow x^{\grm}+\grj^{\grm}$ leads to the conservation ($\nabla_{\grm}J^{\grm}=0$) of the Noether current 
\be
J^{\grm}=-2\nabla_{\grn}(P^{\grm\grs\grn\grr}+P^{\grm\grr\grn\grs})+ 2P^{\grm\grn\grr\grs}\nabla_{\grn}\nabla_{\grr}\grj_{\grs}- 4\grj_{\grs}\nabla_{\grn}\nabla_{\grr}P^{\grm\grn\grr\grs},
\ee
\be
J^{\grm\grn}=2P^{\grm\grn\grr\grs}\nabla_{\grr}\grj_{\grs}-4\grj_{\grs}(\nabla_{\grr}P^{\grm\grn\grr\grs}).
\ee
The antisymmetric tensor $J^{\grm\grn}$ is introduced by the equation $J^{\grm}=\nabla_{\grn}J^{\grm\grn}$.
The expression for the Noether current $J^{\grm}$ reduces to
\be
J^{\grm}=2E^{\grm\grn}\grj_{\grn}+L\grj^{\grm}
\ee
if $\grj^{\grm}$ is a Killing vector that satisfies the conditions $\nabla_{(\grm}\grj_{\grn)}=0$ and $\nabla_{\grm}\nabla_{\grn}\grj_{\grr}=R_{\grr\grs\grm\grn}\grj^{\grs}$. The integral of $J^{\grm}$ over a spacelike surface defines the conserved Noether charge $\mathcal{N}$.

One associates with an horizon the entropy (called the Wald entropy \cite{Waldentropy}) 
\be\label{noetherentropy}
S_{\text{Noether}}\equiv \grb\mathcal{N}=\grb\int \! \rmd^{D-1}\Sigma_{\grm}J^{\grm}=
\frac{\grb}{2}\int \! \rmd^{D-2}\Sigma_{\grm\grn}J^{\grm\grn},
\ee
where $\grb=\grk/2\pi$ is the temperature of the horizon. The integral is over any $(D-2)$ dimensional surface that is a spacelike cross-section of the Killing horizon on which the norm of $\grj^{\grm}$ vanishes. 
The above expression for the entropy allows the interpretation of the term $\grb_{\text{loc}}J^{\grm}$ as an entropy density associated with the horizon, with $\grb_{\text{loc}}$ being the redshifted local temperature near the horizon. In the case of Einstein's theory of gravity, if one  choose $\grj^{\grm}$ to be a timelike Killing vector in a Schwarzschild spacetime, the entropy (\ref{noetherentropy}) reduces to the standard result $S_{\text{Noether}}=A/4$ .  However, in general, the entropy of the horizon is not proportional to its area, but depends on the theory.

We introduce at any point $\mathcal{P}$ of the spacetime a local inertial frame. Let $k^{\grm}$ be a future directed null vector at the point $\mathcal{P}$. By accelerating along $x-$axis with an acceleration $\grk$, one introduces a local Rindler frame from the usual transformations. Let $\grj^{\grm}$ be an approximate Killing vector corresponding to translation in the Rindler time such that the vanishing of $\grj^{\grm}\grj_{\grm}=-N^2$--- $N$ is the lapse function--- characterises the location of the local Rindler horizon $\mathcal{H}$. We consider also a timelike surface infinitesimally away from $\mathcal{H}$ with $N=\text{constant}$. This surface is usually referred as the stretched horizon. Let the spacelike unit normal to the stretched horizon be $r_{\grm}$, pointing in the direction of increasing $N$.

When matter with an amount of energy $\grd E$ gets close to the horizon (within a few Planck lengths), a local Rindler observer attributes a loss of entropy $\grd S=(\grk/2\pi)\grd E$. Furthermore, the local Rindler observer moving along the orbits of the Killing vector field $\grj^{\grm}$ with four velocity $u^{\grm}=\grj^{\grm}/N$ associates an energy $\grd E=u^{\grm}(T_{\grm\grn}\grj^{\grn})\rmd V_{\text{prop}}$ with a proper volume $\rmd V_{\text{prop}}$--- $T_{\grm\grn}\grj^{\grn}$ is the energy-momentum density. If energy is transfered across the horizon, the corresponding entropy transfer is  $\grd S_{\text{matter}}=\grb_{\text{loc}}\grd E$, where $\grb_{\text{loc}}=\grb N=(2\pi/\grk)N$ is the properly redshifted local Tolman temperature. Then, since $\grb_{\text{loc}}u^{\grm}=\grb\grj^{\grm}$, 
\be
\grd S_{\text{matter}}=\grb\grj^{\grm}\grj^{\grn}T_{\grm\grn}\rmd V_{\text{prop}}.
\ee
In addition, since one interprets $\grb_{\text{loc}}J^{\grm}$ as a local entropy density, $\grd S= \grb_{\text{loc}}u_{\grm}J^{\grm}\rmd V_{\text{prop}}$ is interpreted as the gravitational entropy associated with a volume $\rmd V_{\text{prop}}$ as measured by an observer with four-velocity $u^{\grm}$. Thus, for observers moving along the orbits of the Killing vector field $\grj^{\grm}$
\be
\grd S_{\text{grav}}=\grb [\grj^{\grr}\grj^{\grm}(2E_{\grm\grr})+L\grj_{\grr}\grj^{\grr}]\rmd V_{\text{prop}}.
\ee
Approaching the horizon, $\grj_{\grr}\grj^{\grr}\rightarrow 0$. Consequently,
\be
\grd S_{\text{grav}}=\grb [\grj^{\grr}\grj^{\grm}(2E_{\grm\grr})]\rmd V_{\text{prop}}.
\ee
In the same limit, $\grj^{\grr}\rightarrow \grk\grl k^{\grr}$, where $\grl$ is an affine parameter associated with the null vector $\grj^{\grm}$. 

Then, the entropy balance condition $\grd S_{\text{grav}}=\grd S_{\text{matter}}$ yields
\be\label{generalgrequat}
(2E^{\grm\grn}-T^{\grm\grn})k_{\grm}k_{\grn}=0.
\ee
The equation holds for all null vectors for all events in the spacetime. Hence, using the conditions $\nabla_{\grm}E^{\grm\grn}=0$ and $\nabla_{\grm}T^{\grm\grn}=0$, one gets $2E^{\grm\grn}-\grl g^{\grm\grn}=T^{\grm\grn}$, where $\grl$ is some constant. We note that the equation (\ref{generalgrequat}) has an additional symmetry. It is invariant under the shift $T^{\grm\grn}\to T^{\grm\grn}+\grm g^{\grm\grn}$, where $\grm$ is some arbitrary constant.

The Lanczos-Lovelock theory of gravity to the lowest order (the lowest order of the four rank tensor $P^{\grm \grn}_{\grr\grs}$ is given by equation (\ref{lowestorder})) corresponds to the Einstein's gravity. In this case, the entropy balance condition (\ref{generalgrequat}) takes the form
\be
(G^{\grm\grn}-8\pi T^{\grm\grn})k_{\grm}k_{\grn}=0,
\ee
where $G^{\grm\grn}$ is the Einstein tensor.

One can also interpret the above result in a different way. We consider a displacement of a local patch of the stretched horizon in the direction of $r_{\grm}$ by an infinitesimal proper distance $\epsilon$. This displacement changes the proper volume by $\rmd V_{\text{prop}}=\epsilon \sqrt{\grs}d^{D-2}x$, where $\grs$ is the determinant of the metric in the transverse space. The flux of energy through the surface is $T^{\grm}_{\grn}\grj^{\grn}r_{\grm}$. The corresponding entropy flux is obtained by multiplying the energy flux by $\grb_{\text{loc}}=\grb N$. Hence, the loss of the matter entropy to the outside observer, caused because the virtual displacement of the horizon has engulfed some matter, is 
\be
\grd S_{\text{matter}}=\grb_{\text{loc}}\grd E=\grb_{\text{loc}}T^{\grm\grr}\grj_{\grm}r_{\grr}\rmd V_{\text{prop}}.
\ee
Again, recalling that one interprets $\grb_{\text{loc}}J^{\grm}$ as a local entropy density, the change of gravitational entropy is $\grd S_{\text{grav}}= \grb_{\text{loc}}r_{\grm}J^{\grm}\rmd V_{\text{prop}}$. Equivalently,
\be
\grd S_{\text{grav}}=\grb N[r_\grn \grj^\grm (2E_\grm^\grn)+r_\grm\grj^\grm L]\rmd V_{\text{prop}}.
\ee
As the stretched horizon approaches the true horizon, $Nr_{\grm}\to\grj^{\grm}$ and $\grj^{\grm}\grj_{\grm}\to 0$. Thus, the entropy balance condition $\grd S_{\text{grav}}=\grd S_{\text{matter}}$ leads to equation (\ref{generalgrequat}).

%=============================================================

\subsection{Extremisation of spacetime's entropy functional}

In thermodynamics, one determines the equations that govern the equilibrium state of a system from the extrimisation of a suitable thermodynamic potential (e.g., entropy, free energy, enthalpy). This potential is a function of appropriate thermodynamic variables (e.g., volume, temperature). If gravity is a thermodynamic phenomenon, the definition of a suitable thermodynamic potential for the spacetime--- the thermodynamic system--- may be possible.  The extrimisation of this potential should lead to the gravitational field equations.

The motivation for the definition of a thermodynamic potential for the spacetime --- in the sense of determining the form of its function--- comes from elasticity theory. In elasticity theory, one introduces the notion of the displacement vector field $\grj^{i}(x)$ through the transformation $x^{i}\rightarrow x^{i}+\grj^{i}(x)$. The displacement vector field describes the deformation of a solid caused by an applied force. Then, the thermodynamic potentials are quadratic
in the gradient of the displacement vector field $\nabla \boldsymbol{\grj}$\footnote{For instance, the general expression for the free energy of a deformed isotropic body at some constant temperature is 
\bes
F=\frac{1}{2} K\Theta^2+\grm \Sigma_{ij} \Sigma_{ij},
\ees
where $\Theta=\nabla \cdot \boldsymbol{\grj}$ is the body's expansion, $\Sigma_{ij}=1/2(\nabla_i\grj_j+\nabla_j\grj_i)$ is the body's shear, $K$ is the bulk modulus, and $\grm$ is the shear modulus \cite{elasticity}.}. Their extremisation allows one to determine the equations that govern the elastic deformation.

Next, we consider that gravity is an emergent phenomenon, like elasticity. The spacetime is viewed as the coarse grained limit of some microscopic structure. Then, the diffeomorphism $x^{\grm}\rightarrow x^{\grm}+\grj^{\grm}(x)$ is analogous to the elastic deformation of the spacetime solid. In the latter case, however, one works in $D$-dimensions. In analogy with elasticity theory, one wishes to attribute a thermodynamic potential with a given spacetime deformation. Let this thermodynamic potential be the entropy (the reasoning for this choice will be apparent later). In elasticity theory, the extremisation of the entropy  leads to an equation for the displacement field. On the contrary, the extremisation of the entropy functional in the case of gravity should lead to the equations governing the background metric.

One expects the entropy functional of the spacetime to be an integral over a local entropy density. In the case of an elastic solid, a constant $\grj^{\grm}$, i.e., in the absence of external fields, does not contribute to the expression for the entropy density because of translation invariance. As a consequence, one expects the entropy density to be quadratic, to the lowest order, in the scalars constructed from derivatives of the deformation field $\grj^{\grm}$. Now, expecting this to be true in the case of gravity as well, the entropy density should have the form  $P^{\ \ \grr\grs}_{\grm\grn}\nabla_{\grr}\grj^{\grm}\nabla_{\grs}\grj^{\grn}$. The fourth rank tensorial object $P^{\ \ \grr\grs}_{\grm\grn}$ is made of the metric and other geometrical quantities like curvature tensor.  However, the presence of non-gravitational matter distribution in spacetime breaks the translation invariance. Hence, the entropy density can have quadratic terms $\grj^{\grm}$ as well. Let this contribution be denoted as $T_{\grm\grn}\grj^{\grm}\grj^{\grn}$. The second rank tensor $T_{\grm\grn}$ is determined by the matter distribution. It vanishes in the absence of matter. One also assumes that $T_{\grm\grn}$ is symmetric. Thus, one suggests that the entropy functional has the form
\bea\label{spacetimesentropy}
S[\grj^{\grm}]&=&S_{\text{grav}}[\grj^{\grm}]+S_{\text{matt}}[\grj^{\grm}]\nonumber \\
&=&-\int_{\mathcal{V}}\!\sqrt{-g}\rmd^Dx\left(4P^{\ \ \grr\grs}_{\grm\grn}\nabla_{\grr}\grj^{\grm}\nabla_{\grs}\grj^{\grn}-T_{\grm\grn}\grj^{\grm}\grj^{\grn}\right),
\eea
where $\mathcal{V}$ is a $D$-dimensional region in the spacetime with boundary $\partial \mathcal{V}$. The additional factors and signs in the above expression are introduced with hindsight.
 
In the case of elasticity theory, the coefficients of the quadratic terms are constants (e.g., the bulk and the shear modulus). One argues that the analogues of these coefficients in the case of gravity are denoted by divergence free quantities, i.e., one postulates the conditions
\be\label{jppl}
\nabla_{\grn}P_{\grm}^{\ \, \grn\grr\grs} =0, \qquad \nabla_{\grm}T^{\grm\grn}=0.
\ee
The choice for the first of conditions (\ref{jppl}) also ensures that the equations resulting from entropy extremisation contain derivatives of the metric only up to second order.
Furthermore, one assumes  that the tensor $P^{\grm \grn\grr\grs}$ has the algebraic index symmetries of the Riemann curvature tensor. It is antisymmetric in its first two $(P^{\grm \grn\grr\grs}=-P^{\grn\grm \grr\grs})$ and last two indices $(P^{\grm \grn\grr\grs}=-P^{\grm \grn\grs\grr})$ respectively.  In addition,  $P^{\grm \grn\grr\grs}$ is symmetric under the interchange of the first pair of indices with the second ones $(P^{\grm \grn\grr\grs}=P^{\grr\grs\grm \grn})$.

In a complete theory, the explicit form of $P^{\grm \grn\grr\grs}$ is determined by the long wavelength limit of the microscopic theory, just as the elastic constants are determined from the microscopic theory of the lattice. Since such a theory is unknown, motivated by the approaches of the renormalization group, one expands $P^{\grm \grn\grr\grs}$ in powers of the derivatives of the metric
\be
P^{\grm \grn\grr\grs}(g_{\gra\grb},R_{\gra\grb\grg\grd})=c_1\overset{(1)}{P}\!\,^{\grm \grn\grr\grs}(g_{\gra\grb})+c_2\overset{(2)}{P}\!\,^{\grm \grn\grr\grs}(g_{\gra\grb},R_{\gra\grb\grg\grd})+ \dots,
\ee
where $c_1, c_2, \dots$ are coupling constants. The lowest order term is made only of the metric. The only fourth rank tensor $P^{\grm \grn\grr\grs}$ that has the symmetries of the curvature tensor, is divergence free, and at the same time is built of the metric tensor only is the
\be\label{lowestorder}
\overset{(1)}{P}\!\,^{\grm \grn}_{\grr\grs}=\frac{1}{16\pi}\frac{1}{2}\grd^{\grm\grn}_{\grr\grs}=\frac{1}{32\pi}\left(\grd^{\grm}_{\grr}\grd^{\grn}_{\grs}-\grd^{\grm}_{\grs}\grd^{\grn}_{\grr}\right).
\ee
With hindsight, a constant coefficient is added. The next term in the expansion, besides being made from $g_{\gra\grb}$, depends linearly on the curvature tensor as well. This term gives the Gauss-Bonnet correction. One expects the third term to be quadratic in curvature and so on. In the most general case, the m-th order term that satisfies the imposed constraints is given by
\be
\overset{(m)}{P}\!\,_{\grm \grn}^{\grr\grs}\propto\grd_{\grm \grn\grn_3\ldots\grn_{2m}}^{\grr\grs\grm_3\ldots\grm_{2m}}R^{\grn_3\grn_4}_{\grm_3\grm_4}\cdots R^{\grn_{2m-1}\grn_{2m}}_{\grm_{2m-1}\grm_{2m}}=\frac{\partial \mathcal{L}_m}{\partial R_{\ \ \grr\grs}^{\grm\grn}},
\ee
where $\grd_{\grm \grn\grn_3\ldots\grn_{2m}}^{\grr\grs\grm_3\ldots\grm_{2m}}R^{\grn_3\grn_4}_{\grm_3\grm_4}$ is the alternating tensor.  One can also express this term as a derivative of the m-th order Lanczos-Lovelock Lagragian
\be
\mathcal{L}=\sum_{m=1}^Kc_m\mathcal{L}_m, \quad \mathcal{L}_m=\frac{1}{16\pi}2^{-m}\grd^{\grm_1\grm_2\ldots\grm_{2m}}_{\grn_1\grn_2\ldots\grn_{2m}}R^{\grn_1\grn_2} _{\grm_1\grm_2}R^{\grn_{2m-1}\grn_{2m}}_{\grm_{2m-1}\grm_{2m}}.
\ee
The term $m=1$ leads to Einstein gravity. The $m=2$ term gives the Gauss-Bonnet correction.

A particular feature of the null surfaces in spacetime is that they act as a one-way membranes that block informations to a certain class of observers. Characteristic examples of such null surfaces are the event horizons of the black holes and the Rindler horizons perceived by uniformly accelerated observers in the Minkowski spacetime. One introduces such local Rindler horizons in any point of spacetime. One, then, expects that any deformation of a local patch of a null surface changes the amount of information --- and the amount of entropy--- accessible to this class of observers. The above fact motivates one to associate an entropy functional (\ref{spacetimesentropy}) with any null hypersurface in the spacetime, with $\grj^{\grm}$ being the normal to these hypersurfaces. In the followings, we extremise the expression $S[\grj^{\grm}]$ with respect to variations of the null vector field $\grj^{\grm}$.

The variation of the entropy functional $S$ with respect to the null vector field $\grj^{\grm}$, after adding a Lagrange multiplier $\grl$ for the constraint $\delta(\grj_{\grm}\grj^{\grm})=0$, gives
\be
-\delta S=2\int_{\mathcal{V}} \! \rmd^Dx\sqrt{-g}\left[4P^{\ \ \grr\grs}_{\grm\grn}\nabla_{\grr} \grj^{\grm}\left(\nabla_{\grs}\delta \grj^{\grn}\right)-T_{\grm\grn}\grj^{\grm}\delta\grj^{\grn}-\grl g_{\grm\grn}\grj^{\grm}\delta\grj^{\grn}\right],
\ee
where the symmetries of $P^{\grm\grn\grr\grs}$ and $T^{\grm\grn}$ are used. Integrating by parts and using the condition $\nabla_{\grn}P_{\grm}^{\ \, \grn\grr\grs} =0$ one gets
\bea
-\delta S= && 2\int_{\mathcal{V}} \! \rmd^Dx\sqrt{-g}\left[-4P^{\ \ \grr\grs}_{\grm\grn}\left(\nabla_{\grs}\nabla_{\grr}\grj^{\grm}\right)-\left(T_{\grm\grn}+\grl g_{\grm\grn}\right)\grj^{\grm}\right]\delta \grj^{\grn}\nonumber \\
&& +8\int_{\partial\mathcal{V}} \! \rmd^{D-1}x\sqrt{h}\left[n_{\grs}P^{\ \ \grr\grs}_{\grm\grn} \left(\nabla_{\grr}\grj^{\grm}\right)\right]\delta \grj^{\grn},
\eea
where $n_{\grs}$ is the $D$-dimensional vector field normal to the boundary $\partial \mathcal{V}$, and $h$ is the determinant of the intrinsic metric on the boundary. As done in the usual process of the calculus of variations, one requires the variation $\delta \grj^{\grn}$ of the null vector field to vanish on the boundary. Then, the extremum principle $\grd S/\grd \grj^{\grn}=0$ impliesparallely
\be
2P^{\ \ \grr\grs}_{\grm\grn}\left(\nabla_{\grr}\nabla_{\grs}-\nabla_{\grs}\nabla_{\grr}\right)\grj^{\grm}
-\left(T_{\grm\grn}+\grl g_{\grm\grn}\right)\grj^{\grm}=0,
\ee
where one uses the antisymmetry of $P^{\ \ \grr\grs}_{\grm\grn}$ to rewrite the first term. Using the commutator of the covariant derivatives $[\nabla_{\grm}\nabla_{\grn}-\nabla_{\grn}\nabla_{\grm}]V^{\grr}=R^{\grr}_{\ \grs\grm\grn}V^{\grs}$, one writes
\be
\left(2P_{\grn}^{\ \grk\grl\grz}R^{\grm}_{\ \grk\grl\grz}-T^{\grm}_{\grn}+\grl\grd^{\grm}_{\grn}\right)\grj_{\grm}=0.
\ee
The requirement the above equation to hold for arbitrary null vectors $\grj^{\grm}$ implies that 
\be
2P_{\grn}^{\ \grk\grl\grz}R^{\grm}_{\ \grk\grl\grz}-T^{\grm}_{\grn}=F(g)\grd^{\grm}_{\grn},
\ee
where $F(g)$ is some scalar functional of the metric, with the $\grl$ being absorbed in its definition.  To the lowest order in the derivative expansion, $P^{\grm \grn\grr\grs}$ is given by equation (\ref{lowestorder}). The substitution of (\ref{lowestorder}) in the equation above gives
\be\label{gleichung}
\frac{1}{8\pi}R^{\grm}_{\grn}-T^{\grm}_{\grn}=F(g)\grd^{\grm}_{\grn}.
\ee
This equation is written in the form $(G^{\grm}_{\grn}-8\pi T^{\grm}_{\grn})=Q(g)\grd^{\grm}_{\grn}$, with $Q=8\pi F-(1/2)R$. Then, using $\nabla_{\grm}G^{\grm}_{\grn}=0$ and the condition $ \nabla_{\grm}T^{\grm}_{\grn}=0$, one gets $\partial_{\grn}Q=\partial_{\grn}[8\pi F-(1/2)R]$. Hence, $Q$ is an undetermined constant, say $\Lambda$. The function $F$ has the form $8\pi F=(1/2)R+\Lambda$.  Equation (\ref{gleichung}) takes the form
\be
R^{\grm}_{\grn}-\frac{1}{2}R\grd^{\grm}_{\grn}=8\pi T^{\grm}_{\grn}+\Lambda\grd^{\grm}_{\grn}.
\ee
This is the Einstein's gravitational equation with an undetermined cosmological constant $\Lambda$, if one identifies $T^{\grm}_{\grn}$ with the energy momentum tensor. Finally, we note that in the general case one obtains the equation 
\be
16\pi\left[P_{\grn}^{\ \grk\grl\grz}R^{\grm}_{\ \grk\grl\grz}-\frac{1}{2}\grd^{\grm}_{\grn}\mathcal{L}_m^{(D)}\right]=8\pi T^{\grm}_{\grn}+\Lambda \grd^{\grm}_{\grn}, 
\ee
which is identified with the field equations of Lanczos-Lovelock gravity. A cosmological constant arises as an integration constant as well.

It is worth pointing out that in the above approach one does not vary the metric tensor in order to obtain the gravitational equations. This is important, since in a thermodynamic interpretation of gravity, $g_{\grm\grn}$ is a derived macroscopic quantity and not a fundamental dynamical variable. This quantity provides a coarse grained description of the spacetime at macroscopic scales. 

Using the symmetries of $P^{\grm\grn\grr\grs}$ and the condition $\nabla_{\grn}P_{\grm}^{\ \, \grn\grr\grs} =0$, one finds that
\bea
4P^{\ \ \grr\grs}_{\grm\grn}\nabla_{\grr}\grj^{\grm}\nabla_{\grs}\grj^{\grn}&=&4\nabla_\grr[P^{\ \ \grr\grs}_{\grm\grn}\grj^\grm\nabla_\grs\grj^\grn]-4\grj^\grm P^{\ \ \grr\grs}_{\grm\grn}\nabla_\grr\nabla_\grs\grj^\grn\nonumber\\
&=& 4\nabla_\grr[P^{\ \ \grr\grs}_{\grm\grn}\grj^\grm\nabla_\grs\grj^\grn]-2\grj^\grm P^{\ \ \grr\grs}_{\grm\grn}\nabla_{[\grr}\nabla_{\grs]}\grj^\grn\nonumber\\
&=&4\nabla_\grr[P^{\ \ \grr\grs}_{\grm\grn}\grj^\grm\nabla_\grs\grj^\grn]-2\grj^\grm P^{\ \ \grr\grs}_{\grm\grn}R^\grn_{\ \,\grl\grr\grs}\grj^\grl\nonumber\\
&=&4\nabla_\grr[P^{\ \ \grr\grs}_{\grm\grn}\grj^\grm\nabla_\grs\grj^\grn]+\grj^\grm E_{\grm\grl}\grj^\grl.
\eea
Then, replacing the above expression in (\ref{spacetimesentropy}) and integrating, one writes the entropy functional as
\be
S[\grj^\grm]=-\int_{\partial \mathcal{V}}\! d^{D-1}x\sqrt{h}k_\grr(4P^{\ \ \grr\grs}_{\grm\grn}\grj^\grm\nabla_\grs\grj^\grn)-\int_{\mathcal{V}}\! d^Dx\sqrt{-g}[(2E_{\grm\grn}-T_{\grm\grn})\grj^\grm\grj^\grn],
\ee
where in the last equation the expression (\ref{emn}) for $E_{\grm\grn}$  and the condition  $\grj_{\grm}\grj^{\grm}=0$ are used. The above expression for the entropy functional implies that when the field equations hold, the total entropy of a bulk region is entirely on its boundary. 

The interpretation of the thermodynamic potential $S_{\text{grav}}$ as the gravitational entropy is justified as follows. When the potential is evaluated on-shell, it gives (we refer to the case of Einstein's theory of gravity) 
\bea
S|_{\text{on-shell}}&=&4\int_{\partial \mathcal{V}}\! d^{D-1}x\sqrt{h}n_{\grm}(P^{\grm\grn\grr\grs}\grj_{\grr}\nabla_{\grn}\grj_{\grs})\nonumber\\&\to& \frac{1}{8\pi}\int_{\partial \mathcal{V}}\! d^{D-1}x\sqrt{h}n_{\grm}(\grj^{\grm}\nabla_{\grn}\grj^{\grn}-\grj^{\grn}\nabla_{\grn}\grj^{\grm})\nonumber\\
&=& -\frac{1}{8\pi}\int_{\partial \mathcal{V}}\! d^{D-1}x\sqrt{h}n_{\grm}(Ku^{\grm}+a^{\grm}).
\eea
This expression has the familiar structure of  the surface term of (\ref{action3+1}), where $a^{\grm}=\grj^{\grn}\nabla_{\grn}\grj^{\grm}$ is the acceleration associated with the vector field $\grj^{\grm}$ and $K\equiv -\nabla_{\grn}\grj^{\grn}$ is the trace of the extrinsic curvature. One interprets the matter term $S_{\text{matt}}$ as the matter entropy transferred across a horizon.

\subsection{Summary and remarks}

In this chapter, we presented Padmanabhan's thermodynamic programme on gravity. The main results of this programme are summarised as follows.
\begin{itemize}

\item
In the case of a static and spherically symmetric spacetime, one interprets the Einstein's equation evaluated on the horizon
\bes
\underbrace{\frac{\hbar cg'(a)}{4\pi}}_{k_B T}\underbrace{\frac{c^3}{\hbar G}d \left(\frac{1}{4}4\pi a^2\right)}_{k_B ^{-1}d S}\underbrace{-\frac{1}{2}\frac{c^4d a}{G}}_{-d E}=\underbrace{Pd\left(\frac{4\pi}{3}a^3\right)}_{Pd V}
\ees
 as the thermodynamic relation $TdS=dE+PdV$ arising from virtual radial displacements of the horizon. This result is also demonstrated for a wide class of models, such as the stationary axisymmetric horizons and the evolving spherically symmetric horizons in Einstein's gravity, the static spherically symmetric horizons in Lanczos-Lovelock gravity and the Friedmann-Robertson-Walker cosmological models in various gravity theories.

\item
The Einstein-Hilbert Lagrangian for gravity is decomposed into a bulk and a surface term
\bea
\sqrt{-g}R&=&\sqrt{-g}g^{\gra\grb}\left(\Gamma_{\grr\gra}^{\grm}\Gamma_{\grm\grb}^{\grr}
-\Gamma_{\gra\grb}^{\grm}\Gamma_{\grm\grr}^{\grr}\right)
+\partial_{\grs}\left[\sqrt{-g}\left(g^{\grm\grn}\Gamma_{\grm\grn}^{\grs}-g^{\grm\grs}\Gamma_{\grm\grl}^{\grl}\right)\right] \nonumber \\
&\equiv&\sqrt{-g}L_{\text{bulk}}+L_{\text{sur}} \nonumber
\eea
that are related by the holographic relation
\bes
\sqrt{-g}L_{\text{sur}}=-\partial_{\grs}\left[g_{\grm\grn} \frac{\partial \sqrt{-g}L_{\text{bulk}}}{\partial(\partial_{\grs}g_{\grm\grn})}\right].
\ees
The term ``holographic'' is used to denote that given the form of the action on a surface $\partial\mathcal{V}$, there is a way for one to obtain the full action on the bulk $\mathcal{V}$. The surface term in the gravitational action, when evaluated on the horizon, represents the entropy of the horizon. Furthermore,  in the (3+1) formalism of gravity, in any static spacetime with periodicity in Euclidean time, the gravitational action represents the free energy of the spacetime. In addition, employing the general covariance, the principle of equivalence, the quantum field theory, and the postulate 
that the entropy of a Rindler horizon is proportional to its area, one uniquely determines the Einstein-Hilbert action for gravity.

\item
Gravitational field equations imply the law of equipartition
\bes
E=\frac{1}{2}k_B\int_{\partial \mathcal{V}}\! \frac{\sqrt{\grs}\rmd^2x}{\ell ^2_p}\left\{\frac{Na^in_i}{2\pi}\right\}\equiv \frac{1}{2}k_B\int_{\partial \mathcal{V}}\! \rmd nT_{\text{loc}}
\ees
in any static spacetime, allowing the determination of the density of the microscopic degrees of freedom. The result shows that gravity is holographic, in the sense that the microscopic degrees of freedom scale as the area of the boundary of a region and not as the volume.

\item
Local entropy balance condition $\grd S_{\text{grav}}=\grd S_{\text{matter}}$, in terms of thermodynamic variables perceived by local Rindler observers, leads to the field equations of gravity
\bes
(G^{\grm\grn}-8\pi T^{\grm\grn})k_{\grm}k_{\grn}=0,
\ees
for all null vectors $k_{\grm}$. The addition of a cosmological constant through the transformation $T^{\grm\grn}\to T^{\grm\grn}+\Lambda g^{\grm\grn}$ leaves the equations invariant. Gravity ignores the bulk vacuum energy. The cosmological constant arises as an integration constant. One can set any value to this constant, as a feature of the solution to the field equations.

\item
In the emergent perspective of gravity, the field equations are obtained by the extremisation of the entropy functional
\bea
S[\grj^{\grm}]&=&S_{\text{grav}}[\grj^{\grm}]+S_{\text{matt}}[\grj^{\grm}]\nonumber \\
&=&-\int_{\mathcal{V}}\!\sqrt{-g}\rmd^Dx\left(4P^{\ \ \grr\grs}_{\grm\grn}\nabla_{\grr}\grj^{\grm}\nabla_{\grs}\grj^{\grn}-T_{\grm\grn}\grj^{\grm}\grj^{\grn}\right) \nonumber
\eea
with respect to the variations of the null vector field $\grj^{\grm}$.
\end{itemize}
We note that the thermodynamic results above are not restricted only to the case of Einstein's theory of gravity, but are extended to the more general Lanckzos-Lovelock theory of gravity as well. Finally, we conclude with some remarks:
\begin{itemize}

\item
The results of this chapter imply that the really important degrees of freedom in gravity for a certain volume $\mathcal{V}$ reside in its boundary $\partial \mathcal{V}$. In fact, the surface term of the gravitational action dictates the bulk dynamics of gravity. Thus, we conclude that gravity is intrinsically holographic already at the classical level. For a similar conclusion see also \cite{anastopoulos}.

\item
Only the fact that one associates a temperature with spacetime ---in the sense of attributing temperature to a horizon--- is sufficient  to draw the conclusion that spacetime has microstructure. This implies that the general relativity should not be considered as a fundamental theory, but rather as an emergent one, obtained by averaging over some underlying microscopic degrees of freedom. The idea that spacetime is formed by the interactions of some underlying micro-constituents is not a new perspective in (quantum) gravity. It is believed that these constituents may be strings or loops for example. However, it was shown that one can describe these microscopic degrees of freedom--- whatever they may be and yet unknown--- thermodynamically, employing the usual laws and expressions of thermodynamics. Thus, any candidate theory of quantum gravity must be consistent with the thermodynamic description and explain the way that the macroscopic theory (and spacetime) arises from the interactions of the underlying microstructures.

The thermodynamic (or more generally emergent) description of gravity may enable us to gain some important insights into the nature of quantum gravity. If gravity is an emergent phenomenon, like thermodynamics or hydrodynamics, the dynamic variables like metric tensor should not be considered as fundamental variables, but  as macroscopic (collective) variables. These variables may not have any relevance in quantum gravity. Then, the quantisation of general relativity (in the sense of quantisation of the metric) makes no sense, since this only gives a theory of the quantised collective degrees of freedom, like in phonon physics. This quantisation does not lead to the quantum structure of spacetime. Finally, studying the thermodynamics of gravity and spacetime, one may find possible residues of the microscopic effects in the macroscopic theory. These residues may give us some clues about the nature of the microscopic theory. 

\item
Another conclusion that is drawn from the discussions until this point of the thesis is that thermodynamic variables are observer dependent. This is apparent from the fact that while an inertial observer attributes a zero temperature and a zero entropy to the Minkowski vacuum, a uniformly accelerated observer associates a finite temperature and entropy to the same vacuum. Already in quantum field theory in curved spacetime, particles are considered to be an observer dependent notion (see section \ref{Santa}). 
\end{itemize}
%===================CHAPTER 4====================================
%=============================================================
\section{Gravity as an entropic force}\label{chapter4}

An entropic force is a macroscopic force that originates in a system by the statistical tendency to increase its entropy. Recently, Verlinde argued that gravity is interpreted as an entropic force. In particular, considering the holographic principle to be a valid concept, Verlinde assumed that the information concerning the motion of the bodies are stored on  surfaces that cover the whole spacetime, the so-called holographic screens. Then, he showed that one obtains the law of the Newtonian gravity as a dimensional result from the relation $F\Delta x=T\Delta S$ of the entropic force and the equipartition theorem, if the temperature is taken to be the Unruh temperature. In addition, Verlinde showed that the Newton's law of inertia $F=ma$ is also obtained as a dimensional result from the relation of the entropic force. In this sense, the forces and generally the motion result from the existence of an entropy's gradient. Finally, he also provided a generalisation of the entropic interpretation of gravity in the case of general relativity.

In this section, we present Verlinde's proposal on the interpretation of gravity as an entropic force \cite{Verlinde,Sabine}. 

\subsection{Entropic force }

An entropic force is an effective macroscopic force that originates in a system with many degrees of freedom by the statistical tendency to increase its entropy. Typical examples of entropic forces are elasticity of polymers, osmotic forces, and depletion forces in suspensions, resulting from excluded volume effects.

We consider some polymer molecule, for example a rubber, a DNA molecule or some protein. The simplest model employed to describe a single polymer molecule is to consider it as a chain of repeated monomers of finite length that are joined together. Any kind of interaction among the monomers is neglected. Hence, each monomer is free to rotate around the points of attachment and orientate in any direction. Consequently, such a freely jointed polymer chain is found in various configurations, each of which has the same internal energy. A stretched polymer has less entropy than a coiled one, since the occupied volume of the configuration space for the coiled polymer is larger. 

\begin{figure}[t!]
\centering
\includegraphics[scale=0.4]{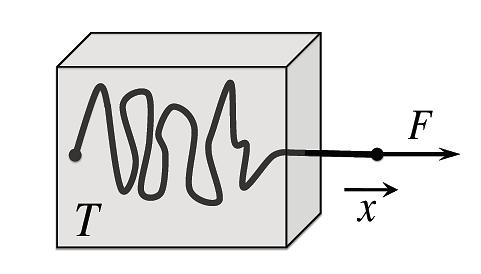}
\caption{A free jointed polymer is immersed into a heat bath with temperature $T$ and pulled out of its equilibrium state by an external force $F$. The entropic force points the other way.} 
\label{entropicforce}
\end{figure}

When a polymer is immersed into a heat bath, it forms randomly coiled configurations. These states are entropically favored. Its tendency to return to a maximal entropy state--- a direct consequence of the second law of thermodynamics--- gives rise to a macroscopic force, the entropic force. One brings the polymer out of its equilibrium state by exerting an external force $F$ on it. For simplicity, let the force be exerted on its one endpoint, in the direction of $x$-axis. One keeps the other endpoint kept fixed.  The entropic force points to the opposite direction (see figure \ref{entropicforce}). 

The entropy $S$ of the system is 
\be
S(E,x)=k_B\log \Omega(E,x),
\ee
where $\Omega(E,x)$ is the volume of the configuration space. The quantity $x$ denotes different polymer's configurations. In the micro-canonical ensemble given by $\Omega(E+Fx,x)$, one determines the entropic force imposing the extremal principle for the entropy
\be
\frac{d}{dx}S(E+Fx,x)=0.
\ee
The entropic force $F$ is then given by
\be
F=T\left(\frac{\partial S}{\partial x}\right)_E,
\ee
where the temperature is defined as
\be
\frac{1}{T}=\frac{\partial S}{\partial E}.
\ee
By the balance of forces, the external force $F$ is equal to the entropic force.  The entropic force tries to restore the polymer to its equilibrium position. In the case of a polymer, the entropic force is identified with the elastic force and has the form of the Hooke's law. An entropic force points in the direction of increasing entropy and it is proportional to the temperature. 

%================================================================
\subsection{Emergence of Newton's second law}\label{inertia}

We consider the holographic principle to be a valid concept. Hence, one assumes that the information (e.g., the position of a particle)  concerning the bulk of a region  is stored on surfaces or screens. On the one side of a screen, one considers that there is the part of the space that has already emerged. This part of space is described in the usual way. For example, one defines a coordinate system there. On the other side, there is the part that has not yet emerged--- there is no space yet. One considers that this part of space is described by some unknown microscopic degrees of freedom. The information associated with these degrees of freedom is stored holographically on the screen.

One assumes that the notion of time is well defined in the microscopic theory. The dynamics of the theory is then time translation invariant. Hence, the notion of the energy is defined by the Noether's theorem. The temperature is defined as the conjugate variable to the energy. From the number of the states, a canonical partition function is constructed. Hence, one derives the first law of thermodynamics. Next, one introduces some arbitrary macroscopic variable $x$. A number of microstates $\Omega(E,x)$ is then defined. A variable $F$ is also introduced as the thermodynamic dual to $x$. The first law of thermodynamics is $dE=TdS-Fdx$. Spacetime is emergent means that the space coordinate $x$ can be viewed as an example of such a macroscopic variable. If the number of the states depends on $x$, there is an entropic force when there is a finite temperature. 

\begin{figure}[b!]
\centering
\includegraphics[scale=0.3]{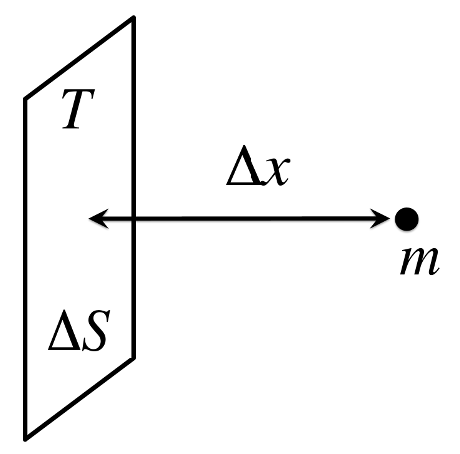}
\caption{A particle with mass $m$ approaches a patch of a holographic screen. The screen bounds the emerged part of space that contains the particle. The screen also stores the data that describe the part of the space that has not yet emerged and some part of the emerged space.} 
\label{verlinde1}
\end{figure}

We consider a particle of mass $m$ residing in the emerged part of the spacetime. This mass approaches a small piece of a holographic screen (see figure \ref{verlinde1}). Motivated by Bekenstein's derivation of the black hole entropy formula \cite{Bekenstein73}, one assumes that when the particle is one Compton wavelength 
\be
\Delta x =\frac{\hbar}{mc}
\ee
away from the screen, it causes a change of the screen's information by one bit. The corresponding change in entropy is
\be
\Delta S=2\pi k_B.
\ee
The normalization factor $2\pi$ is used for later convenience. Next, assuming that the change in entropy is linear in displacements, one writes
\be\label{entropychange}
\Delta S=2\pi k_B\frac{mc}{\hbar}\Delta x .
\ee

A force arises if one uses the analogy with osmosis across a semi-permeable membrane. When a particle has an entropic reason to be on the one side of the membrane, and the membrane is endowed with a temperature, it experiences an effective force, the entropic force, equal to
\be
F\Delta x=T\Delta S.
\ee
Then, if one takes the temperature of the screen to be the Unruh temperature
\be\label{screentemper}
k_BT=\frac{1}{2\pi}\frac{\hbar a}{c},
\ee
the Newton's second law 
\be
F=ma
\ee
is recovered. In this way, one interprets the Unruh temperature as the temperature $T$ required to cause an acceleration equal to $a$.

When the particle reaches the screen, it merges with the microscopic degrees of freedom on it. Hence, the particle is made up of the same bits as those that reside on the screen. Considering that each bit carries an energy $\frac{1}{2}k_B T$, as implied by the equipartition theorem, the number of the bits $n$ follows from
\be
mc^2=\frac{1}{2}nk_BT.
\ee
Then, employing the equations (\ref{entropychange}) and (\ref{screentemper}) one gets the relation 
\be\label{changeentropy}
\frac{\Delta S}{n}=k_B\frac{a\Delta x}{2c^2}
\ee
that concerns the entropy changes.
We conclude that there is a direct connection between the acceleration and the entropy gradient $\Delta S/\Delta x$. The absence of an entropy gradient implies a zero acceleration for some particle. Consequently, the law of inertia is viewed as follows:{\em A particle at rest, will stay at rest if there are no entropy gradients}.

Introducing the Newton's gravitational potential $\Phi$ (it turns out that the function $\Phi$ is indeed the gravitational potential), one writes
\be
a=-\nabla \Phi.
\ee
Thus, the relation (\ref{changeentropy})  takes the form
\be\label{keeptrack}
\frac{\Delta S}{n}=-k_B\frac{\Delta \Phi}{2c^2}
\ee
Another conclusion is drawn. The potential $\Phi$ monitors the depletion of the entropy per bit.

We consider a holographic screen with some amount of information (microscopic degrees of freedom) associated with the spacetime stored on its surface. Applying a proper coarse graining process, one gets a coarse grained version of the original screen with less information and greater entropy (entropy increases with coarse graining). The coarse graining process is repeated successive times. At each step, one gets a further coarse grained version of the original microscopic data. Hence, there is an emerging direction in space that corresponds to a coarse graining variable. Equation (\ref{keeptrack}) implies that this variable, which measures the amount of the coarse graining on the screens, is naturally identified with the potential $\Phi$ Then, the information on the screens is coarse grained in the direction of decreasing values of the $\Phi$. Thus, the screens correspond to equipotential surfaces. The spacetime manifold is foliated by a series of non-intersecting closed holographic surfaces. The time coordinate is defined microscopically on the screen.

%================================================================
\subsection{Newton's law of gravity}
We consider a spherically symmetric holographic screen. Taking into account the holographic principle, the number of the used bits on the screen is
\be
N=\frac{Ac^3}{G\hbar}.
\ee
An undetermined constant $G$ is introduced for dimensional reasons. This constant turns out to be the Newton's gravitational constant. However, at this point, one cannot make any speculation about the existence of gravity.

One assumes that the total energy of the system is $E$. One also assumes that this energy is divided evenly over the $N$ bits. Then, the temperature is determined by the equipartition law 
\be
E=\frac{1}{2}Nk_BT
\ee
as the average energy per bit. Furthermore, if $M$ is the mass that would emerge in the part of spacetime surrounded by the closed holographic surface, the energy is also $E=Mc^2$.

\begin{figure}[t!]
\centering
\includegraphics[scale=0.25]{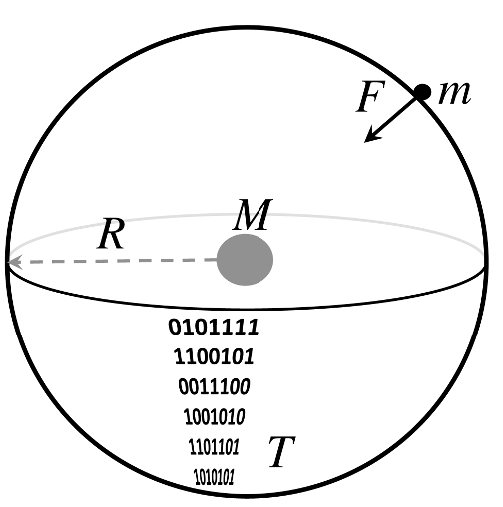}
\caption{A particle with mass $m$ approaches a spherically symmetric holographic screen. The energy is evenly distributed over the occupied bits. The energy is also equal to the mass $M$ that would emerge in the part of spacetime enclosed by the screen.} 
\label{sphersymm}
\end{figure}

Eventually, using equation (\ref{entropychange}) for the change of entropy caused by a mass $m$ (see figure \ref{sphersymm}) and $A=4\pi R^2$, one obtains
\be
F=G\frac{Mm}{R^2},
\ee
i.e., the Newton's law of gravity. The gravitational force is in fact an {\em entropic force}.

%================================================================
{\em General matter distributions}. We consider a general mass distribution that is enclosed by a holographic screen $\mathcal{S}$. This  screen is identified with an equipotential surface $\Phi$. Let $\grr (\mathbf{r})$ to be the mass density that describes the distribution. Taking the system to be in a local equilibrium one defines the temperature as
\be\label{omfg}
T=\frac{\hbar \nabla \Phi}{2\pi k_Bc}.
\ee
Temperature (\ref{omfg}) is obtained if one takes a test particle, moves it close to the screen and measures the local acceleration.  The density of the bits on the screen, which are uniformly distributed on it, are
\be
dN=\frac{c^3dA}{G\hbar}.
\ee
Then, employing the equipartition law of energy
\be
E=\frac{1}{2}k_B\int_{\mathcal{S}}\! TdN
\ee
and expressing it in terms of the total enclosed mass $M$, one takes
\be\label{kt}
M=\frac{1}{4\pi G}\int_{\mathcal{S}}\! \nabla \Phi dA.
\ee
This is the Gauss's law in its integral form. Equation (\ref{kt}) should hold generally for arbitrary equipotential surfaces. Thus, the potential $\Phi$ satisfies the Poisson equation
\be
\nabla^2 \Phi (\mathbf{r})=4\pi G \grr (\mathbf{r}).
\ee
Now, it seems natural for one to identify $\Phi$ with the Newtonian potential. All equations describing Newtonian gravity are recovered.

Finally, we consider a collection of test particles with masses $m_i$ located at arbitrary positions $\mathbf{r}_i$ outside a screen. The screen contains the general mass distribution. Bearing in mind equation (\ref{keeptrack}), one assumes that the change in the entropy density $\grd s$ locally on the holographic screen $\mathcal{S}$, due to infinitesimal displacements $\grd\mathbf{r}_i$ of particles, is
\be
\grd s=k_B\frac{\grd \Phi}{2c^2}dN.
\ee
The corresponding change $\grd \Phi$ in the Newtonian potential is determined by the variation of the Poisson equation
\be
\nabla^2\grd \Phi (\mathbf{r})=4\pi G\sum_i m_i \grd \mathbf{r}_i \nabla_i \grd (\mathbf{r}-\mathbf{r}_i).
\ee
Then, the work done by all the entropic forces on the particles is 
\be
\sum_i \mathbf{F}_i \cdot \grd\mathbf{r}_i=\int_{\mathcal{S}}\! T\grd s,
\ee
where $T$ is the local temperature.

%================================================================
\subsection{Newtonian gravity as an entropic force }\label{sabine}

The aforementioned emergent interpretation of gravity proposed by Verlinde can be recast in the following way \cite{Sabine}. The $\mathbb{R}^3$ space is covered by a continuous set of non-intersecting surfaces $\mathcal{S}$, the so-called holographic screens. One defines two scalar quantities $S$ and $T$ on the holographic screens; they are interpreted as the entropy and the temperature respectively.  Then, the theory is defined by the relation
\be\label{gfjg}
2G\int_{(\mathcal{S})}\! \grr dV=\int_{\mathcal{S}}\! TdA,
\ee
where $(\mathcal{S})$ is the volume enclosed by an arbitrary holographic surface $\mathcal{S}$ and $M=\int_{(\mathcal{S})}\! \grr dV$ is the total mass contained in this volume. The force acting on a test-mass $m$ is given by
\be
\mathbf{F}\cdot\grd \mathbf{r}=\int_{\mathcal{S}}\! T\grd dS,
\ee
where the integral is taken over a screen that does not include the test-mass, and $\grd x$ is a virtual displacement of the test-mass from its position that induces a change of the screen's entropy. In this way, Newtonian gravity results from the formula of the entropic force.

The inverse argument that Newtonian gravity can be viewed as an entropic force, is demonstrated as well. To this end, we consider a scalar field $\phi$ that obeys the Poisson equation. Then, there are surfaces where the field $\phi$ remains constant. These surfaces are identified with the holographic screens. Let, also, $\mathbf{n}$ be the normal vector in each point of an equipotential surface. Furthermore, let the corresponding surface $A(\grf)$ be assigned to every value of $\grf$ on $\mathcal{S}$. One normalises it to the unit area $A_0=G$. One defines the scalar functions $S$ and $T$ as
\be\label{entrpdefinition}
S(\mathbf{r}):=-\grf (\mathbf{r}) \frac{A}{2G}+S_0,
\ee
where $S_0$ some additive constant, and 
\be
T(\mathbf{r}):=\frac{1}{2\pi} \nabla_n \grf
\ee
respectively. In the case of gravity the scalar $S$ reduces to the usual black hole entropy, while $T$ to the horizon temperature of a black hole. On any equipotential surface the Gauss's law implies that
\be
\int_{(\mathcal{S})}\! \grr dV=\frac{1}{4\pi G}\int_{\mathcal{S}}\!  \nabla_n \grf dA= \frac{1}{2 G}\int_{\mathcal{S}}\! T dA.
\ee

Next, we consider  a mass $M$, with potential $\grf_M$, contained in a compact volume $(S)$ that is bounded by a surface $S$. We consider also a test particle with mass $m\ll M$ and with potential $\grf_m$,  located at some position $\mathbf{r}$ outside the volume. Let the outside volume, which is separated from the inside volume by the surface $S$, be denoted by $\mathbb{R}^3\backslash S$. The potential energy of the system is
\be
U=-\int \! \grr \grf dV.
\ee
 Consequently, the work needed to be done by some force $F$ in order for the test-mass $m$ to be displaced by $\grd \mathbf{r}$ is
\be
\mathbf{F}\grd \mathbf{r}=\grd U =-\int \! \grf_M \grd \grr dV=-\frac{1}{4\pi G}\int \! \grf_M\nabla^2\grd\grf_m dV,
\ee
where the integral is taken over some volume outside. 

Next, one writes the volume integral as an integral over all space $\mathbb{R}^3$ minus the integral over the inside $(S)$ and uses Green's second identity\footnote{ The Green's second identity is
\be
\int_{V}(\grc\nabla^2\grf-\grf\nabla^2\grc)dV=\int_{\partial V}(\grc\nabla_n\grf-\grf\nabla_n\grc)dA,
\ee
where $\mathbf{n}$ in the normal to surface $\partial V$ vector, while $\grc$ and $\grf$ are  two scalar fields.
} 
to write the volume integral over the inside as a surface integral 
\bea\label{dgfg}
-4\pi G\grd U=&&\int_{\mathbb{R}^3}\! \grf_M\nabla^2\grd\grf_m dV-\int_{(\mathcal{S})}\! \grd \grf_m \nabla^2 \grf_M dV\nonumber \\ &+& \int_{\mathcal{S}}\! (\grd\grf_m\nabla\grf_M-\grf_M\nabla\grd\grf_m)dA.
\eea
One employs the Gauss's law to write the second term of the integral over the equipotential surface $\grf_M$
as an integral over the volume $(S)$. Since there are no sources of $\grf_m$ inside the volume this integral vanishes. The integral over the volume $(S)$ is written as an integral over all space $\mathbb{R}^3$ minus an integral over the outside volume $\mathbb{R}^3\backslash S$. Since there are no sources of $\grf_M$ there, this integral vanishes as well. Equation (\ref{dgfg}) takes the form
\be
-4\pi G\grd U=\int_{\mathbb{R}^3}\! (\grf_M\nabla^2 \grd \grf_m-\grd\grf_m\nabla^2\grf_M)+ \int_{\mathcal{S}}\grd \grf_m\nabla \grf_MdA.
\ee

Then, one employs the Green's second identity to write the volume integral as a surface integral. Moving the surface to infinity this integral vanishes. Therefore,
\be
\mathbf{F}\grd \mathbf{r}=\frac{1}{4\pi G}\int_{\mathcal{S}}\grd\grf_m\nabla \grf_MdA.
\ee
At the equipotential surface $S$, equation (\ref{entrpdefinition}) implies that $2G\grd S=-A\grd\grf_m$. Hence, $2G\grd (dS)=\grd \grf_m dA$ for a surface element. Finally, one concludes that 
\be
\mathbf{F}\grd \mathbf{r}=\int_{\mathcal{S}}\! T\grd dS.
\ee
The Newtonian gravitational force is realized as an entropic force. 

The above arguments are reversed as follows. We define a holographic screen $\mathcal{S}$ as a surface of constant entropy that obeys the relation $2G\int_{(\mathcal{S})}\grr dV=\int_\mathcal{S} TdA$. The Newtonian potential is then defined as $\grf =-2GS/A$ , where $A$ is the area of the screen. A small change of $S$ on a constant surface implies $\grd\grf=-2GS/A(\mathcal{S})$. Finally, if $4\pi G\int_{(\mathcal{S})}\grr dV=\int_\mathcal{S} \nabla_n \grf dA$ holds for every surface $\grS$ with normal vector $\mathbf{n}$, one concludes that the density $\grr$ must obey the Poisson equation. The Newtonian gravity follows from an entropic force law. We note that unlike Verlinde's derivation, the number $N$ of the bits on the screens and  the equipartition theorem are not used here.

We note that the above discussion holds in the electrodynamics as well if instead of having test masses one has test charges. In this case, unlike gravity, the force is attractive between opposite charges. Then, temperature can be negative and entropy can decrease without one having to do work. Thus, it does not make sense the interpretation of the scalar quantities $S$ and $T$ as thermodynamic functions. However, the existence of negative gravitational charges is not excluded. In this case, one has to do work to bring to opposite charges closer, since in this process $dS<0$ .

%================================================================
\subsection{Relativistic generalisation}

We consider a static spacetime with a global timelike Killing vector field $\grj^{\grm}$. The generalization of the Newton's potential in the theory of general relativity is 
\be
\grf=\frac{1}{2}\ln (-\grj^{\grm}\grj_{\grm}).
\ee
The potential $\grf$ is used to define the foliation of space. The holographic screens are identified with surfaces of constant redshift.
In a way analogous to the Newtonian case, one defines the local temperature as
\be\label{Vtemper}
T=\frac{\hbar}{2\pi}e^{\grf}N^{\grm}\nabla_{\grm}\phi,
\ee
where $N^{\grm}$ is a unit outward pointing vector normal to the screen. A redshift factor $e^{\grf}$ is inserted since the temperature is measured with respect to a reference point at infinity. The acceleration is expressed as $a^{\grm}=-\nabla^{\grm}\grf$ .

Next, as in section \ref{inertia}, the change of entropy at the screen, for a displacement of a particle by one Compton wavelength normal to the screen, is $2\pi$ , i.e.,
\be
\nabla_{\grm}S=-2\pi \frac{m}{\hbar}N_{\grm},
\ee
where the minus sign comes from the fact that the entropy increases when we cross from the outside to the inside. The generalization of the entropic force is  
\be
F_{\grm}=T\nabla_{\grm}S=-me^{\grf}\nabla_{\grm}\phi,
\ee
which is the relativistic analogue of Newton's law $F=ma$.

We consider now a static mass configuration of total mass $M$ enclosed by a holographic screen of constant redshift $\grf$. As in the Newtonian case, the equipartition relation implies that
\be
M=\frac{1}{2}\int_{S}\! TdN.
\ee
The density of the bits on the screen is.
\be
dN=\frac{dA}{G\hbar}
\ee
Employing relation (\ref{Vtemper}) for the temperature one gets 
\be\label{komar}
M=\frac{1}{4\pi G}\int_{\mathcal{S}}e^{\phi}\nabla\phi dA,
\ee
which is the generalisation of Gauss's law to the case of general relativity. The right hand size of the above equation is identified  with the Komar mass.

The right hand size of (\ref{komar}) is expressed in terms of the Killing vector field $\grj^{\grm}$ and the Ricci tensor.  The left hand side is related to the stress energy tensor $T_{\grm\grn}$. Equation (\ref{komar}), eventually, takes the form \cite{Wald}
\be
M=2\int_{\grS}\! \left(T_{\grm\grn}-\frac{1}{2}Tg_{\grm\grn}\right)n^{\grm}\grj^{\grn}dV= \frac{1}{4\pi G} \int_{\grS}\! R_{\grm\grn}n^{\grm}\grj^{\grn}dV,
\ee
where $\grS$ is the three dimensional volume bounded by the holographic screen $\mathcal{S}$, and $n^\grm$ is its normal. The requirement this equation to hold for arbitrary screens and for all the Killing vectors in a local region of spacetime--- in a similar reasoning to Jacobson--- allows one to obtain the Einstein's equation.
%================================================================
\subsection{Summary and remarks }

In this section we presented Verlinde's interpretation of gravity as an entropic force caused by changes in the information ---and thus the entropy--- associated with the positions of material bodies. At first, we demonstrated the equivalence between the Newton's theory of gravity and an entropic force. Next, the entropic interpretation of gravity was generalised to the case of general relativity.

We conclude with some remarks:
\begin{itemize}

\item
The main problem concerning Verlinde's derivation of gravity from thermodynamic arguments is that the  expressions such as those of temperature and entropy lack of physical motivation. One should consider these expressions as general postulates. In any case, these expressions are results of general relativity. They demand, at first, a better understanding before one considers them as the starting point for the derivation of gravity from first principles. In addition, we cannot consider that the Einstein's equation is derived from first principles. The relativistic generalisation of the gravitational potential is used. This generalisation demands the already existing knowledge of general relativity. However, we find the interpretation of gravity as a force caused by entropy changes an intriguing idea that demands more investigation. We note that already from the results of the previous sections one could have concluded that gravity is driven by entropy changes. The close relation between gravity and entropy is explained according to Verlinde from the fact gravity is an entropic force.

\item
Some criticisms concerning Verlinde's interpretation rely on the fact that the equipartition theorem--- a key assumption in Verlinde's derivation--- is valid only in high temperatures.  However, we showed in section \ref{sabine} that the equipartition theorem is not needed in order for one to describe the Newton's law of gravity as an entropic force. In addition, other criticisms focus on the fact that the definition of temperature used in Verlinde's derivation is allowed  to take negative values. However, it is known that systems with an upper limit to their possible energy of their allowed states can have negative absolute temperatures \cite{negativetemperature,nat}. Systems that have negative temperature are hotter than positive temperature systems, i.e., if a system with negative temperature is in thermal contact with a system with positive temperature heat will flow from the negative to the positive temperature system.

\item After the publication of Verlinde's idea, a great amount of work has been done in an attempt possible cosmological implications of the entropic force scenario to be examined. Among these works it is worth mentioning the \cite{entropicaccelerating}. In this, it is argued that the observed accelerated expansion of the universe is due to an entropic force acting on its apparent horizon by virtue of its intrinsic temperature and entropy. The entropy is associated with the information holographically stored on its surface. We note that this approach is distinguished from the idea of gravity being itself an entropic force. Besides, in other works, as for instance \cite{RGCai,Shu}, the holographic principle, the equipartition law and the Unruh temperature are used in order for the Friedmann equation to be derived. However, in their majority, the papers found in the literature concerning cosmological implications of the entropic force scenario remain controversial.

\end{itemize}
%===================CHAPTER 5====================================
%=============================================================
\section{Conclusions}\label{chapter5}

The analogy between the laws of black hole mechanics and the laws of thermodynamics led Bekenstein and Hawking, in the 1970s decade, to argue that black holes should be considered as real thermodynamic systems that are characterised by entropy and temperature. In particular, Bekenstein argued that the entropy of a black hole equals to $S=(k_BAc^3)/(4G\hbar)$, where $A$ is the area of its horizon. In addition, Hawking showed  that the temperature of a black hole is $T=(\hbar\grk)/(2\pi ck_B)$, where $\grk$ is its surface gravity. 

Black hole thermodynamics shows  a deeper connection between thermodynamics and gravity. This perspective motivated several ideas that suggest an interpretation of gravity as a thermodynamic phenomenon. In this thesis, we examined the arguments of Jacobson, Padmanabhan and Verlinde that suggest an interpretation of gravity as a thermodynamic theory.

In section \ref{chapter2}, we examined the interpretation proposed by Jacobson of Einstein's equation as an equation of state. The idea is the following. In any point of spacetime, one introduces local Rindler horizons, as they are perceived by uniformly accelerated observers. A thermodynamic system is defined as the degrees of freedom residing in the region of the spacetime just beyond one of these horizons. The Einstein's equation is obtained from the demand the Clausius relation $\grd Q=T\rmd S$ to hold for all local Rindler horizons, and the conservation of energy. One takes the entropy $S$ to be proportional to the horizon's area. Furthermore, the heat $\grd Q$ and the temperature $T$ are the energy flux and the Unruh temperature respectively, as these are perceived by an accelerated observer just beyond the horizon. In this way, the Einstein's equation can be viewed as an equation of state.  If one assumes that the entropy is also proportional to a function of the Ricci scalar, the approach of the non equilibrium thermodynamics is required. One obtains the field equation of $f(R)$ gravity from the entropy balance condition $\rmd S=\grd Q/Τ+\rmd^i S$, where $\rmd^i S$ is the entropy produced inside the system. Such an entropy production term is allowed in the case of Einstein's gravity as well.

In section \ref{chapter3}, we presented Padmanabhan's programme  on the interpretation of gravity as a thermodynamic and by extension an emergent theory. At first, it is shown that in static spherically symmetric spacetimes, the Einstein's equation, evaluated on the horizon, is viewed as the thermodynamic identity. Then, one notices that the Einstein-Hilbert Lagrangian for gravity is decomposed into a surface and a bulk term that are holographically related. This means that there is a way for one to obtain the full Lagrangian of the bulk, only from the knowledge of the boundary term. It is demonstrated that the full Einstein-Hilbert action represents the free energy of the spacetime, while the surface term of the action, when evaluated on a horizon, represents its entropy. Furthermore, it is shown that (i) if one introduces Rindler horizons everywhere in spacetime and (ii) demands the entropy to be proportional to their horizons' area, then the gravitational action is determined in uniquely way. It is also shown that the microscopic degrees of freedom residing on an horizon obey the equipartition law of energy, and that the field equations of gravity can be viewed as an entropy balance condition.  Finally, assuming that the spacetime is compared to a solid, it is demonstrated that one obtains the Einstein's equation from the extremisation of spacetime's entropy function. The definition of this function is motivated by the standard elasticity theory of solids.

Finally, in section \ref{chapter4} we examined Verlinde's interpretation of gravity as an entropic force. The idea is the following.  Considering the holographic principle to be a valid concept, one assumes that the informations concerning the motion of the bodies are stored on some surfaces that cover the whole spacetime, the so-called holographic screens. Then, it is shown that one obtains the law of the Newtonian gravity as a dimensional result from the entropic force's relation $F\Delta x=T\Delta S$ and the equipartition theorem, if the temperature is taken to be the Unruh temperature. In addition, one shows that the Newton's law of inertia $F=ma$ is also obtained as a dimensional result from the relation of the entropic force. In this sense, the forces and generally the motion result from the existence of an entropy's gradient. Finally, one provides a generalisation of the entropic interpretation of gravity in the case of general relativity.

The study of the several thermodynamic aspects of gravity brings out various intrinsic features of gravity. Such features had not been pinpointed until nowadays, while their interpretation is not possible in the standard approaches of gravity. The conclusions that one can draw from the thermodynamic interpretation of gravity may offer a new window in the understanding of the nature of a possible quantum theory of gravity. 

We summarise the main conclusions of this thesis:

\begin{itemize}

\item
Gravity is intrinsically holographic already at the classical level of general relativity.

\item
Unruh effect and the mathematically similar phenomenon of thermal emission from black holes constitute a fundamental ingredient of theories suggesting a thermodynamic interpretation of gravity. For this reason, we strongly believe that the acceleration temperature requires a deeper conceptual understanding, which may be revealed by its experimental observation.

\item
Bearing in mind the results of the works of Jacobson, Padmanabhan and Verlinde, we can conclude that gravity is driven by entropy changes in a properly defined thermodynamic system.

\item
Since one attributes a temperature (and an entropy) to the spacetime, we draw the conclusion that spacetime is made of some--- yet unknown--- microscopic degrees of freedom. These degrees of freedom obey the usual laws and expressions of thermodynamics.
The thermodynamic interpretation of gravity clearly suggests that we should not consider gravity as a fundamental theory, but rather as an emergent one, obtained in the continuum limit of some underlying theory. 
Furthermore, we should not consider the dynamical variables like metric as fundamental variables, but as macroscopic collective variables. These variable  may not have any relevance in a theory of quantum gravity. Then, the canonically quantisation of the gravitation field makes no sense. This quantisation only gives a theory of the quantised collective degrees of freedom and does not lead to the quantum structure of spacetime (remind the case of phonons in the condensed matter physics). For instance, the quantisation of the--- collective--- variables in the Navier-Stokes equations of hydrodynamics makes no sense and does not lead to the quantum theory of matter, despite the fact that the microscopic degrees of freedom (molecules, atoms) are described quantum mechanically. Thus, as quantum gravity should be defined the theory that describes the microscopic structure of spacetime and matter (in a similar sense to statistical mechanics), and should not be referred, as usually, to the quantisation of a classical field.

\end{itemize}

%============================================

%=======================================================================
\begin{appendices}

\section{Classical Irreversible Thermodynamics}\label{appendix a}

In this appendix, we give a brief review of the so-called {\em classical irreversible thermodynamics}, used to describe non-equilibrium thermodynamics. We mainly focus on fluid systems. Further details are found in \cite{Mazur,Lebon}.

In non-equilibrium thermodynamics, a thermal system is usually inhomogeneous. The various physical quantities are functions of position and time. In order for a system to be described one makes the {\em local equilibrium hypothesis}. The local equilibrium hypothesis is formulated as follows. In a system out of equilibrium, there are sufficiently small regions (elemental volumes), so that the thermal equilibrium is accomplished in each of them. The elemental volumes are also sufficiently large, so that the microscopic fluctuations are negligible. The local and instantaneous relations between the thermodynamic quantities of the system are the same as the corresponding relations of a uniform system in equilibrium. For example, in the case of a $n$-component fluid system, the local equilibrium hypothesis implies that the specific entropy function $s(\mathbf{r},t)$ of the system is defined, i.e., $s(\mathbf{r},t)=s[\gry (\mathbf{r},t), u(\mathbf{r},t), c_k(\mathbf{r},t)]$. The specific entropy is a function of the specific volume $\gry (\mathbf{r},t)$, the internal energy $u(\mathbf{r},t)$, and the mass fraction $c_k(\mathbf{r},t)$ of the substance $k$. Then, the local first law of thermodynamics is
\be\label{1law}
T\frac{\rmd s}{\rmd t}=\frac{\rmd u}{\rmd t}+p\frac{\rmd \gry}{\rmd t}-\sum_{k=1}^n\grm _k \frac{\rmd c_k}{\rmd t},
\ee
where $T$ is the absolute temperature, $p$ is the hydrostatic pressure, $c_k=m_k/m$ is the mass fraction of substance $k$, and $\grm _k$ is the chemical potential of the substance. The specific volume $\gry$ is related to the mass density by $\gry=1/\grr$.

We consider a macroscopic system (continuum medium) with total mass $m$ and volume $V$, bounded by a surface $\Sigma$. We suppose that the system is out of equilibrium. The total entropy of the system at a time $t$ is $S$. The variation of the entropy is written as the sum
\be\label{entropybalance}
\rmd S=\rmd ^e S +\rmd ^i S,
\ee
where $\rmd ^e S$ is the exchange of entropy with the environment, and $\rmd ^i S$ is the entropy produced inside the system by several irreversible processes. In classical irreversible thermodynamics, the second law of thermodynamics takes the form $\rmd ^i S\geq 0$. In the case of a closed system, the exchange of entropy is $\rmd ^e S=\grd Q/T$, where $\grd Q$ is the heat supplied to the system. In the case of open systems, the $\rmd ^e S$ contains an additional term related to the transfer of matter (see equation (\ref{mattertrans})).

Next, we introduce the notion of the entropy flux $\mathbi{J} ^{\,s}$ (i.e., the entropy crossing the boundary surface per unit area and unit time), the rate of the entropy production $\grs ^s$ (i.e., the entropy produced per unit volume and unit time inside the system),  and the specific entropy $s$ (i.e., the entropy  per unit mass). Then, one writes
\be
\frac{\rmd ^e S}{\rmd t}=-\int_{\partial V} \! \mathbi{J} ^{\,s} \cdot \mathbi{n}\, \rmd \Sigma,
\ee 
\be
\frac{\rmd ^i S}{\rmd t}=\int_V \! \grs\, \rmd V,
\ee
\be
S=\int_V \! \grr s \,\rmd V,
\ee
where $\mathbi{n}$ is the unit normal pointing outwards to the volume of the system. Hence, the local entropy balance equation (\ref{entropybalance}) is
\be\label{leb}
\grr \frac{\rmd s}{\rmd t}=-\nabla\cdot \mathbi{J} ^{\,s} +\grs^s.
\ee
To obtain equation (\ref{leb}), we assumed that the local entropy balance equation is valid for any volume $V$ and used the Gauss's and the Reynolds' theorems. The second law of thermodynamics implies that
\be\label{scndlaw}
\grs ^s \geq 0,
\ee
where the equality holds for reversible processes. Equations  (\ref{leb}) and (\ref{scndlaw}) compose the generalised second law of thermodynamics for non-equilibrium thermodynamics.

In general, the entropy production term has the bilinear form
\be
\grs ^s= \sum_{\gra} J_{\gra} X_{\gra},
\ee
where $J_a$ are called the thermodynamic fluxes and $X_a$ the thermodynamic forces. The latter are related to the gradients of the intensive variables. The fluxes and the forces can be scalars, vectors or tensors. Furthermore, for a large class of irreversible processes, the fluxes are linear functions of the forces, i.e.,
\be
J_{\gra}=\sum_{\grb}L_{\gra\grb}X_{\grb},
\ee
where $L_{\gra\grb}$ are phenomenological coefficients. These coefficients depends on the intensive variables. Their values are restricted by the second law and various symmetry laws of the system. Such relations between fluxes and forces are called phenomenological relations. 

We consider a multi-component fluid system that exchanges not only heat but also matter with the environment (i.e., open system); $r$ chemical reactions occur among the system's constitutions. The system is also subjected to an external force. The first law of thermodynamics for this system is given by equation (\ref{1law}). The balance equations for the mass and the internal energy are  \footnote{The {\em material} or {\em substantial time derivative}
\begin{equation}
\frac{\rmd}{\rmd t}=\frac{\partial}{\partial t}+ \boldsymbol{\gry}\cdot \nabla
\end{equation}
is used to describe the rate of change of a variable (scalar or vector) in a velocity field $\boldsymbol{\gry}(\mathbf{r},t)$.}
\be\label{balancequat1}
\grr \frac{\rmd \gry}{\rmd t}=\nabla\cdot \boldsymbol{\gry},
\ee
\be\label{balancequat2}
\grr \frac{\rmd c_k}{\rmd t}=-\nabla \cdot \mathbi{J} ^{\,k}+\sum\limits^r_{j=1}\grn_{kj}\mathbi{J}, ^{\,j}
\ee
\be\label{balancequat3}
\grr \frac{\rmd u}{\rmd t}=-\nabla\cdot \mathbi{J} ^{\,q}-\mathbf{P}^{\text{T}}:\nabla \boldsymbol{\gry}+2(\mathbf{P}^{\text{v}})_{\text{a}}\cdot \grw+\sum\limits^n_{k=1} \mathbi{J} ^{\,k}\cdot \mathbi{F} ^{\,k},
\ee
where $\boldsymbol{\gry}(\mathbf{r},t)$ is the center of the mass velocity of the elemental volume, $\mathbi{J} ^{\,k}$ is the diffusion flux of the substance $k$, $\grn_{kj}\mathbi{J} ^{\,j}$ is the production of $k$ per unit volume in the $j$ chemical reaction, $\mathbi{J} ^{\,j}$ is called the chemical reaction rate of reaction $j$, $\mathbi{J} ^{\,q}$ is the heat flux, $\mathbf{P}^{\text{T}}$ is the transpose of the pressure tensor $\mathbf{P}$, $\grw$ is the mean angular velocity of the constituents at each point in the fluid, and $\mathbi{F} ^{\,k}$ is the external force per unit mass on the component $k$. The colon denotes double contraction, i.e., $\mathbf{A}:\mathbf{B}=A_{ij}B^{ij}$ for two second order tensors $\mathbf{A}$ and $\mathbf{B}$. 

The pressure tensor $\mathbf{P}$ is split into a reversible hydrostatic pressure $p\mathbf{I}$ ($\mathbf{I}$ is the identity tensor) and an irreversible viscous pressure tensor $\mathbf{P}^{\text{v}}$, i.e., $\mathbf{P}=p\mathbf{I}+\mathbf{P}^{\text{v}}$. The viscous pressure tensor, as a second order tensor, is decomposed into a symmetric part $(\mathbf{P}^{\text{v}})_{\text{s}}$ and an antisymmetric part $(\mathbf{P}^{\text{v}})_{\text{a}}$. The symmetric part is further decomposed into a trace part $p^{\text{v}}=\frac{1}{3}\mathrm{tr}(\mathbf{P}^{\text{v}})_{\text{s}}$ and a trace-free part $(\mathring{\mathbf{P}}^{\text{v}})_{\text{s}}$, i.e., $(\mathbf{P}^{\text{v}})_s=p^{\text{v}}\mathbf{I}+(\mathring{\mathbf{P}}^{\text{v}})_{\text{s}}$. Thus, the pressure tensor is written as
\be
\mathbf{P}=(p+p^{\text{v}})\mathbf{I}+(\mathring{\mathbf{P}}^{\text{v}})_{\text{s}}+ (\mathbf{P}^{\text{v}})_{\text{a}}.
\ee
Similarly, the velocity gradient tensor $\nabla \boldsymbol{\gry}$ is written as
\be
\nabla \boldsymbol{\gry}=\frac{1}{3}(\nabla \cdot \boldsymbol{\gry})\mathbf{I}+\mathring{\mathbf{V}}_{\text{s}}+\mathbf{V}_{\text{a}}.
\ee
The term $(\nabla \cdot \boldsymbol{\gry})$ is the trace of the velocity gradient tensor and describes the fluid's rate of expansion. The term $\mathring{\mathbf{V}}_{\text{s}}$ is the trace-free symmetric part of the velocity gradient tensor and describes the rate of shear. The term $\mathbf{V}_{\text{a}}$ is the antisymmetric part of the velocity gradient tensor and describes the rate of rotation. 

The substitution of the equations (\ref{balancequat1})-(\ref{balancequat3}) into the first law (\ref{1law}) yields the entropy balance equation (\ref{leb}). Then, the expressions for the entropy flux $\mathbi{J} ^{\,s}$ and the entropy  production $\grs ^s$ are
\be\label{mattertrans}
\mathbi{J} ^{\,s}=\frac{1}{T}\left(\mathbi{J} ^{\,q}-\sum\limits^n_{k=1} \grm _k\mathbi{J} ^{\,k}\right),
\ee
\bea\label{entropyproduction}
\grs ^s&=&\mathbi{J} ^{\,q} \cdot \nabla T^{-1}-\frac{1}{T}\sum\limits^n_{k=1}\mathbi{J} ^{\,k}\cdot \left[T\,\nabla \left(\frac{\grm _k}{T}\right)-\mathbi{F} ^{\,k}\right] -T^{-1}p^{\text{v}}(\nabla \cdot \boldsymbol{\gry})\nonumber \\&&-T^{-1}(\mathring{\mathbf{P}})_{\text{s}}^{\text{v}}:\mathring{\mathbf{V}}_{\text{s}}-\frac{1}{T}\sum\limits^r_{j=1} \mathbi{J} ^{\,j}A_j-T^{-1}(\mathbf{P}^{\text{v}})_{\text{a}}\cdot\left(\nabla \times \boldsymbol{\gry}-2\grw\right),
\eea
where $A_j=\sum\limits_{k=1}^n\grn_{kj}\grm_k$ is the chemical affinities of the $j$th reaction. In the special case of a single component isotropic fluid that (i) exchanges only heat with the surroundings and (ii) is not subjected to any external force, equations (\ref{mattertrans}) and (\ref{entropyproduction}) take the form
\be
\mathbi{J} ^{\,s}=\frac{1}{T}\mathbi{J} ^{\,q}
\ee
\be\label{entropyproductiocterm}
\grs ^s=\mathbi{J} ^{\,q} \cdot \nabla T^{-1}-T^{-1}p^{\text{v}}(\nabla \cdot \boldsymbol{\gry})-T^{-1}(\mathring{\mathbf{P}})_{\text{s}}^{\text{v}}:\mathring{\mathbf{V}}_{\text{s}}-T^{-1}(\mathbf{P}^{\text{v}})_{\text{a}}\cdot\left(\nabla \times \boldsymbol{\gry}-2\grw\right)
\ee
The fluxes $\mathbi{J} ^{\,q}$, $p^{\text{v}}$, $(\mathring{\mathbf{P}}^{\text{v}})_{\text{s}}$, $(\mathbf{P}^{\text{v}})_{\text{a}}$ of the entropy production term and the corresponding thermodynamic forces $\nabla T^{-1}$, $T^{-1}(\nabla \cdot \boldsymbol{\gry})$, $T^{-1}\mathring{\mathbf{V}}_{\text{s}}$, $\left(\nabla \times \boldsymbol{\gry}-2\grw\right)$ are related by the phenomenological equations
\bea
\mathbi{J} ^{\,q}&=&-\grl \nabla T,\\
p^{\text{v}}&=&-\grz \nabla \cdot \boldsymbol{\gry},\\
\mathring{\mathbf{P}}^{\text{v}}&=&-2\grh \mathring{\mathbf{V}}_s,\\
(\mathbf{P}^{\text{v}})_{\text{a}}&=&-\grh _r\left(\nabla \times \boldsymbol{\gry}-2\grw\right),
\eea
where the phenomenological coefficients are the heat conductivity $\grl$, the bulk viscosity $\grz$, the shear viscosity $\grh$ and the rotational viscosity $\grh _r$. The second law of thermodynamics implies that these coefficients have positive values.
\section{Static spherically symmetric spacetime}\label{appendix b}

We consider an arbitrary static and spherically symmetric spacetime with line element
\be\label{metr}
ds^2=-f(r)dt^2+h(r)dr^2+r^2(d\gru^2+\sin^2\gru d\grf^2).
\ee
The Christoffel symbols given by $\grG^{\grs}_{\grm\grn}=\frac{1}{2}g^{\grs\grr}\left(\partial_{\grm}g_{\grn\grr}+\partial_{\grn}g_{\grr\grm}-\partial_{\grr}g_{\grm\grn}\right)$ for the metric (\ref{metr}) are (only the non-vanishing ones are referred)
\bea
\grG^{r}_{rr}&=& \frac{h'}{2h}, \qquad \grG^{r}_{\gru\gru}=-\frac{r}{h}, \qquad \grG^{r}_{\grf\grf}=-\frac{r\sin^2\gru}{h}, \nonumber\\
\grG^{r}_{tt}&=&\frac{f'}{2h},  \qquad \grG^{\gru}_{\gru r}=\frac{1}{r}, \qquad \grG^{\gru}_{\grf\grf}= -\cos\gru\sin\gru, \\
\grG^{\grf}_{\grf r}&=&\frac{1}{r}, \qquad \grG^{\grf}_{\grf\gru}=\cot \gru, \qquad \grG^{t}_{tr}=\frac{f'}{2f}, \nonumber
\eea
where the prime stands for the derivative with respect to the coordinate $r$. The components of the Ricci tensor $R_{\grm\grn}=R^{\grl}_{\ \grm\grl\grn}$, where the Riemann tensor is $R^{\grr}_{\ \grs\grm\grn}=\partial_{\grm}\grG^{\grr}_{\grn\grs}-\partial_{\grn}\grG^{\grr}_{\grm\grs}+ \grG^{\grr}_{\grm\grl} \grG^{\grl}_{\grn\grs}-\grG^{\grr}_{\grn\grl} \grG^{\grl}_{\grm\grs}$, are
\be
\begin{aligned}
R_{tt}&= \frac{f''}{2h}+\frac{f'}{rh}-\frac{f'}{4h}\left(\frac{h'}{h}+\frac{f'}{f}\right), \\
R_{rr}&= \frac{f'}{4f}\left(\frac{h'}{h}+\frac{f'}{f}\right)+\frac{h'}{rh}-\frac{f''}{2f}, \\
R_{\gru\gru}&= R_{\grf\grf}= 1-\frac{1}{h}-\frac{r}{2h}\left(\frac{f'}{f}-\frac{h'}{h}\right).
\end{aligned}
\ee
The Ricci scalar $R=R^{\grm}_{\ \grm}$ is
\be
R=\frac{2}{r^2}\left(1-\frac{1}{h}\right)-\frac{f''}{fh}+\frac{f'}{2fh}\left(\frac{f'}{f}+\frac{h'}{h}\right)-\frac{2}{rh}\left(\frac{f'}{f}-\frac{h'}{h}\right).
\ee
The components of the Einstein tensor $G_{\grm\grn}=R_{\grm\grn}-\frac{1}{2}Rg_{\grm\grn}$ are 
\be
\begin{aligned}
G_{tt}&=f\left[\frac{1}{r^2}\left(1-\frac{1}{h}\right)-\frac{1}{r}\frac{d}{dr}\left(\frac{1}{h}\right)\right] \overset{g(r)\equiv1/h(r)}{=}f\left[\frac{1}{r^2}\left(1-g\right)-\frac{g'}{r}\right], \\
G_{rr}&=\frac{1}{r^2}(1-h)+\frac{f'}{rf},\\
G_{\gru\gru}&=\frac{r^2}{2h}\left(\frac{f''}{f}+\frac{f'}{rf}-\frac{f'^2}{2f^2}-\frac{h'}{rh}-\frac{f'h'}{2fh}\right),\\
G_{\grf\grf}&=\sin^2\gru \ \!G_{\gru\gru}.
\end{aligned}
\ee
Finally, the components of the stress-energy tensor $T_{\grm\grn}=(\grr+P)u_{\grm}u_{\grn}+g_{\grm\grn}P$, where $P(r)$ is the radial pressure and $\grr(r)$ the energy density, for a perfect fluid are
\be
\begin{aligned}
T_{tt} &= f\grr, \qquad T_{rr}=hP, \\
T_{\gru\gru}&=r^2P, \qquad T_{\grf\grf}=\sin^2\gru \ \!T_{\gru\gru}.
\end{aligned}
\ee

%================================================================
\section{Imaginary time, periodicity and horizon temperature}\label{imaginarytime}

We consider a general static spherically spacetime described by the line element
\be\label{sphericalspacetime}
ds^2=-f(r)c^2dt^2+\frac{1}{g(r)}dr^2+r^2d\Omega^2.
\ee
The Taylor expansion of the functions $f(r)$ and $g(r)$ near the horizon $r=a$ is
\be
f(r)\approx f'(a)(r-a), \qquad g(r)\approx g'(a)(r-a).
\ee
Hence, near the horizon, the line element (\ref{sphericalspacetime})  takes the form
\be
ds^2\approx -f'(a)(r-a)c^2dt^2+\frac{dr^2}{g'(a)(r-a)}+a^2d\Omega^2.
\ee
Next, we perform the change of variable
$
\grj=2\sqrt{\frac{r-a}{g'(a)}}
$
to write the line element above as
\be
ds\approx -\frac{f'(a)g'(a)}{4}\grj^2c^2dt^2+d\grj^2+a^2d\Omega^2.
\ee
Equivalently,
\be
ds^2\approx \underbrace{-\grk^2\grj^2dt^2+d\grj^2}_{2-\text{dimensional Rindler spacetime}}+\underbrace{a^2d\Omega^2}_{2-\text{sphere of radius}\ a},
\ee
where we set
\be\label{surfgravspherical}
\grk=\frac{\sqrt{f'(a)g'(a)}c}{2},
\ee
as the surface gravity of the horizon. Performing a Wick rotation $t=-i\grt$, we obtain the  Euclidean form of the metric
\be
ds_E^2\approx \grj^2d(\grk \grt)^2+d\grj^2+a^2d\Omega^2.
\ee
The singularity at the origin $\grj=0$ is a coordinate singularity provided that the imaginary time $\grt$ is periodic with period $\frac{2\pi}{\grk}$. We note that in this form of the metric the horizon is mapped to the origin.

Next, we consider a quantum scalar field $\grf(x)$ at the region near the horizon. The field is described by a Lagrangian $L$. The path integral for this field is defined as
\be\label{pathintegral}
\langle \grf_2(\mathbf{x})|e^{-iH(t_2-t_1)}|\grf_1(\mathbf{x})\rangle=\int \! \mathcal{D} \grf \exp\left[\frac{i}{\hbar}\int \! dt L\right],
\ee
where the integral is taken over field configurations restricted to $\grf_1(\mathbf{x})$ at $t=t_1$ and $\grf_2(\mathbf{x})$ at $t=t_2$. One writes the partition function of a canonical ensemble consisting of the field ($H$ is the Hamiltonian) at inverse temperature $\grb=(k_BT)^{-1}$ as the path integral 
\be\label{pathpartition}
Z=\text{tr} \left(e^{-\grb H}\right)=\int \! \mathcal{D}\grf \exp{\left[-\frac{1}{\hbar}\oint_{0}^{\grb\hbar}\! d\grt L_E\right]},
\ee
where $L_E$ is the Euclidean Lagrangian. The integral is taken over fields that are periodic in $\grt$, with period $\grb\hbar$. Comparing the two expressions above, we note that the path integral (\ref{pathintegral}) is equal to the partition function (\ref{pathpartition}), if one performs a Wick rotation (then it is $L_E\equiv -L(t=-i\grt)$) and takes the field configurations to be periodic in the imaginary time $\grt$ with period $\grb\hbar$. 

We saw that the time $\grt$ has also a period $\frac{2\pi}{\grk}$. Thus, one concludes that the fields are in equilibrium with the horizon at temperature\footnote{For a rigorous proof see \cite{horizons,UnruhWeiss}.}
\be
k_BT=\frac{\hbar \grk}{2\pi}.
\ee
According to (\ref{surfgravspherical}), one associates with the horizon of a spherically symmetric spacetime a temperature 
\be
k_BT=\frac{\hbar c\sqrt{f'(a)g'(a)}c}{4\pi}.
\ee

\end{appendices}
%======================================================================
%============BIBLIOGRAPHY=================================================

\end{document}